\documentclass{aa}  

\usepackage{graphicx}
\usepackage[colorlinks,urlcolor=blue,citecolor=blue,linkcolor=blue]{hyperref}
\usepackage{adjustbox}

\usepackage{txfonts}
\usepackage{subfigure}
\usepackage{graphicx,caption}

\def\R23{\mbox{$\rm R_{23}$}}

\begin{document}

 \title{Morpho-kinematics of MACS J0416.1-2403 low mass galaxies}


   \author{Ciocan, B. I.
          \inst{1}
         \and
          Ziegler, B. L.
          \inst{1}
          \and
         B{\"o}hm, A.
            \inst{1}
        \and
            Verdugo, M.
            \inst{1}
            \and 
            Maier, C.
           \inst{1}
}
   \institute{Institute for Astronomy (IfA), University of Vienna,
              T\"urkenschanzstrasse 17, A-1180 Vienna\\
              \email{bianca-iulia.ciocan@univie.ac.at} }
   \date{}

\abstract { We use optical integral field spectroscopy from VLT/MUSE, as well as photometric observations from  Hubble Space Telescope and VLT/HAWK-I, to study the morpho-kinematics of 17 low mass  ($\rm{log(M/M_{\odot})<9.5}$) MACS J0416.1-2403 cluster galaxies at $\rm{R_{200}}$ and 5 field galaxies with a redshift of z $\sim$0.4. \\ By measuring fluxes of strong emission lines from the MUSE data, we have recovered the star formation rates, gas-phase metallicities, spatially resolved gas kinematics and also investigated the ionising mechanisms. We have analysed the structure and morphology of the galaxies from the optical and infrared photometric data, performing a multi-component decomposition into a bulge and a disk. The spatially resolved gas velocity fields of the cluster members and field galaxies were modelled using a 3D approach, which allowed us to retrieve their intrinsic gas kinematics, including the maximum rotation velocity and velocity dispersion. This enabled us to study scaling relations such as the Tully--Fisher and the stellar mass -- $\rm{S_{0.5}}$ relation for low mass galaxies in different environments and to search for signatures of cluster-specific processes using disturbed gas velocity fields as tracers. \\
Most galaxies from our sample fall in the star forming and composite region in the BPT diagnostic diagram, which allows the ionising sources in a galaxy to be disentangled.  The cluster and field population can be classified as star forming main-sequence galaxies, with only a sub-sample of four quenched systems. We observe significant scatter for the cluster galaxies in the mass-metallicity plane, and the lowest mass systems deviate from the predictions of the fundamental metallicity relation, showing higher metallicities, whereas the higher mass ones are in accordance with the model predictions. This might hint at the cut-off of pristine gas inflow and/or the removal of the hot halo gas as the mechanisms driving these offsets. 
Our morpho-kinematic analysis reveals a sub-sample of  dwarfs with  maximum velocities $\rm{v_{max}} < $50 km/s and $\rm{v_{max,\:gas}/\sigma_{gas}<1}$, which depart from the Tully--Fisher relation. 
This might indicate that their interstellar medium is affected by external environmental processes, such as RPS. However,  $\sim 30\%$ of the cluster galaxies have rotation-dominated gas disks and follow the Tully--Fisher relation within $1\sigma$. Using the $\rm{S_{0.5}}$ parameter, which links the dynamical support of ordered motions with that of random motions, we can separate between galaxies affected by gravitational processes and hydrodynamical ones.  In the stellar mass -- $\rm{S_{0.5}}$ plane, both cluster and field galaxies follow a tight sequence, with only a sub-population of 5 galaxies strongly departing ($>4\sigma$) from this relation, showing high $\rm{\sigma_{gas}}$ values.  Both the morphology and kinematics of the outlier galaxies hint at a combination of pre-processing and cluster-specific interactions affecting their stellar and gas disks.  Thus, in our sample of low-mass galaxies in a rich cluster at $z\sim0.4$, 65\% exhibit disordered motions in their velocity fields, in contrast to previous studies of field dwarf galaxies, which find that $\sim$ 90\% have settled disks. }


\maketitle



\setcounter{section}{0}
\section{Introduction}
It is well established that the environment in which a galaxy resides has a great impact on its evolution. Observations have demonstrated that the local galaxy number density and the morphological type of galaxies are correlated, leading to the phenomenon of morphological segregation, first proposed by \cite{dress}, with early-type galaxies  dominating in high-density environments, especially in cluster cores and late-types dominating in the field. Many studies have also found that the fraction of lenticular galaxies in clusters, as well as that of dwarf ellipticals (dEs), has increased by a significant factor between z = 0.5 and today, while the fraction of star forming spiral galaxies experienced a decrease (\citealt{dress2}, \citealt{desai}). These are considered to be the clearest observational signatures that the processes that govern the formation and the evolution of galaxies have an environmental dependence.\\
A large number of studies have tried to explain the predominance of both massive and dwarf early-type galaxies in clusters as the result of physical processes that quench star formation and eventually reshape the morphology of a given galaxy (\citealt{boselli06}, \citealt{peng10},  \citealt{wetzel}, \citealt{lin}). These mechanisms include gravitational interactions between the cluster members and/or with the overall gravitational potential of the cluster, such as cannibalism, dynamical friction, tidal interactions and harassment  (e.g. \citealt{harassment}).  Hydrodynamical cluster-specific interactions between the interstellar medium (ISM) of galaxies and the intracluster medium (ICM) include ram pressure stripping  (RPS, \citealt{rps}), the removal of the hot halo gas \citep{larson}, viscous stripping and thermal evaporation (\citealt{cowie}). 
These cluster-specific processes require different ambient conditions like high ICM gas densities or low relative velocities between galaxies, imprinting specific features in the stellar and gas component of the cluster members, which can be observed, for example in their light distribution (\citealt{her}), structural parameters (\citealt{uli}), velocity fields (\citealt{jose1},   \citealt{asmus}, \citealt{jose}),  oxygen abundances (\citealt{eu}, \citealt{maier22}).  Studies of intermediate redshift cluster galaxies also find evidence that environmental quenching does not only depend on the ICM density but also on the stellar mass of the galaxy, with less massive systems being more strongly affected by environmental mechanisms (e.g. \citealt{pc}, \citealt{tan}). 
 Also, both simulations and observations point to the fact that galaxies are affected by the dense environment well beyond the virial radius of the cluster (\citealt{behroozi}, \citealt{miguel}).  For this reason, low mass galaxies located in galaxy clusters and their outskirts represent ideal laboratories to investigate the interplay between galaxy evolution and the environment, as such systems are expected to be the most susceptible to environmental quenching mechanisms.\\
The dynamical properties of galaxies provide key constraints on models of galaxy evolution. Scaling relations such as the Tully--Fisher relation (TFR, \citealt{t-f}), which links the maximum rotation velocity to the luminosity or the stellar mass,  have been used to study the evolution of star forming galaxies across different environments.
Pioneering studies focusing on the local universe have used regular rotating disk galaxies to investigate the TFR.  Many of these studies have found that at high stellar masses in the order of  $\rm{log(M /M_{\odot} )} > 9.5$, late-type, undisturbed galaxies are rotationally-supported and morphologically disk-like and form a relatively tight TFR (\citealt{Haynes}). Studies focusing on the higher redshift TFR  have found that this relation is well-established up to z$\sim$1 for massive, unperturbed disks (\citealt{miller11}, \citealt{kassin12}).\\ 
The use of galaxy kinematics to study the evolution of star forming galaxies in high-density environments has been also traditionally linked to scaling relations such as the TFR, however, most of these analyses are restricted to high-mass galaxies.
In the local universe, no difference was found in terms of the baryonic TFR (BTFR) of star forming cluster and field galaxies (\citealt{Verheijen}).
Studies focusing on the intermediate redshift universe have also claimed to have found no difference between the TFRs of cluster and field galaxies (\citealt{bodo}), whereas others found that late-type galaxies in dense environments have slightly higher luminosities than their field counterparts (\citealt{Bamford}).  \cite{Moran}  demonstrated that the scatter of the TFR for cluster galaxies is higher than the TFR scatter for field galaxies, and similarly, \cite{jose2} found a higher fraction of galaxies with irregular gas kinematics in the cluster environment than in the field.  \\
On the other hand, the TFR is less well constrained in the low mass regime, both for field and especially for cluster galaxies.
\cite{mcgaugh} have examined the adherence of the Local Group dwarf satellites of the Milky Way and M31  to the BTFR, down to very low stellar masses. They find that most of the brighter dwarfs are consistent with the extrapolation of the BTFR, while the fainter dwarfs do not follow this relation. They argue that the dwarfs that deviate from the BTFR  seem to be lacking luminosity for their velocity dispersion, and this may be explained by the removal of cold gas by stellar feedback and RPS.\\
Kinematic studies of low mass galaxies in higher-density environments mainly focused on the quiescent population of local galaxy clusters.  \cite{tobala} analysed a sample of dwarf early-type galaxies in the Virgo cluster and field, in terms of their kinematical, morphological, and stellar population properties. They found that galaxies in the outskirts of the cluster are mostly rotationally supported systems with disky morphological shapes. The rotationally supported dEs from their sample follow the same TFR as star forming systems. This led to the conclusion that dEs are the descendants of low-luminosity star forming systems that have recently entered the cluster environment and lost their gas due to RPS, quenching their star formation activity and transforming them into quiescent systems, but conserving their angular momentum. \\
At higher redshifts, most kinematic studies of low mass galaxies are restricted to the field population:
 \cite{miller14} used deep spectroscopy from DEIMOS on Keck to study the TFR of  low-mass, star forming field galaxies with 0.2 < z <0.9. They find a similar, but slightly steeper TFR slope and a larger scatter, as compared to the one previously measured by the same authors for a sample of high mass galaxies at similar redshifts.\\
Hence, many of these studies came to similar conclusions, namely that low-mass galaxies show larger residuals relative to the TFR obeyed by higher-mass spirals.\\
Many kinematic studies have made use of data from multi-slit spectroscopic surveys, however, now with the advent of integral-field spectroscopy (IFS), one can obtain a deeper insight into the complex spatially-resolved physical properties of galaxies.
In a more recent study, based on 3D  Multi-Unit Spectrograph Explorer (MUSE) observations from the MAGIC survey, \cite{am} have analysed a sample of late-type galaxies in 8 groups with a redshift range of 0.5 < z < 0.8, down to lower stellar masses in the order of  $\rm{log(M*/M_{\odot})}>8$. They find a significant scatter in the TFR when including dispersion-dominated galaxies in their sample.  They also find that including the velocity dispersion in the velocity budget reduces the scatter in the scaling relation. Their results suggest a significant offset of the TFR zero-point between galaxies in low density and group environments, and they interpret these results to be consistent with a decrease of either stellar mass by $\sim$ 0.05 - 0.3 dex or an increase of rotation velocity by  $\sim$ 0.02 - 0.06 dex for galaxies in groups.\\
The literature still lacks a comprehensive study of kinematic scaling relations for a morphologically unbiased sample of late-type cluster galaxies with low stellar masses and intermediate redshifts.
In this pilot study, we present a morpho-kinematic analysis for a sample of star forming cluster galaxies at $z\sim0.4$ using data observed with the MUSE integral field unit (IFU) spectrograph  \citep{bacon14}.
We identified 17 spatially resolved emission-line MACS J0416.1-2403 cluster members in the MUSE data, as well as 5 field galaxies with similar redshifts.  88\% of the cluster galaxies have masses  $\rm{log(M /M_{\odot} )} < 9.5$, and they all lie at cluster-centric radii $\rm{R/R_{200}\gtrsim 1}$, i.e., outside but close to the virial radius of the galaxy cluster. Owing to their low stellar masses, we expect that cluster-specific quenching mechanisms might be affecting the interstellar medium of these cluster members, as studies suggest that galaxies, especially low-mass ones, might be affected by the cluster environment beyond the virial radius (\citealt{bahe13}). \\
This article is structured as follows: in Sect. \ref{data}, we describe the archival MUSE observations and the additional sources of data used in this work. This section also describes the main steps of the data reduction procedure and the selection of cluster galaxies for our morpho-kinematic study. In Sect. \ref{data ana}, we describe the analysis of the spectroscopic and photometric observations, as well as the kinematic modelling. In Sect. \ref{results}, we present the global properties of our sample,  their morpho-kinematic parameters and maps, as well as different scaling relations, such as the TFR and the stellar mass--$\rm{S_{0.5}}$ relation.
 In Sect. \ref{discussion}, we discuss whether environmental mechanisms can account for the trends we observe. We summarise our conclusions in Sect. \ref{concl}. In the appendix \ref{a1}, we describe the morpho-kinematic maps of each object analysed in this article.\\
Throughout this study,  we use the concordance  $\rm{\Lambda CDM}$ cosmology with $ \rm{H_{0}} = 70\: \rm{km \:s^{-1}\: Mpc^{-1}}$, $\Omega_{0}  = 0.32$, $\Omega_{\Lambda} = 0.68$.  With these cosmological parameters, 1'' corresponds  $\sim 5$ kpc at the redshift of z=0.39 of the MACS J0416.1-2403  cluster.  All magnitudes quoted in this paper are in the AB system.  We assume a  \cite{salpeter} initial mass function (IMF) for all the derived stellar masses and star formation rates (SFRs).

\section{Data}
\label{data}
\subsection{MACS J0416.1-2403  galaxy cluster }
MACS J0416.1-2403 (hereafter M0416, \citealt{macs}) is a massive, X-ray-luminous galaxy cluster, at a redshift z=0.39. The system was observed as part of the Cluster Lensing And Supernova survey with Hubble (CLASH, \citealt{postman12}),  as well as part of the  CLASH-VLT  \citep{rosati14} and the Hubble Frontier Fields (HFF) survey  (\citealt{lotz17}).  More recently, the cluster was also observed with the VLT/MUSE spectrograph. There are two pointings targetting the cluster core, whereas 
the cluster outskirts were observed as part of the MUSE Cosmic Assembly Survey Targeting Extragalactic Legacy Fields (MUSCATEL),  an observing campaign targeting the parallel fields of the HFF clusters.\\
Based on the CLASH-VLT large spectroscopic campaign, \cite{balestra16} have performed a dynamical and structural analysis of M0416. From the phase-space analysis, they derived  a velocity dispersion $\rm{\sigma_{gas}} \sim 1000$ km/s, a virial radius $\rm{R_{200}} \sim 1.8$ Mpc, and a virial mass of $\rm{M_{200}} \sim 1 \cdot 10^{15} \rm{M_{\odot}}$ for the system, as well as a redshift interval for the cluster members of  of 0.373<z<0.413. The same study demonstrated that the dynamical state of the system is complex,  showing a bimodal velocity distribution as well as a projected elongation of the main substructures along the NE-SW direction, which can be explained by a pre-collisional phase scenario. \\
The rich multi-wavelength data available for this galaxy cluster makes M0416 a unique target to study the impact of environmental effects on the cluster members.

\subsection{Observations and data reduction}
The outskirts of the M0416 galaxy cluster were observed in November 2019 with the ESO-VLT MUSE integral field spectrograph \citep{bacon14}  under the  Guaranteed Time Observations  programme  1104.A-0026 (PI: Wisotzki, L.), as part of the MUSCATEL survey. These observations comprise 9 MUSE pointings with a field of view of  1x1 arcmin each, targeting fields outside the virial radius of the cluster. The pointings consist of four exposures of t=1500 s each, yielding a total integration time of $\rm{t}\sim1.66$ hours.
We used the raw data from the  ESO Science Archive Facility and reduced it using the standard calibrations provided by the ESO-MUSE pipeline, version 1.2.1 \citep{pipeline}.  To reduce the sky residuals, we used an additional tool,  the \texttt{Zurich Atmosphere Purge version 1.0} (\texttt{ZAP}, \citealt{soto}), on the calibrated cubes. 
The next step was to align and combine the individual exposures using the \texttt{MUSE Python Data Analysis Framework}  (\texttt{MPDAF}, \citealt{mpdaf}) into a single data cube per pointing. The final calibrated data cubes have a spatial sampling of $0.2"$ in the wavelength range $4750 -9350 \:\AA$ (rest frame: 3417-6727 $\:\AA$ for z$\sim0.39$) and a spectral resolution of $\sim2.5 \:\AA$. \\
The optical IFU data is supplemented by optical Hubble Space Telescope (HST) F814W  imaging with the  Advanced Camera for Surveys (ACS)  from the HFF campaign \citep{lotz17}. We used the already reduced data with a total exposure time of 29.3h. In addition to the optical photometric data, we also used infrared VLT/HAWK-I Ks observations of the M0416 galaxy cluster and its outskirts (\citealt{brammer}). We made use of the already reduced data from the ESO archive, with a total exposure time of 25.82h.\\

\subsection{Selection of the cluster galaxies for the morpho-kinematic study}
\label{cl}
\begin{figure}[t]
  \centering
  \captionsetup{width=0.4\textwidth}
    \includegraphics[width=0.4\textwidth,angle=0,clip=true]{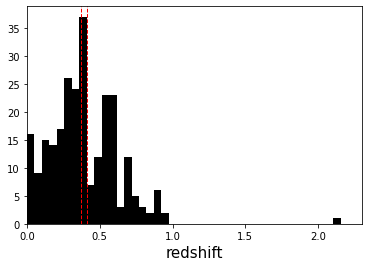}
   \centering   
    \caption{Redshift histogram for the $\sim$ 260 objects which fall in the field of view of the 9 MUSE pointings targeting the parallel field outside the virial radius of the M0416 galaxy cluster. The redshift interval of the M0416 cluster members is shown by the red dotted lines, and is estimated to be 0.373<z<0.413 according to \citealt{balestra16}.}  
\label{hist}
\end{figure}

\begin{figure}[t]
  \centering
  \captionsetup{width=0.4\textwidth}
    \includegraphics[width=0.4\textwidth,angle=0,clip=true]{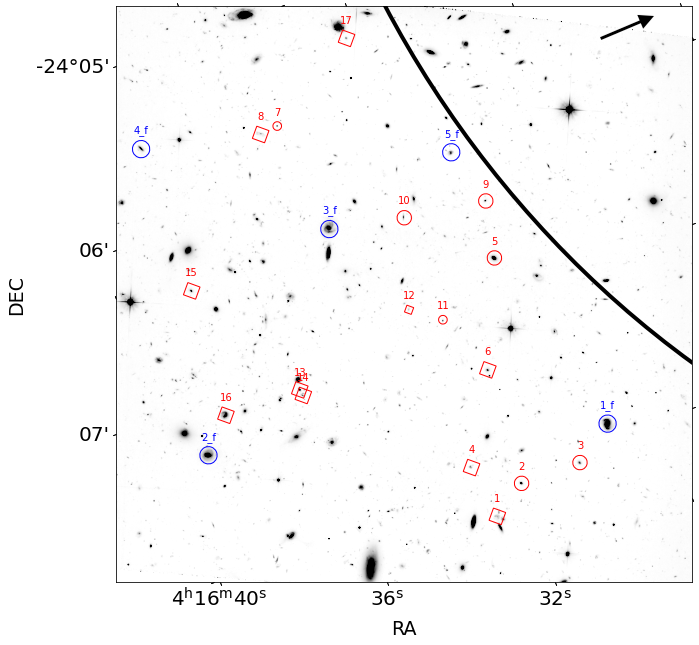}
   \centering   
    \caption{ HST ACS F814W image, showing the location of the M0416 cluster and field galaxies analysed in this study. The morphologically regular cluster galaxies are encircled with red circles, while the morphologically disturbed cluster galaxies  are marked with red boxes. The different sizes of the symbols indicate whether the systems are resolved (large symbols) or marginally resolved in the MUSE data. The field galaxies are encircled with blue circles. The labels indicate the galaxy IDs. The arrow shows the direction toward the M0416 cluster centre, and the solid black curve depicts the virial radius of the cluster of $R_{200}=1.8$ Mpc. }
\label{hst}
\end{figure}

We investigated all of the 9 reduced MUSE pointings targetting the regions outside $\rm{R_{200}}$ of the M0416 galaxy cluster, in search of emission-line galaxies.  These MUSE observations comprise approximately 260 objects: stars, fore- and background galaxies as well as cluster members. The first step was to extract the integrated spectra of all the sources from the field of view of these observations and compute their redshifts. 
The redshift determination was done using  \texttt{MARZ} \citep{marz}, a tool that provides powerful automatic matching capabilities of spectra with model template spectra to produce spectroscopic redshift measurements. We generate the top five \texttt{MARZ} solutions for each spectrum rather than a single best-fit redshift, to visually investigate the different results and to check whether there are any line mismatches or contamination from neighbours.\\
The redshift distribution of all the sources from the field of view of the 9 MUSE pointings is shown in Fig. \ref{hist}, displaying an over-density of galaxies at z$\sim0.39$.
 \cite{balestra16} calculated a redshift interval of 0.373<z<0.413 for the M0416 cluster members, which we also adopted for this study. A total number of 35  galaxies within this redshift range are in the FoV of the 9 MUSE pointings, out of which  24 are emission-line galaxies.  Only 17 galaxies have a high enough signal-to-noise ratio (SNR) to allow flux measurements of their strongest emission lines. These 17 cluster members with clustercentric radii $\rm{R \gtrsim R_{200}}$ represent the main sample for our further analysis. 
We detected only these few galaxies as the MUSCATEL survey is targeting fields outside the virial radius of the cluster, where the number density of cluster galaxies decreases rapidly. 
We also examine 5 galaxies at  $0.29\lesssim z \lesssim0.36$, but outside the caustic region in the phase space and, thus, are called field galaxies here. However, we stress that these serendipitous objects do not serve as a reference sample in our analysis since they have higher masses.\\
It is worth mentioning that we have also investigated the two archival  MUSE pointings available for the M0416 cluster core, in search of emission-line cluster members. However, as expected from the morphology-density relation (\citealt{dress}), cluster cores are dominated by early-type, quiescent galaxies, while later-types populate less dense regions. Therefore, we could not find any star-forming cluster members with strong emission lines, which we could use for our morpho-kinematic study, in the two MUSE pointings targeting the cluster core, despite the Butcher-Oemler-effect \citep{BO}.\\
Fig. \ref{hst} shows the ACS F814W HST image of the regions outside the virial radius of the M0416 galaxy cluster. The cluster and field galaxies analysed in this study are encircled with red and blue symbols, respectively. 

\section{Data analysis}
\label{data ana}
In the following section, we describe the analysis of the MUSE and photometric data. \\

\subsection{Emission line flux measurements with \texttt{FADO} and \texttt{MPDAF} }
Using both the population spectral synthesis code \texttt{Fitting Analysis using the Differential Evolution Optimisation} (\texttt{FADO}, \citealt{fado}), as well as \texttt{MPDAF}, we reliably measured the fluxes of strong emission lines in the optical spectra of our galaxies, such as  $\rm{[O\textsc{ii}]}\:\lambda 3727$, $\rm{H\beta}$, $\rm{[O\textsc{iii}]}\:\lambda 5007$,  $\rm{H\alpha}$, $\rm{[N\textsc{ii}]}\:\lambda 6584$. \\
From the 9 MUSE pointings, we extracted sub-cubes centred on each cluster and field galaxy of interest, using different \texttt{MPDAF} routines. From these sub-cubes, we have extracted both the integrated spectra, as well as one spectrum for each spectral pixel (spaxel), which we then fit with the \texttt{FADO} pipeline. This is a tool specially designed to perform population spectral synthesis, with the additional capability of automatically deriving emission line fluxes and equivalent widths,  assuming consistency between the best-fitting star formation history and the observed nebular emission characteristics of a star forming galaxy. We have used the library of  SSP spectra from \cite{bc03}, with SSP ages between $10^5$ and $10^{10}$ yrs, with a resolution of 3 $\AA$ across the wavelength range 3200 to 9500 $\AA$ and a wide range of metallicities for Padova 1994 \citep{96} evolutionary tracks.  For the fitting routine, we have used the Calzetti extinction law extended to the FUV \citep{calz}. The fluxes offered by \texttt{FADO} are corrected for underlying stellar absorption.
We note, however, that  \texttt{FADO} encounters problems in measuring the $\rm{H\alpha}$ emission line in some z$\sim 0.4$ galaxies because this spectral feature lies exactly at the edge of the spectrum. \\ For this reason, we have also used   \texttt{MPDAF}  for the measurements of the emission line fluxes.  We developed a series of \texttt{PYTHON} codes performing for each spaxel simultaneous Gaussian line fitting for the emission lines of interest after subtracting the stellar continuum. The fit to each emission line is automatically weighted by the variance of the spectrum. The free parameters of the code are the peak position of the Gaussians, their standard deviation,  and the amplitude.\\
We find a  very good agreement between the flux measurements yielded by the two different tools, within the errors. Nevertheless, we chose to use the \texttt{MPDAF} measurements to study the properties of the ionised gas. \\ For our subsequent analysis, we considered only the spaxels which have an SNR larger than 10 to 15 in the emission line of interest. 
The observed $\rm{H\alpha}$ SNR, flux, velocity and velocity dispersion maps derived from the  Gaussian fits of the emission lines, offered by \texttt{MPDAF}, are shown in Fig. \ref{ID10} and Figs. \ref{ID1}-\ref{field5}.


\subsection{Stellar Masses and SFRs}
\label{mass}

\begin{figure}[t]
  \centering
  \captionsetup{width=0.4\textwidth}
    \includegraphics[width=0.4\textwidth,angle=0,clip=true]{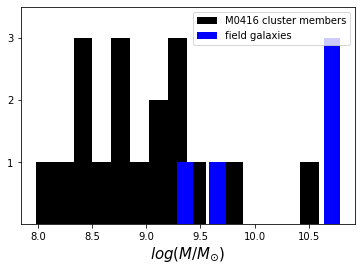}
   \centering   
    \caption{Mass histogram for the M0416 cluster members (black) and field galaxies (blue). The cluster sample lies in the low stellar mass regime, with  $\rm{log(M/M_{\odot})<9.5}$, while only two cluster galaxies show  $\rm{log(M/M_{\odot})>9.5}$.}  
\label{hist_m}
\end{figure}

Stellar masses and SFRs were derived using the code  \texttt{LePhare} of \cite{lephare}, which fits stellar population synthesis models to photometric data. \\ We have employed the HFF-DeepSpace Photometric Catalog of \cite{shipley18}, making use of magnitudes for 6  HST ACS filters (F435W, F606W, F775W, F814W, F850LP, F160W), as well as deep Ks-band imaging from the VLT HAWK-I, and post-cryogenic Spitzer imaging at 3.6 and 4.5 $\mu m$ with the Infrared Array Camera. This aforementioned catalogue was merged with a  second photometric catalogue, namely that from the CLASH survey  \citep{postman12}, containing observations with the Subaru telescope in the B, R and z-bands. All these magnitudes were corrected for Galactic extinction. 
The tool used for the derivation of the stellar masses and SFRs  incorporates the standard $\chi^2$ minimisation method that offers the best match to a reference set of spectral templates for the given photometric data. We have used the galaxy library of simple stellar populations spectra from  \cite{bc03}, and the templates were fitted for SSP ages between 1  and 10 Gyrs. The redshift was kept fixed to the spectroscopic one during the fitting.
\texttt{LePhare} assumes a Chabrier IMF (\citealt{chabrier03}) when deriving stellar masses, which we converted to Salpeter  IMF (\citealt{salpeter}) masses. This was done using the conversion provided by
\cite{pozzetti07}, who found the factor of 1.7 to be a systematic median offset in the masses derived with the two different IMFs. Likewise, the SFRs were converted to be consistent with a Salpeter IMF, following the same prescription as for the stellar masses. \\The total calibration for all bands is estimated to have an accuracy of 0.1 mag, yielding an error budget of $\sim0.15$ dex for the logarithmic stellar masses and SFRs.\\
Fig. \ref{hist_m} shows the stellar mass histogram for the M0416 cluster galaxies in black and the field galaxies in blue. The cluster members lie in the low stellar mass regime, with  $\rm{log(M/M_{\odot})<9.5}$, while only two cluster galaxies show  $\rm{log(M/M_{\odot})>9.5}$. The field galaxies, on the other hand, span a stellar mass   $\rm{9.05<log(M/M_{\odot})<10.56}$.\\

\subsection{Structural parameters of the stellar disk from HST and HAWK-I}
Galaxy structural parameters like inclination, effective radius, and major axis position angle are crucial for a kinematic analysis, in particular for measuring the maximum rotational velocity.\\
For this reason, we have analysed the structure of the M0416 cluster and field galaxies from optical HST ACS F814W photometric data observed as part of the HFF survey. These observations are ideal for such an analysis, due to their high spatial resolution and depth. 
However,  photometric observations taken in redder filters capture the light from the old stellar populations, and these are the populations dominating the structure of a galaxy, with little contamination from bright HII regions.  Therefore, as a consistency check, we have also used IR photometry from VLT/HAWK-I in the Ks band, to analyse the structure of the different systems.  \\
We modelled the surface brightness profile of our targets and measured their structural parameters using the \texttt{GALFIT} tool (\citealt{galfit}), which is a 2-D fitting algorithm designed to extract structural components from galaxy images.
The first step was to create an effective point spread function (ePSF)  from the HST F814W image. For this, we have used the python package \texttt{photutils},  with the \texttt{EPSFBuilder} class, a tool following the prescription of \cite{epsf}.  First, we have identified and extracted cutouts of 16 non-saturated, bright and isolated stars from the FoV of the HST observations, after subtracting the background. Then, we initialised an \texttt{EPSFBuilder} instance and used the cutouts of our selected stars to create the empirical ePSF. The same procedure for creating the ePSF was applied for the HAWK-I imaging, but using 32 non-saturated stars.\\ Using the ePSF, we modelled the surface brightness profiles of all the galaxies in our sample, following a two-component approach, i.e. by fitting two Sersic profiles \citep{sers}, one representative of the bulge and the other one of the disk component. The bulge is described by a de Vaucouleurs profile (Sersic index n=4), whereas the disk is described by an exponential profile  (Sersic index n=1). However, some galaxies in our sample have morphologies that are not well represented by a bulge and a disk, and for these particular cases, we have only fitted an exponential profile to their light distribution. \\
The modelling with \texttt{Galfit} provided us with the position angle (PA) of our objects, the disk scale length $\rm{(r_{d})}$, and the ratio between the apparent minor and major axis, (b/a).  The ratio between the axes can be used to compute the inclination, (i), with respect to the line of sight, using the following equation \citep{incl}:  $$ cos(i)=\sqrt{ \frac{(b/a)^2-q^2}{1-q^2}}, $$ where q represents the ratio between the disk scale height and scale length. This factor was empirically determined to be q = 0.2  for local spiral galaxies, according to \cite{tully}.\\
The HST/ACS F814W galaxy images and/or HAWK-I Ks, the \texttt{Galfit} models and the residual maps are shown in Fig. \ref{ID10} and Figs. \ref{ID1}-\ref{field5}.

\subsection{Disk kinematics modeling}
\texttt{GalPaK3D} \citep{galpak} is a tool to extract the intrinsic galaxy structural parameters and kinematics from 3D IFU observations. This tool directly compares the 3D data cubes to a 3D galaxy disk model that is convolved with the spatial PSF and instrumental line spread function (LSF).  The algorithm uses a disk parametric model with 10 free parameters and the model parameters are optimised using Monte Carlo Markov Chains (MCMC) approach,  with non-traditional sampling laws to efficiently probe the parameter space. 
The shape of the rotation curve is chosen to be an arctangent with: $\rm{v(r)=v_{max}\cdot 2\pi  \cdot arctan \left(\frac{r}{r_{t}}\right)}, $ where $\rm{r_{t}}$ is the turnover radius  and $\rm{v_{max}}$ the maximum rotation velocity reached at the plateau. We have assumed a single-exponential profile for the light distribution of the galaxies' gas disk. 
For the kinematic modelling, we have used a synthetic Gaussian PSF,  with its Full Width Half Maximum (FWHM) equal to the seeing value, and the MUSE  LSF class from \texttt{MPDAF}.  As the pipeline considers the PSF and LSF, it returns the intrinsic deconvolved structural and kinematic parameters (i.e. the ones corrected for instrumental effects and beam-smearing),  such as the galaxy centre, inclination, half-light radius, maximum rotation velocity, turnover radius, and velocity dispersion.\\
Using the \texttt{GalPaK3D} tool, we fit the kinematics directly to the continuum-subtracted data cubes, truncated in wavelength to be centred on the strongest emission line in the galaxy spectrum, which is usually $\rm{H\alpha}$ unless otherwise noted, with a 50 $\AA$ window, blue and red-wards of this emission-line of interest. For every galaxy, we ran the algorithm for at least 10,000 iterations and checked that the MCMC converged for each of the parameters. The uncertainties were estimated from the last $60\%$ of the iterations. \\
Extensive tests presented in \cite{galpak} show that this tool works well, irrespective of the seeing, on data with an SNR> 3 in the brightest pixel and when the ratio of the half-light diameter of the object to the seeing is 0.5, or greater.  For the high SNR and more extended galaxies from our sample, all morpho-kinematic parameters could be well recovered with \texttt{GalPaK3D}.\\
The convolved flux, velocity and velocity dispersion maps computed with \texttt{GalPaK3D} are shown in Fig. \ref{ID10} and Figs. \ref{ID1} -\ref{field5}.

\section{Results}
We have classified the cluster galaxies analysed in this study according to the regularity of their stellar morphologies and observed gas velocity fields.\\
After performing a structural decomposition into a bulge and a disk, we have subtracted the \texttt{Galfit} models from the HST/HAWK-I images and we have analysed the residual maps of the individual objects. A galaxy is considered to be morphologically regular if it does not show strong asymmetries, as revealed by the model residuals. We consider a galaxy as morphologically disturbed when it displays signs of gravitational interactions such as warps, tidal arms, asymmetric, bent, displaced or extremely patchy stellar disks and/or elongated or double-peaked cores. A system is considered to be kinematically regular, if it shows rotation in its observed velocity field. Kinematically disturbed galaxies are considered to be the systems which show no signs of rotation or minor-axis rotation in their observed  gas velocity fields.\\
In the subsequent plots, we use the following symbols: morphologically regular M0416 cluster galaxies are represented by diamonds, whereas morphologically disturbed cluster galaxies are depicted as stars.  Kinematically regular systems are depicted by filled symbols, whereas kinematically disturbed objects are represented by open symbols.   The size of the symbols depicts whether the object is resolved (big symbols) or just marginally resolved (small symbols) in the MUSE data. We consider an object to be only marginally resolved in the IFU data if we can measure fluxes of strong emission lines in less than 10 spaxels. The field galaxies are represented by circles in the subsequent plots. We stress, however, that the field galaxies are not considered to be a comparison sample to the cluster members, and are not included in the statistics, as we analyse this sample separately.  All symbols are colour-coded according to the $\rm{v_{max,\: gas}/\sigma_{gas}}$ value measured from the 3D kinematic modelling.  We stress that $\rm{v_{max,\: gas}/\sigma_{gas}}<1$ does not directly imply that the systems are pressure supported and, thus, dynamically hot. 
See section \ref{morphoandkin} and appendix \ref{a1}, as well as Figs.  \ref{ID10} and \ref{ID1} -\ref{field5} for more details.

\label{results}
\subsection{Global properties: ionising mechanisms, SFRs and  (O/H)s of cluster and field galaxies}
The measurement of strong emission-line fluxes from the integrated spectra of the galaxies allowed us to investigate the ionising mechanisms, as well as to derive the SFRs and (O/H) gas metallicities. However, we could measure the fluxes of all emission lines of interest only in 10 out of 17 cluster galaxies.  In some cases, the  $\rm{H\beta}$ and $\rm{[O\textsc{iii}]}\:\lambda 5007$ emission lines are fairly weak and/or hampered by sky-lines, and hence, these spectral features have a low SNR, making their flux measurements problematic.

\subsubsection{Ionising mechanisms}
\begin{figure}[t]
  \centering
  \captionsetup{width=0.5\textwidth}
    \includegraphics[width=0.5\textwidth,angle=0,clip=true]{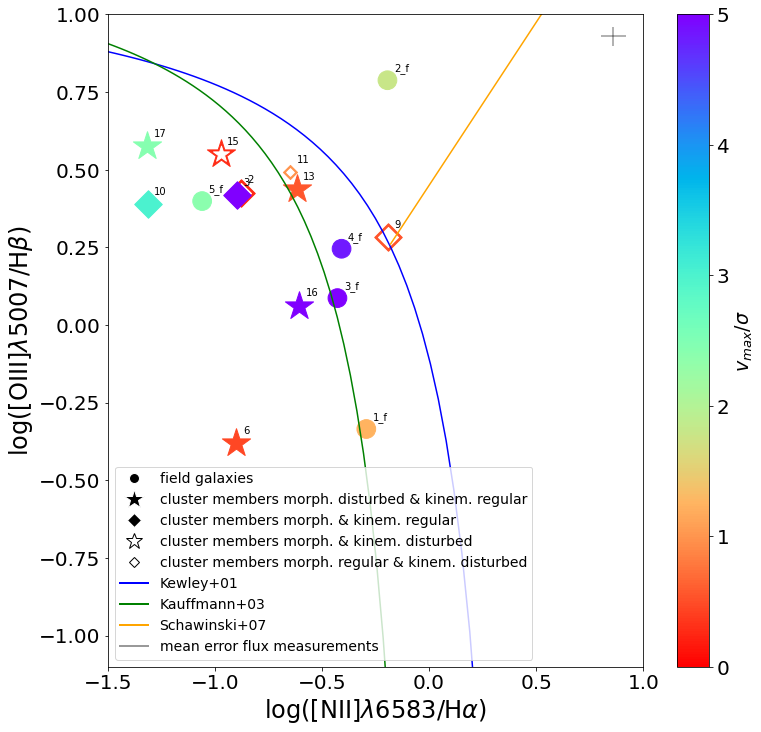}
   \centering   
    \caption{BPT diagnostic diagram  \citep{bald81}  to distinguish the ionisation mechanism of the nebular gas. The blue curve represents the theoretical curve of \cite{kewley01}  and the green  one the empirical curve of \cite{kaufm03}, which separate SF galaxies from AGNs. The orange curve of  \cite{sw17}  depicts the separation line between Seyfert II galaxies and LINERs.    Each data point represents one galaxy from our sample,  which has an SNR>10 in the emission lines used for the diagnostic. The diamonds show the cluster members with regular morphologies, whereas the stars depict the cluster galaxies with disturbed morphologies, i.e. which show signs of gravitational interactions. The filled symbols represent the galaxies with regular velocity fields, and the open symbols show the galaxies with disturbed velocity fields, for which no rotation is evident in the observed velocity maps. The circles show the field galaxies. The symbols are colour-coded according to $\rm{v_{max,\: gas}/\sigma_{gas}}$. The size of the symbols depicts whether the galaxies are resolved (big symbols) or just marginally resolved in the MUSE data, i.e. have measurements only for a few spaxels (small symbols). The labels indicate the galaxy IDs. The cross from the upper right corner shows the mean error of the flux measurements. 
    }  
\label{BPT}
\end{figure}

Using a set of four strong emission lines, one can reliably distinguish between star forming (SF) systems, Seyfert II galaxies, low-ionisation nuclear emission-line regions (LINERs), and composite galaxies. We used the classical BPT diagnostic diagram \citep{bald81} to investigate the ionising mechanisms in the 10  cluster galaxies and 5 field galaxies.\\ Figure \ref{BPT} shows the BPT diagram, which uses   $ \rm{[O\textsc{iii}]/H\beta}$ vs $\rm{[N\textsc{ii}]/H\alpha}$ emission-line flux ratios. 
Seven cluster members and one field galaxy from our sample follow the SF sequence in the BPT diagnostic diagram, meaning that the main source of ionisation in these objects is star formation. Two cluster and three field galaxies fall in the composite region of the diagnostic diagram, and hence, their ISM is ionised by both star formation and possibly an AGN and/or mechanisms that give rise to LINER-like emission. Several hypotheses have been proposed for the source of  LINER-like emission: stellar populations may be responsible for this emission, which could arise due to photoionisation of the gas by young starbursts, or by old pAGB stars (\citealt{olsson10}, \citealt{l13}, \citealt{belfiore}).  LINER-like emission may also be characteristic of shocks \citep{allen08}, starburst-driven outflows \citep{sharp10}, Lyman continuum photon escape through tenuous warm gas \citep{polis13}, or gas heated by the surrounding medium. The latter mechanism includes photoionisation by cosmic rays,  collisional heating by cosmic rays,  conduction from hot gas, suprathermal electron heating from the hot gas, X-ray photoionisation and turbulent mixing layers (e.g.  \citealt{dh11}, \citealt{mc12}, \citealt{sparks12}).\\
One cluster galaxy lies exactly at the intersection of the three separation curves, and given the flux measurement uncertainties, we can not securely classify this object as an AGN / LINER. However, one of the galaxies from our field sample shows very high values for both emission-line ratios used for the diagnostic, meaning that field galaxy \#2-f  can be securely classified as a Seyfert II.\\ 

\subsubsection{The stellar mass -- SFR  relation}

\begin{figure}[t]
  \centering
  \captionsetup{width=0.47\textwidth}
    \includegraphics[width=0.47\textwidth,angle=0,clip=true]{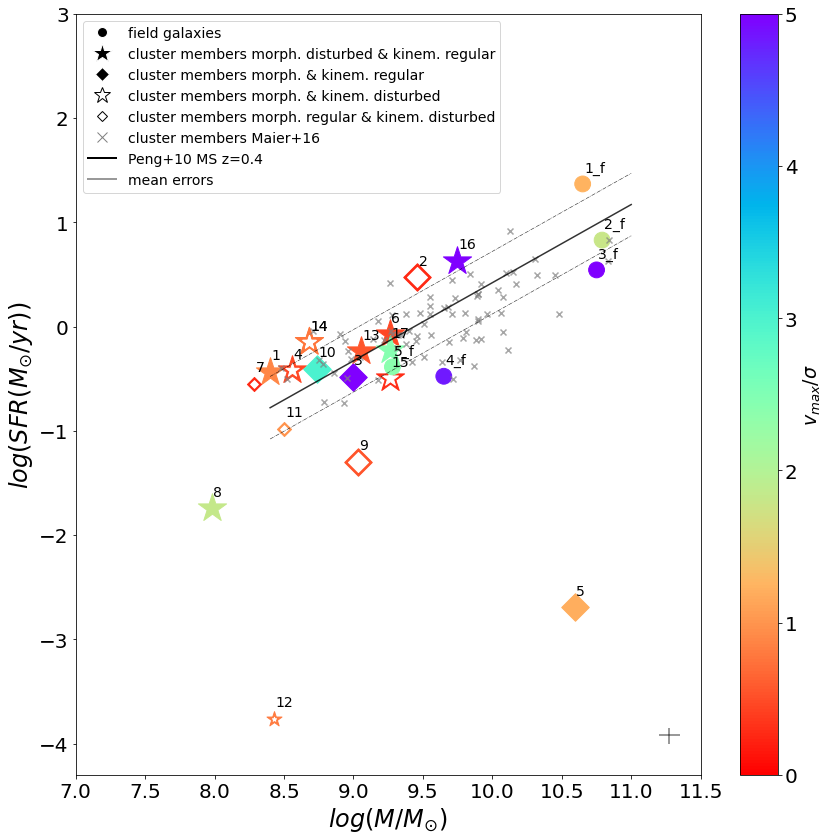}
       \centering   
     \caption{    Stellar mass -- SFR relation, as derived from the photometric data using the \texttt{LePhare} code, for all galaxies analysed in this study. The symbols are the same as in as in Fig. \ref{BPT}, but in addition, we also plot here the M0416  star-forming cluster members from \cite{maier16}, observed with the VLT/VIMOS, as light grey crosses. The black solid and dashed lines represent the SF MS at z=0.4  and its $\sim 0.3$ dex scatter, as derived by  \cite{peng10}. The grey cross in the lower right corner of the plot shows the measurement errors for the SFRs and stellar masses. } 
     \label{sfr_comp}
\end{figure}

The SFRs were computed from the extinction corrected luminosity of the $\rm{H\alpha}$ emission line. 
We make the simplifying assumption that the SFR is nearly constant over the past $\sim100$ Myr and that  case B recombination applies and therefore, the $\rm{H\alpha}$ luminosity can be used for estimating the SFR following the conversion proposed by  \cite{ken98} for solar metallicity and a Salpeter IMF: 
$ \rm{SFR({M}_{_{\odot}} \cdot yr^{-1})=7.9 \cdot 10^{-42} L(H\alpha)}\:  \left(\rm{\frac{ergs}{s}}\right) $.\\
The luminosity of the $\rm{H\alpha}$  emission line was corrected for extinction based on the Balmer decrement following the equations introduced by \cite{brock71}. 
However, we were able to compute the SFRs from spectroscopy only for a sub-sample of our cluster dwarfs, and for this reason, we have also computed the SFRs from the photometric data (see Sect. \ref{mass}), finding very good agreement between the two measurements, within the errors.\\ 
Observations have demonstrated that out to at least $z \sim 2$  there exists a 'main sequence' (MS) of star-forming galaxies in which the SFR is closely linked with the stellar mass of a galaxy (\citealt{brinch}, \citealt{whitaker}). 
Fig. \ref{sfr_comp} shows the stellar mass--SFR relation for the whole sample of galaxies analysed in this study. 
The SFRs of the M0416 cluster and field galaxies, as well as the scatter,  are consistent with the SF MS at z=0.4, given the measurement uncertainties of $\sim 0.15$ dex.
$\sim$40\% of the cluster galaxies from our sample show SFRs which are consistent with the MS relation and with its scatter of $\sim0.3$ dex.  On the  MS, we find a mixed population of galaxies,  showing both regular and irregular morphologies and kinematics. 
$\sim$30\% of the cluster galaxies show $\rm{\Delta_{MS}}>0.3$ dex, meaning that they have SFRs slightly more enhanced than  MS galaxies. 
The systems lying slightly above the MS tend to be either morphologically and/or kinematically disturbed, and they mainly show $\rm{v_{max,\: gas}/\sigma_{gas}}<1$.  $\sim$ 30\% of the sample deviates from the MS relation showing  $\rm{\Delta_{MS}}<-0.3$ dex, out of which 4 galaxies are strongly scattering towards low SFR values, namely the cluster galaxies with the IDs \#5, \#8, \#9, \#12. These objects show a mean $\rm{\Delta_{MS}}\sim-1.7$, meaning that they are currently in the process of quenching. All galaxies which lie below the SF MS show signs of disturbances, either morphological or kinematical, except galaxy \#5, which has a morphology and spectrum typical of an earlier type galaxy. We refer the reader to appendix \ref{a1} for the detailed description of the individual objects.\\
Regarding the field population, these systems can be classified as MS galaxies, given the evolution of the MS with redshift towards lower SFRs (\citealt{peng10}). The only exception is object \#1-f, which shows higher values for the SFR than MS objects at $z\sim0.35$.




\subsubsection{The  Mass -- gas-phase metallicity and Fundamental metallicity relation}

\begin{figure*}
   \centering
    \includegraphics[width=0.4\textwidth,angle=0,clip=true]{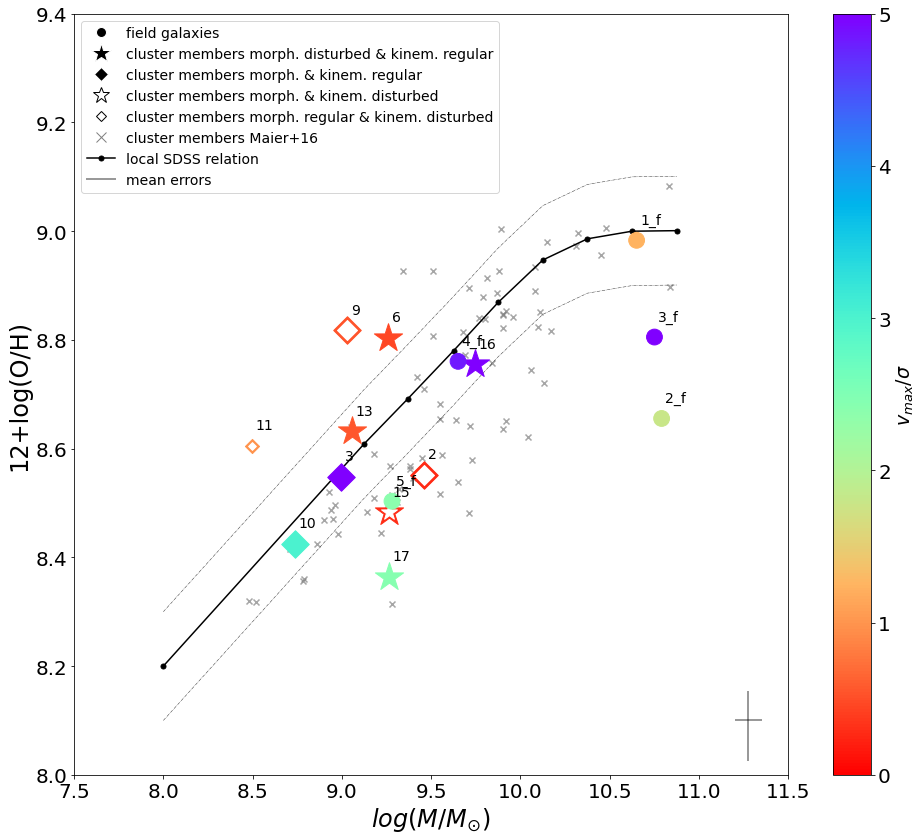}
   \centering   
    \centering
    \includegraphics[width=0.4\textwidth,angle=0,clip=true]{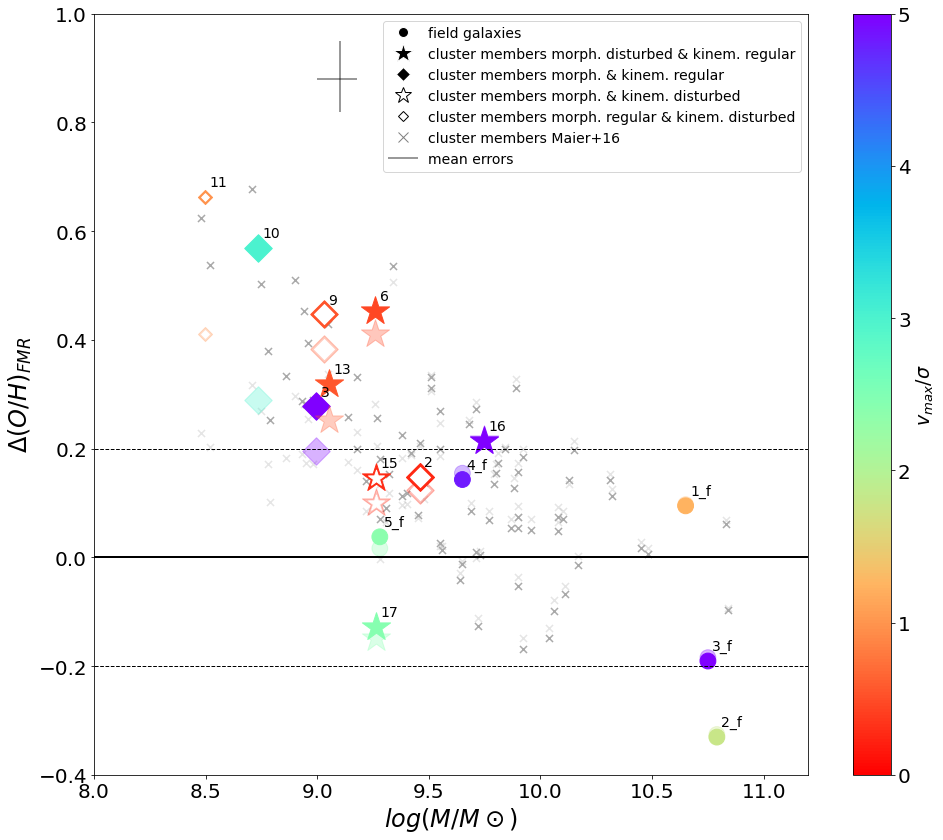}
   \centering   
  
    \caption{\textit{left: }Mass--metallicity relation for M0416 cluster members  and field galaxies using the O3N2 metallicity calibration of  \cite{kewley13a}.   The symbols are the same as in Fig. \ref{BPT} and in addition, we also show the M0416  cluster sample from \cite{maier16}, observed with VLT/VIMOS, as light grey crosses.  The black points and curves represent the local SDSS relation from the latter authors, consistently derived using the O3N2 calibration, with its $1\sigma$ scatter. The MZR for the SDSS sample was extrapolated down to lower masses of $\rm{log(M/{\rm M}_{\odot})=8}$, by assuming that the slope remains constant for $\rm{log(M/{\rm M}_{\odot})< 9.2}$.  The cross in the lower right corner shows the mean value of the measurement errors. \textit{right:} Difference between the measured (O/H)s for  M0416 cluster and field galaxies, using the O3N2 metallicity calibration and the expected (O/H)s from the formulations of \cite{lilly13} for different infall metallicities  $\rm{Z_{0}}$,  relative to the yield, y, as a function of the stellar mass of the systems. The models assuming a primordial metallicity for the inflowing gas, $\rm{Z_{0}/y = 0}$, are depicted by the bright data points, whereas the models assuming a metal-enriched gas inflow, $\rm{Z_{0}/y = 0.1}$, are depicted by the faint data points. The symbols are the same as in the plot from the left-hand side.}  
\label{oh}
\end{figure*}

The oxygen abundances for the M0416 cluster  and field galaxies were computed using the empirically calibrated metallicity estimator of \cite{kewley13a}. 
 By means of stellar evolutionary synthesis and photoionisation models with chemical evolution measurements from cosmological hydrodynamic simulations, the latter authors derived the following empirical metallicity calibration using the O3N2 index: $ \rm{12 + log(O/H)= 8.97 - 0.32 \cdot log\left(\frac{[O III] \lambda5007/H\beta}{[N II] \lambda6583/H\alpha}\right)}$. This calibration relies on ratios of emission lines that are close in wavelength, and hence, no corrections for reddening are needed. \\ 
 A large number of studies focusing on the local universe have shown that there is a tight correlation of 0.1 dex between the stellar mass of a galaxy and its gas phase metallicity, giving rise to the so-called mass-metallicity relation (MZR,  \citealt{trem04}, \citealt{kew2008}). Studies focusing on the MZR in high-density environments have  found that, at fixed stellar mass, galaxies with lower gas fractions typically show higher oxygen abundances. A relationship between gas fraction and metal content was also established, whereby gas-poor galaxies are typically more metal-rich, and this is likely due to the removal of gas from the outskirts of spirals due to environmental mechanisms, which leads to an increase in the observed average metallicity \citep{hughes}.  \\
Fig. \ref{oh} displays the MZR for all galaxies from our sample with available flux measurements for the emission lines needed for this calibration. 
We observe significant scatter around the local SDSS relation for the MUSE M0416 cluster and field sample, both towards lower and higher gas-phase metallicities. 4 cluster galaxies follow the local SDSS relation within $1\sigma$, whereas 6 members scatter both above and below the local MZR. Regarding the field galaxies, they mainly show lower gas-phase metallicities than  SDSS galaxies. We note that we do not observe any trends between the offsets from the local MZ relation, and the stellar morphology and/or regularity of the gas velocity fields.\\For this reason, we also investigate the fundamental metallicity relation $\rm{Z(M, SFR)}$. Observations have shown that oxygen abundances anticorrelate with SFRs, especially at low stellar masses, such that at a given stellar mass, galaxies with high SFRs show lower (O/H)s. This led to the conclusion that the chemical abundance of a galaxy is dependent on both the stellar mass and the SFR, giving rise to the so-called fundamental metallicity relation (FMR;  \citealt{mannu10}).\\  \cite{lilly13}  introduced a simple model of galaxy evolution in which the SFR is regulated by the amount of gas present in a galaxy, in order to explain the dependency of the gas-phase metallicity on M and SFR. We test whether at higher redshift and in a  cluster environment, the  $\rm{Z(M, SFR)}$ is similar to the one found in the local universe. Therefore, we computed the expected (O/H) values from the simple gas regulated model of \cite{lilly13} for each galaxy individually with their respective stellar mass and SFR. This is done by  adopting  their Eq. (40), which can be rewritten, in a simplified version, as: \begin{equation} \rm{Z_{eq}=Z_{0}+y(1-R)SFR/\Phi},\label{FMR} \end{equation}  where $\rm{Z_{eq}}$  stands for the equilibrium value or the metallicity, $\rm{Z_{0}}$ is the metallicity of the infalling gas, y is the yield, R is the fraction of mass returned to the ISM and $\Phi $ the gas inflow rate.  As it is evident from this equation, an increased metallicity of the inflowing gas or a suppressed gas inflow rate can lead to an increase in  $\rm{Z_{eq}}$. We use both a model for primordial gas inflow, with the metallicity of the infalling gas, $Z_{0}$, relative to the yield, y,  being $Z_{0}/y = 0$,  as well as a model for metal-enriched gas inflow, with  $Z_{0}/y = 0.1$.\\ 
The panel from the right-hand side of Fig. \ref{oh} displays the difference between the measured (O/H)s using the O3N2 metallicity calibration and the expected (O/H)s from the FMR  as a function of the stellar mass of the systems. The lowest mass cluster galaxies deviate significantly from the model predictions, by up to $\sim 0.6$ dex for models assuming primordial inflow, and by up to  $\sim 0.4$ dex for models assuming enriched inflow. On the other hand, both cluster and field galaxies from the higher mass end are in fairly good agreement with the predictions of the model, within $\sim 0.2$ dex. We note that field galaxy \# 2-f is a strong AGN, and therefore, the oxygen abundance measured from strong line methods is not robust for this object. \\
To summarise, the lowest mass cluster galaxies have more enhanced metallicities than predicted by models assuming a primordial or a metal-enriched gas inflow. The disagreement between the observed and predicted abundances hint at a suppressed gas inflow rate, $\Phi$, as the culprit for the more enhanced abundances we measure, and not $\rm{Z_{0}}$.

\subsection{Morphology and kinematics of M0416 cluster and field galaxies}
\label{morphoandkin}
In this section, we describe the results from the morpho-kinematic analysis of the photometric and spectroscopic data. 

\subsubsection{ Structure and morphology from HST and HAWK-I}
We have used both optical and IR imaging to study the structure and morphology of the galaxies using \texttt{Galfit}. We observe a substantial population of disturbed cluster galaxies, which show signs of gravitational interactions. The cluster galaxies displaying rather irregular and complex broad-band morphologies make up $\sim$60\% of our sample.\\
It is also worth noting that we observe good agreement between the structural parameters derived from the HST and HAWK-I photometry. The inclinations obtained from the different photometric observations differ by  $15^{\circ}$ at most, within the errors, as do the PAs. The HST imaging is expected to yield more robust inclinations,  given the much better spatial resolution. The disk scale lengths, $\rm{r_{d}}$, obtained from  HST  are slightly larger than the ones obtained from the  HAWK-I imaging, but they agree within the errors. However, this is expected, as the HAWK-I imaging probes the old stellar component, whereas with the HST imaging, we also observe the contribution from younger stellar populations. In the subsequent results, we use the structural parameters obtained from the higher resolution HST data, unless otherwise noted.\\

\subsubsection{Morpho-kinematic maps }
\label{morphokin}
\begin{figure}[t]
  \centering
  \captionsetup{width=0.5\textwidth}
    \includegraphics[width=0.5\textwidth,angle=0,clip=true]{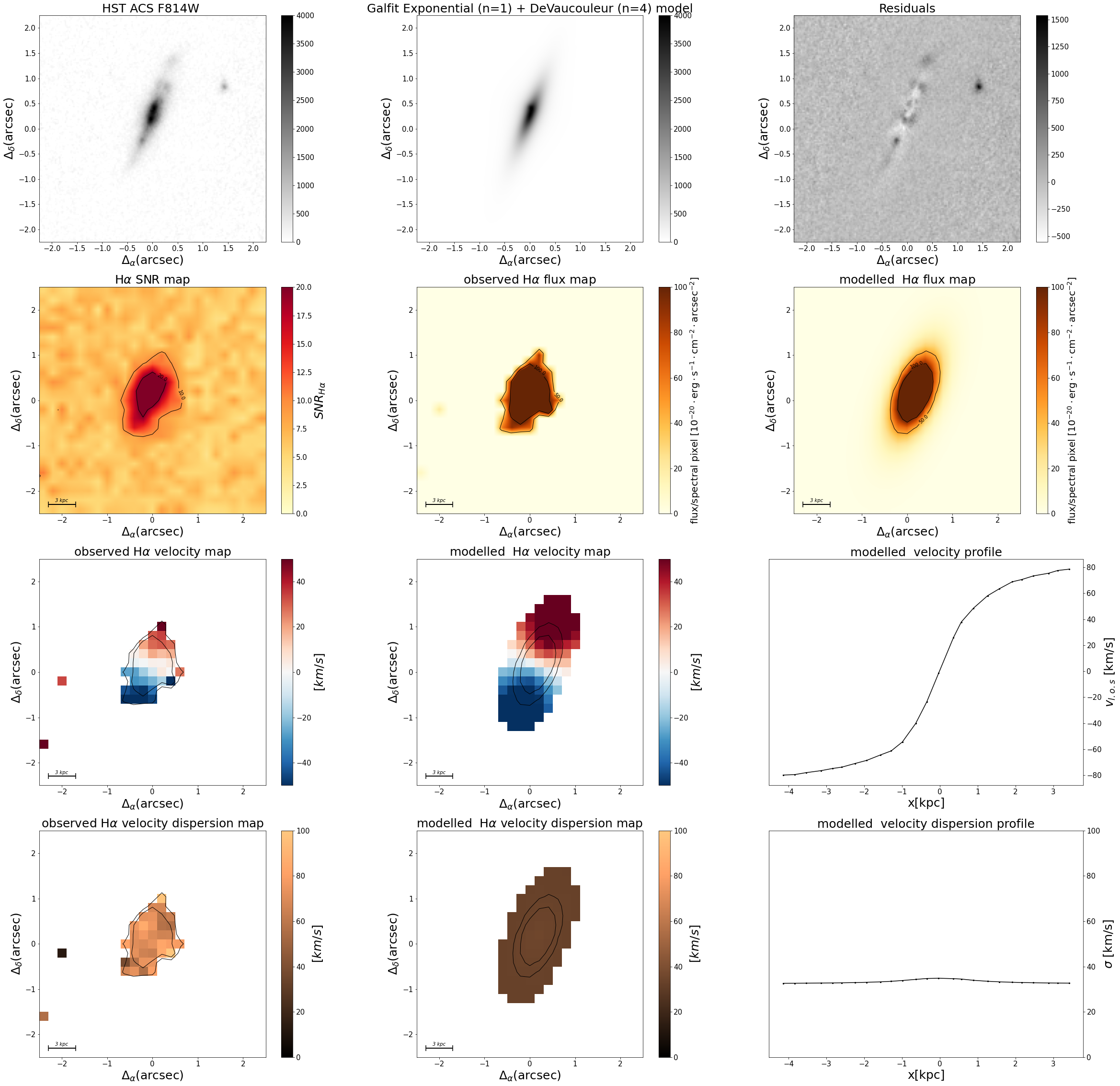}
   \centering   
    \caption{Morpho-kinematic analysis for cluster galaxy  \#10.  \textit{top row left:} HST/ACS F814W image;  \textit{top row middle:} \texttt{Galfit} model (exponential + de Vaucouleurs  profile); \textit{top row right:} \texttt{Galfit} residuals.   \textit{second row left:} MUSE $\rm{H\alpha}$ SNR map, as measured with \texttt{MPDAF};  \textit{second  row middle:}  MUSE $\rm{H\alpha}$ flux map, as measured with  \texttt{MPDAF}, in units of $10^{-20}\: \rm{ erg \:s^{-1}\: cm^{-2}\: arcsec^{-2}}$; \textit{second row right:} modelled $\rm{H\alpha}$ flux map from \texttt{GalPaK3D}, in the same flux units as before. \textit{third row left:} MUSE $\rm{H\alpha}$ velocity map, in units of $\rm{km/s}$ as measured with \texttt{MPDAF};  \textit{third  row middle:}   modelled $\rm{H\alpha}$ velocity map in  $\rm{km/s}$ from  \texttt{GalPaK3D} ; \textit{third row right:} modelled $\rm{H\alpha}$ velocity profile from \texttt{GalPaK3D}.  \textit{last row left:} MUSE $\rm{H\alpha}$ velocity dispersion map, in units of $\rm{km/s}$ as measured with \texttt{MPDAF};  \textit{last  row middle:}   modelled $\rm{H\alpha}$ velocity dispersion map in  $\rm{km/s}$ from  \texttt{GalPaK3D} ; \textit{last row right:} modelled $\rm{H\alpha}$ velocity  dispersion profile from \texttt{GalPaK3D}.  This galaxy is classified as being morphologically and kinematically regular.  It has the following physical parameters: $\rm{\log(M/M_{\odot})}=8.74 $; $\rm{v_{max} }=98$ km/s; $\rm{v_{max,\: gas}/\sigma_{gas}}=3.03 $; offset from SF MS: $\rm{\Delta_{MS}}=0.1$; offset from TFR:  $\rm{\Delta M_{TFR}}= 0.27 $; offset from stellar mass -- $\rm{S_{0.5}}$:  $\rm{\Delta M_{S_{0.5}}}=  1.0$.  }  
\label{ID10}
\end{figure}

Fig. \ref{ID10} shows an example of the morpho-kinematic analysis of the galaxy  \#10. This galaxy is a highly inclined M0416 cluster member at a redshift z=0.395, with a diameter of $\sim 2''$. For the stellar disk, we measure an inclination of  i=$80^{\circ}$ and a $\rm{r_{d}} $= 1.7 kpc. The morphology of this galaxy is well reproduced by a bulge and a disk, with some patchy regions visible in the residuals, which are probably SF regions.\\
For the gas $\rm{H\alpha}$ disk, we measure an inclination of  i=$65^{\circ}$,  in good agreement with the one computed for the stellar disk using the HST data, and a $\rm{r_{d}}=3.3$ kpc. The gas disk, is thus, more extended than the stellar disk, which is expected for late-type, SF galaxies  (\citealt{rd}). The kinematic and photometric PAs agree well, with $\rm{\Delta_{PA}}<5^{\circ}$.\\
The disk kinematic modelling reproduces well the observed velocity fields. The velocity field derived from the $\rm{H\alpha}$ line shows a regular rotation pattern, with $\pm$ 80 km/s, compatible with its morphology. The yielded velocity dispersion is  low, in the order of $\rm{\sigma_{gas}} = 35$ km/s.   The velocity dispersion map is flat, except for the centre along the minor axis.Thus, the gas dynamics of this galaxy are dominated by rotation, with a ratio  $\rm{v_{max,\: gas}/\sigma_{gas}} \sim 3$, and this cluster member follows the TFR and the stellar mass--$\rm{S_{0.5}}$ relation.\\
The description of the morpho-kinematic analysis of all galaxies from our sample can be found in the appendix \ref{a1} and Figs. \ref{ID1} -\ref{field5}.

\subsubsection{Morpho-kinematic parameters }

\begin{figure}[t]
  \centering
  \captionsetup{width=0.5\textwidth}
    \includegraphics[width=0.5\textwidth,angle=0,clip=true]{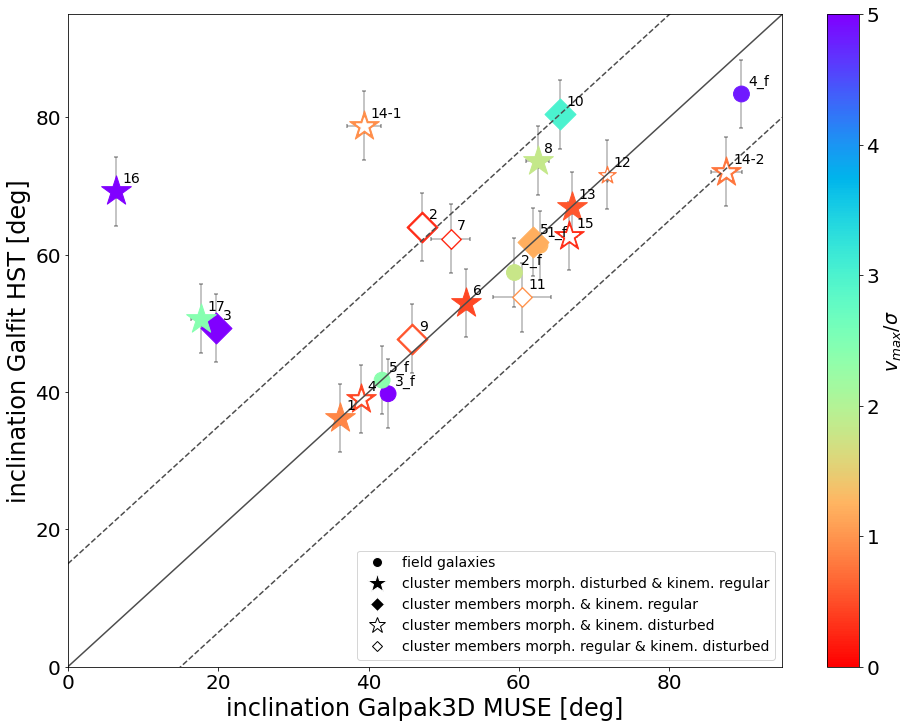}
   \centering   
    \caption{Comparison between the inclination of the gas disk, as measured with \texttt{GalPaK3D} from the MUSE data and the inclination of the stellar disk, as measured with \texttt{Galfit} from the HST/HAWK-I data. The symbols are the same as in Fig. \ref{BPT}.}
\label{incl}
\end{figure}

\begin{figure}[t]
  \centering
  \captionsetup{width=0.5\textwidth}
    \includegraphics[width=0.5\textwidth,angle=0,clip=true]{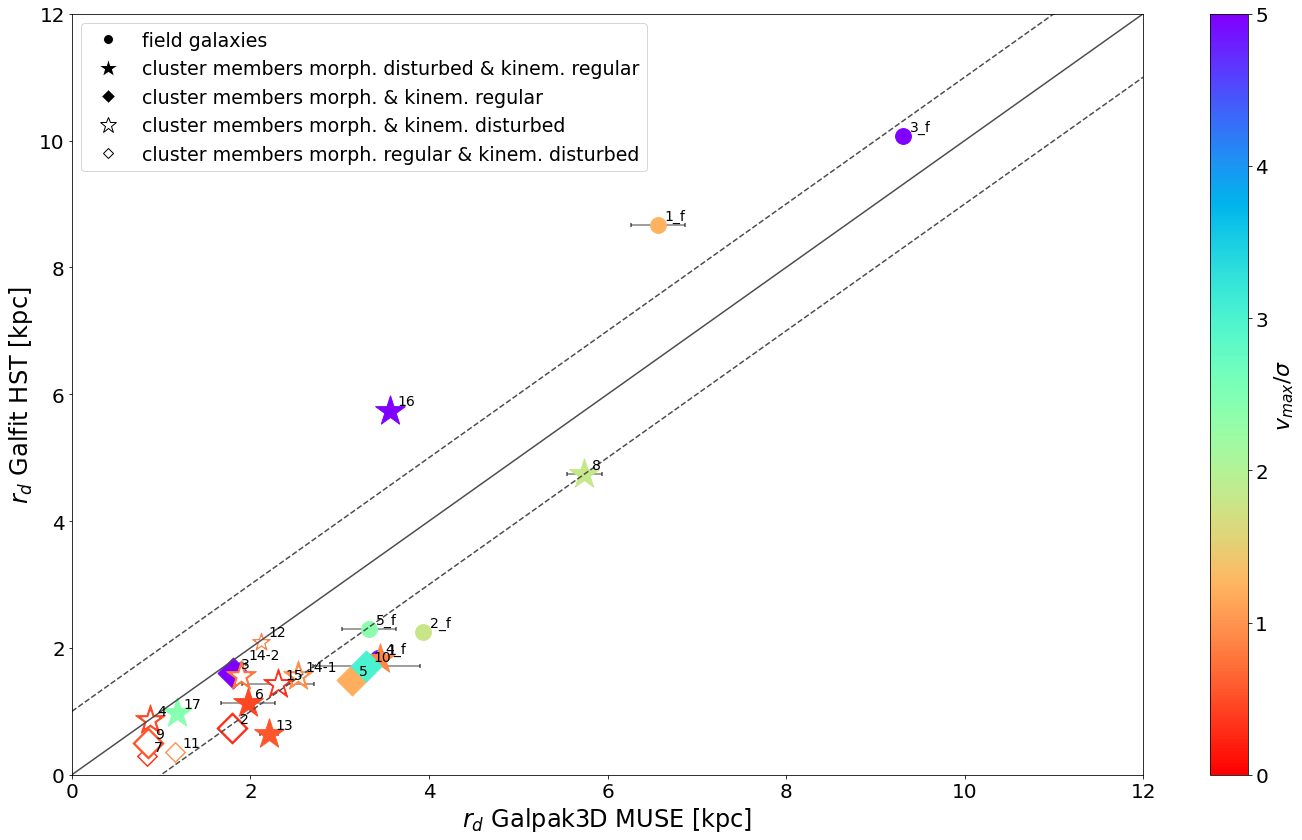}
   \centering   
    \caption{Comparison between the disk scale length of the gas component in kpc , as measured with \texttt{GalPaK3D} from the MUSE data and the disk scale length of the stellar component, as measured with \texttt{Galfit} from the HST/HAWK-I data. The symbols are the same as in Fig. \ref{BPT}. }
\label{rd}
\end{figure}

Fig. \ref{incl} shows the comparison between the inclination of the stellar disk, obtained by using \texttt{Galfit} on the HST/HAWK-I data,  and the inclination of the gas disk, derived with \texttt{GalPaK3D} from the MUSE data-cubes. Taking the intrinsic uncertainties of the different measurements and the errors into account, the inclinations of the stellar and gas disks agree well for 81\% of the galaxies from our sample, within $15^{\circ}$. However, due to convergence problems, we constrained the inclination of the gas disk in \texttt{GalPaK3D} to the one obtained for the stellar disk for a sub-sample of 5  cluster galaxies and one field galaxy, see appendix \ref{a1} for more details. For galaxies with IDs \#3,  \#14, \#16, \#17,  \texttt{GalPaK3D} strongly underestimates the inclination, compared to the one derived from the higher-resolution HST data. However, when constraining the inclination of the gas disk for these galaxies to the one of the stellar disk, \texttt{GalPaK3D} has problems converging and the resulting velocity fields do not agree with the observed ones. Galaxies \#14, \#16 and \#17 are morphologically disturbed, and hence, for such galaxies, we would expect the gas component to possibly not follow the gravitational potential of the stellar component. Therefore, it is plausible that for these systems, the inclination and other structural parameters of the gas disk differ with respect to those of the stellar disk. Furthermore, hydrodynamical cluster-specific interactions can also lead to discrepancies between the structural parameters of the stellar and gas component, as such interactions only affect the ISM of the galaxies. \\
 Fig. \ref{rd} shows the comparison between the disk scale length of the stellar component and the disk scale length of the gas component. Taking the measurement uncertainties into account, the $\rm{r_{d}}$ estimates for the stellar and gas disks agree within 1kpc, with a tendency for the gas disk to be more extended than the stellar disk. This is, however, expected for  SF galaxies (\citealt{rd}).  The only exceptions where the disk scale length of the stellar component is much larger than that of the gas component are for cluster member  \#16 and field galaxy \#1-f. It is worth mentioning that these 2 objects have a highly complex broad-band morphology. For the morphological modelling, we have used Fourier modes to modify the ellipses to better fit the spiral features, and hence, the modelling might not be representative of the true morphology.\\

\subsubsection{Gas dynamics}
\label{dynsup}
The dynamical support of galaxies is often quantified according to the ratio between the maximum stellar circular velocity and the intrinsic stellar velocity dispersion,  $\rm{v_{max}/\sigma}$. 
Many studies at intermediate and high redshift, however, have used  $\rm{v_{max,\: gas}/\sigma_{gas}}$ as a proxy for the dynamical state of the gas disc.  We note, however, that the parameter $\rm{\sigma_{gas}}$ is by no means like the typical pressure-supported velocity dispersion that is measured from stellar absorption lines in early-type galaxies.  $\rm{\sigma_{gas}}$  is tracing the warm ionised gas which can radiate, and a high dispersion system cannot remain in equilibrium after a crossing time. Hence, this parameter is effectively measuring velocity gradients below the seeing limit,  probing disordered components to the velocity field (\citealt{Covington}).\\ In order to tentatively quantify the dynamical state of the gas disk for the sample of M0416 cluster and field galaxies, we assume that the ionised gas in these galaxies shows some rotation pattern (albeit with a small velocity gradient for some cluster members) and that this rotation is suitably well-described by a rotating-disk model.  The velocity dispersion of the stars, $\rm{\sigma_{star}}$, is in general considered to be a direct tracer of the gravitational potential of the galaxy, but this assumption is not valid for $\rm{\sigma_{gas}}$.  Indeed, additional support against gravitational collapse also comes from the pressure component of $\rm{\sigma_{gas}}$.  However, in contrast to $\rm{\sigma_{star}}$,  $\rm{\sigma_{gas}}$ gets additional contributions from non-gravitational effects.  Important sources of non-gravitational effects in low mass galaxies is turbulence induced by stellar feedback, as well as hydrodynamical environmental mechanisms. \\
We stress that a major limitation in measuring gas velocity dispersions is the finite spectral resolution of the MUSE IFU, as the  $\rm{\sigma_{gas}}$ that we measure are often below the instrumental resolution. At the wavelength of  $\rm{H\alpha}$, which, out of the Balmer lines in our spectral range is the dominant line for the determination of the kinematics, the instrumental resolution is in the order of 40-50 km/s, and velocity dispersions below $\sim$ 35 km/s can not be accurately measured (\citealt{den Brok}, \citealt{boselli2}). Therefore,  all the $\rm{\sigma_{gas}}<35$ km/s values should be regarded as  upper limits.\\
We consider the  ionized gas dynamics of a galaxy disk to be "rotation-dominated" if $\rm{v_{max,\: gas}/\sigma_{gas}}>1$, and "dominated by random motions", or  better said, kinematically disordered  if $\rm{v_{max,\: gas}/\sigma_{gas}}<1$. However, this description is intrinsically better justified for rotators  (e.g. \citealt{contini16}).  We stress that whenever the gas velocity dispersion has a significant contribution from non-gravitational effects, the parameter $\rm{v_{max,\: gas}/\sigma_{gas}}$ does not reflect kinematical anisotropy.
Hence, we do not claim that all the systems in our sample, showing $\rm{v_{max,\: gas}/\sigma_{gas}}<1$ are intrinsically dynamically hot. The broadening of the emission lines and, thus, the increased $\rm{\sigma_{gas}}$ can also be an effect of disturbances, i.e, of the larger-scale, but still, spatially unresolved movements of the gas stirred by gravitational and/or hydrodynamical interactions. Also, given the rather small disk scale lengths of some of the cluster dwarfs with  $\rm{v_{max,\: gas}/\sigma_{gas}}<1$, the $\rm{\sigma_{gas}}$ might be artificially inflated by beam-smearing and the contribution from unresolved l.o.s. velocities.  To confirm the dynamical nature of the galaxies would require studying their stellar kinematics, which is not possible with the current data at hand.\\
For $\sim65\%$  of the cluster galaxies from our sample, we measure  $\rm{v_{max,\: gas}/\sigma_{gas}}<1$, hence, these low mass cluster galaxies have a high contribution of turbulent, disordered motions to their velocity field.  All galaxies with  $\rm{v_{max,\: gas}/\sigma_{gas}}<1$  show signs of gravitational interactions and/or do not show any signs of rotation in their observed velocity fields.  However, we note that the spectral resolution of the MUSE instrument is limiting us from accurately measuring velocity dispersions below 35 km/s, and hence, this might be biasing our inferred  $\rm{v_{max,\: gas}/\sigma_{gas}<1}$ values for dwarfs with low rotational velocities, in the order of $\rm{v_{max}}<50 km/s$. For 35\% of the cluster galaxies we measure $\rm{v_{max,\: gas}/\sigma_{gas}} >1$, meaning that their gas dynamics are dominated by rotation. \\ Regarding the field population, all systems have $\rm{v_{max,\: gas}/\sigma_{gas}} >1$, and hence, they are rotation-dominated.

\subsubsection{Difference between the kinematic and morphological position angles}
\begin{figure}[t]
  \centering
  \captionsetup{width=0.5\textwidth}
    \includegraphics[width=0.5\textwidth,angle=0,clip=true]{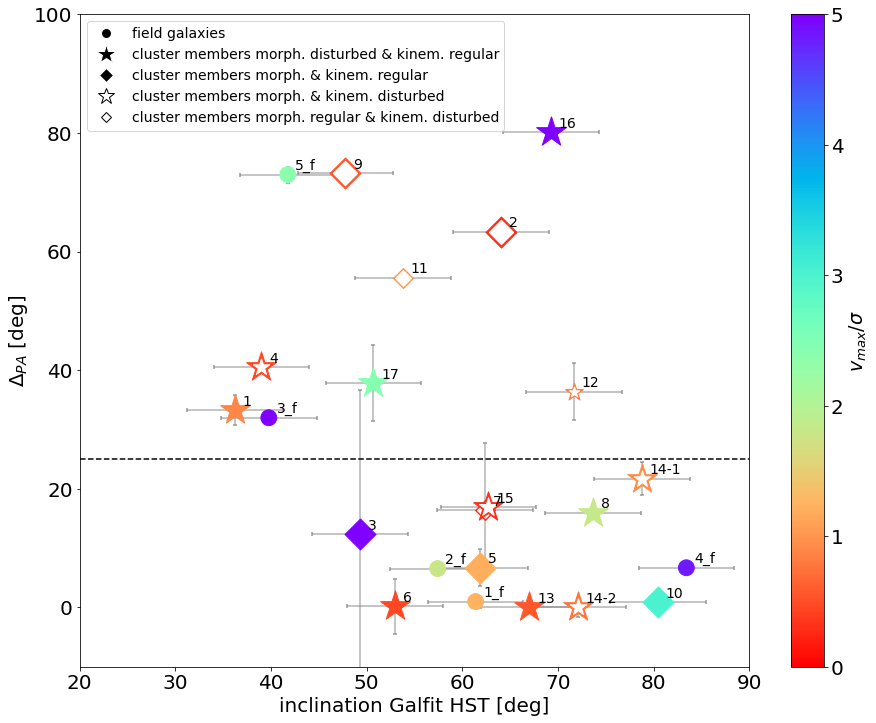}
   \centering   
    \caption{Difference between the morphological and kinematic position angles, $\rm{\Delta_{PA}}$,  as a function of the disk inclination measured from the HST/HAWK-I data. The symbols are the same as in Fig. \ref{BPT}. The dashed line shows the  $\Delta_{PA}=25^{\circ}$ irregularity criterium, as defined by \cite{kut2008}.}
\label{dpa}
\end{figure}

One way to quantitatively determine perturbations is to look for differences between the PAs of the stellar and gas disks in galaxies.
 A large offset between the photometric and kinematic PAs can be caused by interactions or mergers of galaxies and, hence, it can be used as an indicator of such disruptive events (\citealt{bloom}, \citealt{pastuff}).  Also, hydrodynamical interactions between the ISM of cluster galaxies and the ICM can lead to a displacement of the gas component with respect to the stellar one, as suggested by several simulations \citep{Kronberger} and IFU observations of local galaxy clusters \citep{boselli2}.\\
 For this reason, we compute $\rm{\Delta_{PA}}$ as the difference between the photometric and kinematic PA, as derived by \texttt{Galfit} from the HST/HAWK-I data and \texttt{GalPaK3D} from the MUSE data, respectively. This parameter is defined to have an absolute value between $0^{\circ} -90^{\circ}$. \\ 
  E.g. \cite{kut2008} studied the effects of the cluster environment on the kinematics of cluster galaxies at $z\sim0.5$. To measure the irregularity in the gas kinematics, they define 3 parameters, and one of them is the average misalignment between photometric and kinematic PAs. Their adopted irregularity criterium for the misalignment between the two PAs is $\Delta_{PA}>25^{\circ}$, and galaxies in their sample showing $\rm{\Delta_{PA}}$ values lower than this threshold are considered to be kinematically regular. We have adopted the same irregularity criterium for our sample of M0416 cluster and field galaxies. \\
Fig. \ref{dpa} shows the difference between the photometric and kinematic PAs, as a function of the inclination of the stellar disk, for the M0416 cluster and field galaxies. 
For $\sim$ 53\%  of the cluster galaxies analysed in this study, as well as for 3 field galaxies, the photometric and kinematic PAs agree within $<25^{\circ}$. For the rest of $\sim$47\% of the cluster galaxies from our sample, as well as for 2 of the field galaxies, we infer  $\rm{\Delta_{PA}}>25^{\circ}$, and all these systems show signs of morphologic and/or kinematic perturbances. From the 8 cluster galaxies with  $\Delta_{PA}>25^{\circ}$, 5 display a disturbed and peculiar morphology. 
Among the galaxies showing large $\rm{\Delta_{PA}}$ values, the misalignment could be attributed to morphological substructures, such as clumps in the low surface brightness disk (ID \#1, \#4), asymmetric, bent or displaced disks (ID \#12, \#17) and large tidal arms (ID  \#16).  Such substructures may introduce significant errors in the photometric PA estimate. 
The morphologically regular cluster galaxies which also show large  $\rm{\Delta_{PA}}$ values, namely ID \#2, \#9, \#11 have a compact and roundish appearance,  and for them, it is quite complicated to define a major axis and to determine the morphological PA. Also, all have $\rm{v_{max,\: gas}/\sigma_{gas}} \lesssim1$ and show no signs of rotation in their observed velocity fields.  We do not rule out the possibility that RPS is the culprit for the large $\Delta_{PA}$ values we measure for a sub-sample of the cluster galaxies under scrutiny.\\  Two field galaxies, namely \#3-f and \#5-f also show high $\rm{\Delta_{PA}}$ values. We note that field galaxy  \#3-f has a rather complex broadband morphology with large spiral arms, which may introduce uncertainties in the disk inclination estimate. Likewise, field galaxy \#5-f also displays a rather irregular morphology, hinting at gravitational interactions.\\

\subsection{Scaling relations}
In this section, we describe the TFR and stellar mass -- $\rm{S_{0.5}}$ relation for the M0416 cluster and field galaxies. We note, however,  that a caveat of this study is the small sample size. For this reason, we chose to also plot the spatially resolved Hubble Deep Field South sample of emission-line galaxies from \cite{contini16} observed with MUSE, with redshifts 0.2 < z < 1.4 in the subsequent analysis. We have converted their stellar masses to Salpeter masses. \\

\subsubsection{Tully--Fisher relation}
\label{tfr}
\begin{figure}[t]
  \centering
  \captionsetup{width=0.5\textwidth}
    \includegraphics[width=0.5\textwidth,angle=0,clip=true]{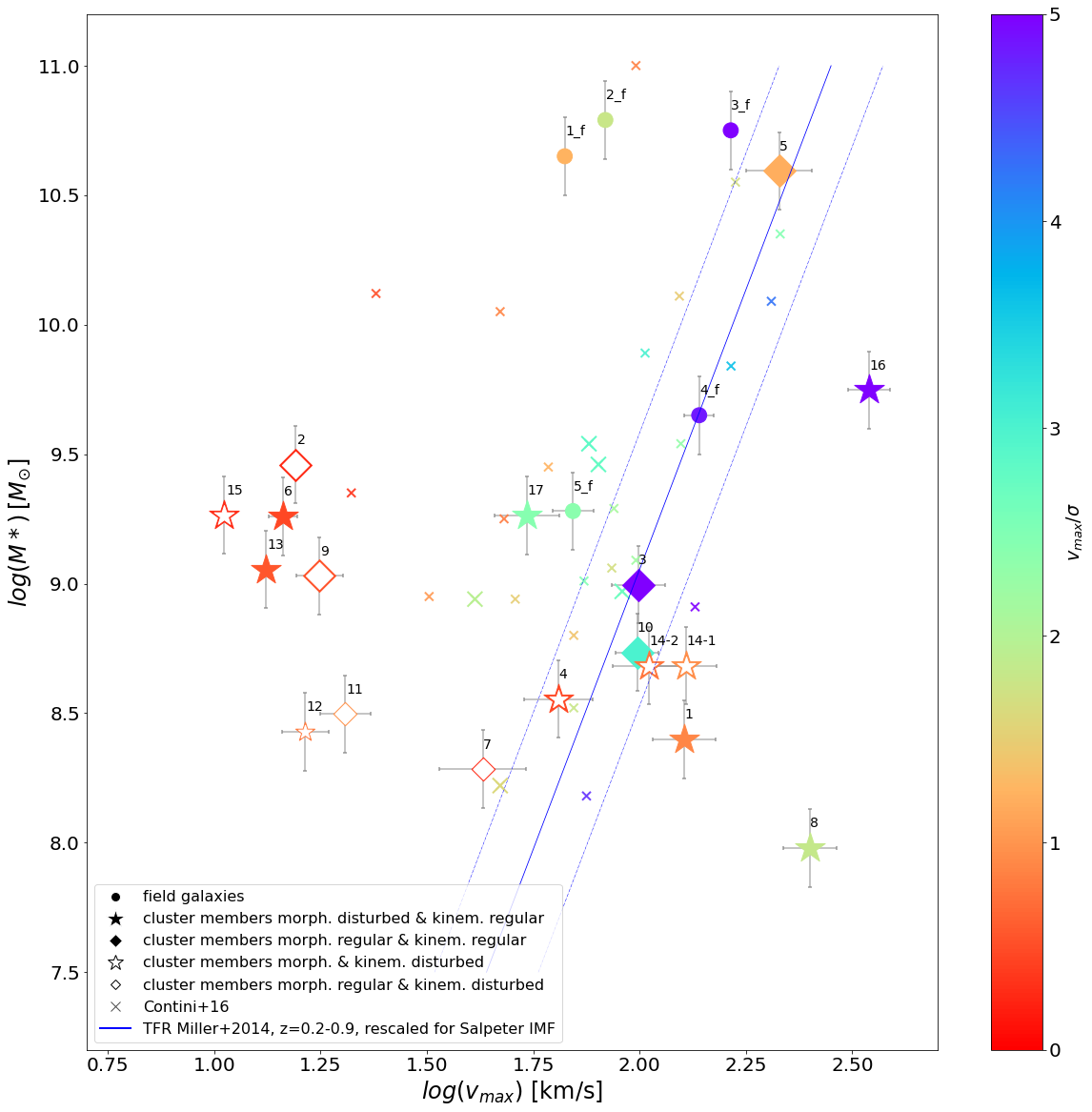}
   \centering   
    \caption{ TFR for the sample of M0416 cluster and field galaxies.  The symbols are the same as in Fig. \ref{BPT}. In addition, we also show here the spatially resolved sample of  MUSE  galaxies from HDFS from \cite{contini16}, with redshifts 0.2 < z < 1.4, as small crosses. The larger, brighter crosses are the galaxies from the latter authors  with 0.3<z<0.5. The blue line shows the TFR  from \cite{miller14}  for low mass galaxies with redshifts  0.2 < z < 0.9.  The dotted lines show the $1\sigma$ scatter around this relation. }
\label{TFR}
\end{figure}

The TFR (\citealt{t-f}) is a tight correlation between the maximum rotation velocity and the stellar mass of late-type galaxies, with a small scatter in velocity (\citealt{cortese}, \citealt{aq18}). \\
Fig. \ref{TFR} shows the TFR for the sample of M0416 cluster and field galaxies.  The blue line shows the TFR for field galaxies with redshifts  0.2 < z <0.9 from \cite{miller14} with its $1\sigma$ scatter ($\sim0.37$ in $\rm{M/M_{\odot}}$ dex), which was rescaled for a Salpeter IMF.   The latter authors find a similar TFR slope to the one previously measured by \cite{miller11}  at higher masses and similar redshifts, albeit with a larger scatter. For this reason, we use the same TFR fit as a reference for the whole mass range probed in this study.\\
Galaxies with gas dynamics dominated  by rotation, defined as having  $\rm{v_{max,\: gas}/\sigma_{gas}}>1$ (blue-purple symbols), mostly follow or are close to the TFR from \cite{miller14}. On the other hand, galaxies with $\rm{v_{max,\: gas}/\sigma_{gas}}<1$ (red-orange symbols), deviate from the TFR towards low $\rm{v_{max}}$ values. Morphologically disturbed galaxies (the stars in Fig. \ref{TFR}), which show signs of gravitational interactions, also mostly depart from the TFR.   However, for the cluster galaxies, there does not seem to be a clear relationship between morphological disturbance and offset from the TFR, as many morphologically undisturbed members (diamonds in Fig. \ref{TFR}) also depart from the relation. However,  these morphologically undisturbed members are kinematically disturbed (open diamonds),  showing no signs of rotation in their observed velocity fields. This could hint at hydrodynamical interactions as the effect driving the offset since such mechanisms leave the stellar body unperturbed.\\ Strikingly, two of the field galaxies, with  $\rm{log(M*/M_{\odot})}>10$ also depart from the TFR. Field galaxy \#1-f is marginally rotation-dominated, with $\rm{v_{max,\: gas}/\sigma_{gas}}=1.2$, and its "S-shaped" morphology hints to tidal interactions, whereas field galaxy \#2-f is a strong AGN.  We do not rule out the possibility that these two galaxies are members of groups (albeit with different z), where gravitational effects are more effective than in a cluster environment. In Fig. \ref{hist}, there are small overdensities of galaxies both at $z\sim0.3$ and $z\sim0.35$, i.e. at the redshifts of the two outlier field galaxies. \\
To sum up, $\sim 30\%$ of the cluster galaxies from our sample follow the TFR of  \cite{miller14} within $1\sigma$, whereas $\sim$12\% of the cluster members fall within $1-2\sigma$. We note that the galaxies following the TFR within $1\sigma$  show $\Delta_{PA}<25$, and hence,  they can be considered kinematically regular. Approximately $ 58\%$ of M0416 galaxies from our sample deviate significantly from the TFR ($>2\sigma$ offset), both towards higher and lower rotational velocities. \\
We measure the highest offsets from the TFR  for objects \#2, \#6, \#13, \#15, which all show $\Delta_{TFR} \: [M_{\odot}] \sim -4$ dex and $\rm{v_{max,\: gas}/\sigma_{gas}}<1$. 
The TFR offsets are not systematically driven by mismatches between photometric and kinematic PA, as many galaxies with $\Delta_{PA}<25^{\circ}$ also depart from the TFR. 
We note that we observe a  trend relating the TFR offsets to the gas dynamics, with galaxies which are more rotation-dominated showing smaller TFR offsets. This indicates that a significant contribution of noncircular, turbulent motions in the gas disk might be contributing to the deviations from the TFR. \\
In conclusion, from Fig. \ref{TFR}, it becomes clear that disturbed galaxies should not be used in a TFR study. For this reason, we analyse the stellar mass -- $\rm{S_{0.5}}$ relation in the following section.\\

\subsection{The stellar mass -- $\rm{S_{0.5}}$ relation}
\begin{figure}[t]
  \centering
  \captionsetup{width=0.5\textwidth}
    \includegraphics[width=0.5\textwidth,angle=0,clip=true]{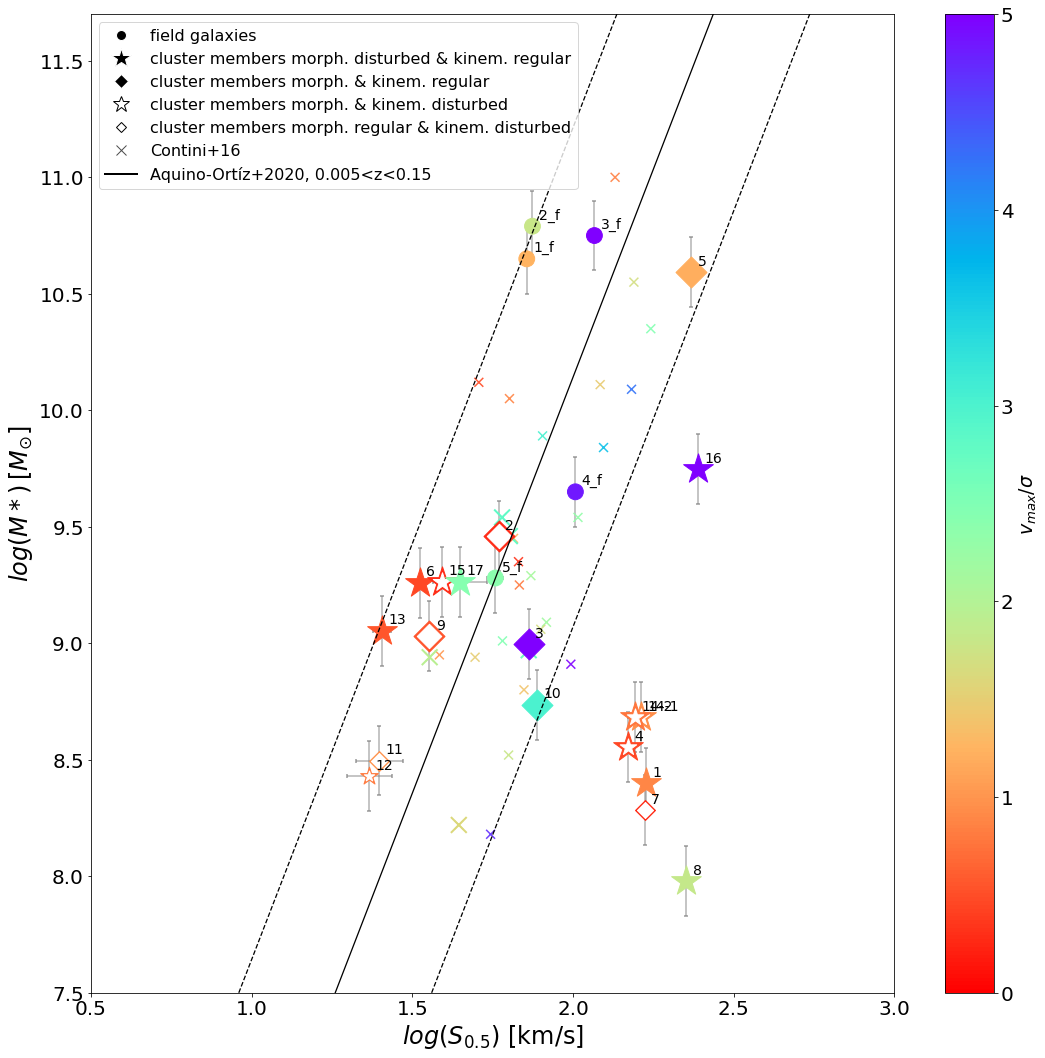}
   \centering   
    \caption{Stellar mass -- $\rm{S_{0.5}}$ relation for the sample of M0416 cluster and field galaxies. The symbols are the same as in Fig. \ref{TFR}. The black line shows the stellar mass -- $\rm{S_{0.5}}$ relation from \cite{aq}, based on a large sample of local galaxies observed as part of the  MaNGA survey. The dashed lines show the 2$\sigma$ scatter around this relation.  }
\label{S05}
\end{figure}

\begin{figure}[t]
  \centering
  \captionsetup{width=0.5\textwidth}
    \includegraphics[width=0.5\textwidth,angle=0,clip=true]{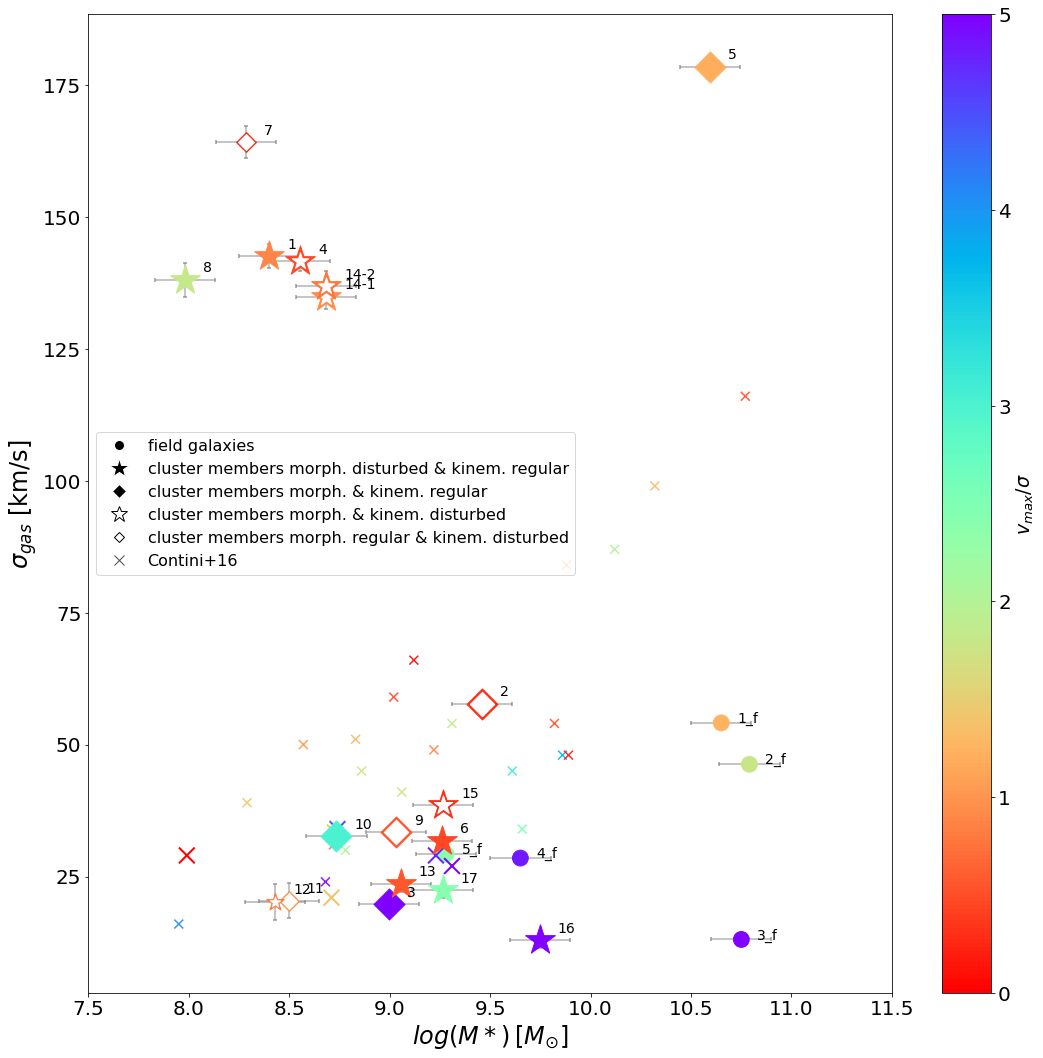}
   \centering   
    \caption{ Stellar mass vs. gas velocity dispersion. The symbols are the same as in Fig. \ref{TFR}. The galaxies showing the highest $\rm{\sigma_{gas}}$ values are the same ones that are strong outlieres in the  stellar mass- $\rm{S_{0.5}}$ plot. }
\label{sigma}
\end{figure}

 \cite{weiner} have introduced a kinematic parameter, $\rm{S_k}$, which combines the dynamical support  of ordered motions given by the maximum stellar rotation velocity  with that from   random motions given by the central stellar velocity dispersion, as:
\begin{equation}
\rm{S_{K}^2=K \cdot v_{max}^2 + \sigma_{*}^2}, 
\end{equation}
where K  is generally assumed to be constant.  Many studies  found that the stellar mass -- $\rm{S_{k}}$  scaling relation has a minimum scatter for K=0.5, and the dispersion around this relation is much smaller than that of the TFR, regardless of the morphology of the galaxies (\citealt{cortese}, \citealt{aq18},  \citealt{barat}).
For a spherically symmetric tracer population,  the velocity dispersion $\rm{\sigma_{gas}}$ is given by $\sigma_{gas}=v_{rot}/\sqrt{\alpha}$ with $K=1/\alpha$, assuming that the motions of the ionised gas trace the gravitational potential of the system (\citealt{binney}). Thus, if a galaxy  (in our case, its gas component) is virialised, then the $\rm{S_{0.5}}$ parameter should provide a better proxy for the kinetic energy of the galaxy, which correlates with its total mass (e.g. \citealt{kassin}). However, one should bear in mind, that the  $\rm{S_{0.5}}$ parameter was devised for the stellar and not for the gas component, and therefore, this parameter is only sensitive to gravity. The gas in a galaxy is, on the other hand, also susceptible to non-gravitational effects. If galaxies strongly depart from the stellar mass-$\rm{S_{0.5}}$   scaling relation, then this might hint at non-gravitational mechanisms affecting their ISM.\\
Fig. \ref{S05} shows the stellar mass -- $\rm{S_{0.5}}$ relation for the sample of M0416 cluster and field galaxies.  The black lines display the stellar mass -- $\rm{S_{0.5}}$  fit from \cite{aq}, based on a large sample of local galaxies observed as part MaNGA survey, together with the 2$\sigma$ scatter around this relation. We have used their fits for the whole sample of 4187  galaxies, including interacting and merging ones, as our small sample of M0416 cluster and field galaxies also contains interacting systems. 
As in \cite{aq}, we define the outliers as objects beyond  2$\sigma$ with respect to the main relation. \\
It is clear that by combining the dynamical support from ordered and unordered motion, the stellar mass -- $\rm{S_{0.5}}$ yields a much tighter relation, with less scatter than the TFR, even for our sample of cluster dwarfs. 
 Approximately $65\%$ of the cluster galaxies  follow the stellar mass -- $\rm{S_{0.5}}$ relation within $2\sigma$, as do the galaxies from the MUSE HDFS sample of \cite{contini16}.  However, 35\% of the cluster members strongly depart from this relation (>4$\sigma$ offset), namely galaxies with IDs \#1, \#4, \#7, \#8 and \#14-1 and\#14-2. Galaxy ID \#16 also departs from this relation, however, for this system, due to strong morphological asymmetries, \texttt{GalPaK3D} converges at a very high rotation velocity, which probably does not reflect the gravitational potential of this galaxy. Regarding the 5 field galaxies, they all follow the mass -- $\rm{S_{0.5}}$ relation within $2\sigma$.\\
To more thoroughly investigate the outliers, we plot in Fig. \ref{sigma} the stellar mass versus the gas velocity dispersion. For the lowest mass cluster galaxies from our sample, we measure the highest values for the gas velocity dispersion. The galaxies that show such high values for the velocity dispersion are the same ones strongly deviating from the stellar mass -- $\rm{S_{0.5}}$ relation. \\
We note that all the outlier galaxies in Figs. \ref{S05} and \ref{sigma} show a peculiar broad-band morphology (except for galaxy \#7). We refer the reader to appendix \ref{a1} for the detailed description of the individual objects.\\


\section{Discussion}
\label{discussion}
The FMR, the morphology and gas velocity fields,  as well as the inferred offsets from the TFR and stellar mass -- $\rm{S_{0.5}}$  relation hint at a combination of environmental mechanisms affecting both the ISM and stellar component of a subsample of the dwarf galaxies analysed in this study. Cluster-specific interactions, both gravitational as well as hydrodynamical are expected to act to some extent on the M0416 cluster members. The M0416 galaxies lie at clustercentric radii $\rm{R\gtrsim R_{200}}$, where such environmental processes are known to be less strong than in the more central regions of the cluster. However, given the small stellar masses of the cluster galaxies, they might be more susceptible to cluster-specific quenching mechanisms.  Recent studies focusing on intermediate redshifts find increasing evidence that environmental quenching does not solely depend on environmental density, but also on the stellar mass of the galaxy, such that quenching is more effective on less massive galaxies  (e.g. \citealt{pc}, \citealt{tan}). Increasing evidence also points to galaxies being affected by the dense environment well beyond the virial radius, the traditionally assumed boundary of the influence of groups and clusters (\citealt{behroozi}). \\E.g. \cite{bahe13} studied the environmental effects which lead to star-formation quenching in cluster galaxies employing a suite of high-resolution cosmological hydrodynamic simulations. They find a systematic reduction of both the hot and cold gas components and a decline in the SF fraction of galaxies with decreasing clustercentric distance. For massive clusters with $\rm{M}\sim10^{15} {\rm M}_{\odot}$  (similar to the mass of the M0416 cluster), the authors estimated the typical values for the ram pressure and concluded that even the most massive galaxies can have their halo gas affected by RPS at cluster-centric distances up to $\sim 2 - 3  \cdot  \rm{R_{200}}$, whereas the lower-mass systems are subject to sufficient RPS of the halo gas even at  $ 5 \cdot \rm{R_{200}}$. As galaxies travel to denser regions of the cluster, the ram pressure will exceed the restoring pressure of the cold gas disk, leading to the complete removal of the cold gas component and a rapid shutoff of star formation.\\
Observational studies also point to such a 'two-phase' quenching mode in galaxy clusters. Both \cite{wetzel} and  \cite{roberts19}  found that cluster galaxies experience a slow quenching phase due to the removal of the hot gas confined in their halos followed by a rapid quenching mode associated with the removal of the cold gas component  as the galaxies approach the cluster centre. \\
The inferred offsets from the FMR hint at a suppressed gas inflow rate as the culprit for the more enhanced abundances we measure for the low mass cluster galaxies from our sample. As is evident from equation \ref{FMR}, an increased metallicity of the inflowing gas $\rm{(Z_{0})}$ or a suppressed gas inflow rate ($\Phi$) can lead to an increase in $\rm{Z_{eq}}$.  We use a model with primordial or enriched gas inflow, as well as our derived  SFRs,  to calculate $\rm{Z_{eq}}$, and we find higher observed gas-phase abundances for our low mass cluster galaxies than predicted by the models. Hence, neither primordial, nor enriched gas inflow can account for the high (O/H)s we measure for a sub-population of cluster galaxies, and therefore, we postulate that a suppressed gas inflow rate, $\Phi$, is leading to the differences between measured and predicted gas-phase metallicities.\\
 The cut-off of inflow of primordial gas from the cosmic web, when the galaxy is accreted into the cluster, together with RPS of the hot halo and/or the removal of the hot halo gas through tidal stripping by the cluster potential might explain why we observe more enhanced metallicities for  low mass cluster galaxies  than predicted by the simple gas-regulated model of \cite{lilly13}. In this scenario,  gas inflow can not dilute the ISM of the galaxy anymore, leading to a successive increase in the gas-phase metallicity.  Also, as RPS is known to act outside-in, this mechanism would first remove the low-metallicity gas confined in the outskirts of the galaxy, also leading to an increased integrated gas-phase metallicity (by up to $\sim 1 $ dex,  \citealt{hughes}). The gas disc is initially left unperturbed by these mechanisms, meaning that star formation can continue until the internal gas reservoir is used up or completely stripped off by RPS as the galaxy approaches denser regions of the cluster, and therefore, there will be a time delay until the system undergoes "rapid" star formation quenching. This might also explain why the cluster galaxies under scrutiny have SFRs fairly consistent with the SF MS at z$\sim0.4$. A similar trend was observed by \cite{maier16} who also analysed the FMR predictions for M0416 cluster galaxies. They find that the cluster members  (the data points depicted as crosses in Fig. \ref{oh}), especially the lower-mass ones, predominantly show more enhanced metallicities than predicted by models and these systems also deviate more strongly from the FMR predictions than galaxies residing in lower density environments. Similar results were also found by \cite{maier19} and \cite{eu} who analysed the gas-phase metallicities of the various cluster members, in comparison to matched samples of field galaxies at similar redshifts. This is also in agreement with \cite{peng15} who studied the stellar metallicities of a large sample of local galaxies, and found that strangulation is the main quenching mechanism in these systems.\\
 From our morpho-kinematic study, we find a large fraction of morphologically and/or kinematically disturbed systems.  By studying the residual maps yielded by the morphological analysis with \texttt{Galfit}, it becomes clear that $\sim 60\%$ of the cluster galaxies display a disturbed broad-band morphology, showing signs of gravitational interactions. However, none of them are observed in close pairs, besides galaxy \#14, hinting at past interactions.  Also, we observe a substantial population of rather small cluster members,  40\% having $\rm{r_{d, stars}}<2$ kpc.
On long time scales, gravitational environment-specific interactions,  such as harassment and tidal interactions with other galaxies as well as with the gravitational potential of the host dark matter halo are known to disrupt the galaxies and lead to an overall reduction of the scale length of the affected systems. At the early stages of harassment, a large fraction (up to 50\%) of the stars are removed, decreasing, thus, the size of the galaxy under scrutiny \citep{boselli06}.\\
 However,  the M0416 cluster members have large line-of-sight velocities relative to the ICM, and at high relative velocities,  gravitational interactions should not be very efficient in the cluster environment.  However, we do not rule out the possibility that some of the cluster galaxies might have been moving as part of an infalling substructure, and hence,  they could have still tidally interacted with other substructure members. 
 Given the complex phase space of the M0416 galaxy cluster, which according to \cite{balestra16} is in a pre-collisional phase, this scenario would be possible. E.g., \cite{gonzalez} measured the surface mass density from a weak-lensing analysis and characterized the overall matter distribution in both the M0416 cluster and parallel fields based on the HST FF data. They find that the surface mass distribution derived for the parallel field shows clumpy overdensities connected by filament-like structures elongated in the direction of the cluster core, reinforcing the idea that, at least some of the cluster galaxies analysed in this study, might be parts of substructures.
Many observations suggest that the physical processes that suppress star formation and eventually reshape the morphology of late-type galaxies in massive haloes begin to act before these galaxies fall into a cluster and that these galaxies may therefore undergo some degree of pre-processing in a group environment (\citealt{miguel}, \citealt{lu}). 
Also, in numerical simulations, massive clusters have been shown to accrete up to 45\% of their galaxies through galaxy groups (\citealt{mcgee}, \citealt{Vijayaraghavan}). E.g. \cite{han} have shown, based on hydrodynamic high-resolution zoom-in simulations of 15 galaxy clusters, that from an observed sample of heavily tidally stripped members in clusters today, nearly three-quarters were previously in a group environment.  They concluded that the visibly disturbed cluster members were more likely to have experienced pre-processing, before entering the cluster.  In galaxy groups, gravitational interactions are much more frequent, due to the lower relative velocities of the group members. Hence, pre-processing might explain the peculiar stellar morphology we observe for some of the cluster members analysed in this study.\\
From the comparison between the structural parameters of the stellar disk and those of the gaseous disk,  we find that 47\% of the cluster galaxies have $\rm{\Delta_{PA}}>25^{\circ}$. Likewise, we find that 23\% of the cluster galaxies have inclinations derived from the MUSE data, which are underestimated with respect to those derived from the photometric data.  Both gravitational and hydrodynamical interactions could lead to such discrepancies.  E.g. \cite{schommer} found that photometric inclination estimates are systematically larger than inclinations derived from kinematics, in accordance with our results. Similarly, \cite{contini16} also found that for some of the galaxies with small disk radii from their sample, the \texttt{GalPaK3D} derived inclinations are underestimated with respect to the values derived with  \texttt{Galfit}.  We do not rule out that the lower spatial resolution of the MUSE data compared to that of the HST observations can also lead to discrepancies between the structural parameters derived by the two different tools. \\
In the TFR, we observe significant scatter, with only  $\sim 30\%$ of cluster galaxies from our sample following the relations defined by \cite{miller14}. This subsample of cluster (and field) galaxies which follow the TFR within $1\sigma$ is represented by the galaxies showing regular morphologies and kinematics, and which have $\rm{v_{max,\: gas}/\sigma_{gas}>1}$.  The TFR offsets, hence,  correlate with $\rm{v_{max,\: gas}/\sigma_{gas}}$. 
Our results are consistent with those of \cite{simons}, who analyzed the stellar mass TFR for a sample of emission-line field galaxies down to low stellar masses at intermediate redshift.  Below a mass of $\rm{log(M*/M_{\odot})}=9.5$, they find significant scatter towards low rotation velocity in the TFR, and galaxies can either be rotation-dominated discs on the TFR or turbulent, asymmetric and/or compact systems departing from the relation towards low $\rm{v_{max}}$ values. Galaxies in their sample which show $\rm{v_{max,\: gas}/\sigma_{gas}}<1$ and depart from the TFR  tend to exhibit quantitative morphologies characteristic of disturbed or compact systems and are statistically distinct from the objects on the TFR. They also find that the residuals from the TFR are a strong function of $\rm{v_{max,\: gas}/\sigma_{gas}}$. \\
In the stellar mass -- $\rm{S_{0.5}}$   plane, we observe the largest offsets from the relation, for the lowest mass galaxies, which show extremely high $\rm{\sigma_{gas}}$ values. All these galaxies also display a peculiar morphology, hinting at pre-processing. However, gravity alone can not explain the high $\rm{\sigma_{gas}}$ we measure for these low mass systems. E.g. \cite{jh} studied the velocity dispersion of a large sample of galaxies at intermediate redshifts. For galaxies from their MUSE sample with low stellar masses ($\rm{log(M/M_{\odot})\sim8}$), they measure a median value of $\rm{\sigma_{gas}} \sim 35$ km/s. This value agrees with the gas velocity dispersion we infer for the non-outlier galaxies from the stellar mass -- $\rm{S_{0.5}}$ relation, but is by far lower than the velocity dispersions we measure for the 5 strong outliers, which have $\rm{\sigma_{gas}} \sim 140$ km/s. \\
RPS of the ISM due to interactions with the ICM leads to an increase in the external pressure, shock formation, thermal instabilities and turbulent motions within the disk, and hence, to an increase in $\rm{\sigma_{gas}}$.  However, we note that the cluster galaxies that depart from the $\rm{S_{0.5}}$-stellar-mass relation are SF galaxies, exhibiting SFRs slightly above the MS with  $\rm{\Delta_{MS}}\sim$ 0.4 dex (except cluster member \#8, which is in the process of quenching). In the early phases of RPS, prior to the removal of a large fraction of the disk gas, this mechanism can boost the overall activity of SF and bring galaxies above the MS, as suggested by several simulations (\citealt{steinh}, \citealt{st}) and observations \citep{vulcani}. Star formation can be triggered by ram pressure depending on the ICM density and the mass of the galaxy and under some conditions,  this hydrodynamical interaction can compress the gas so that it is retained in the central region of the galaxy and forms stars.
This implies that the stellar mass -- $\rm{S_{0.5}}$   outliers might be either experiencing mild RPS of their hot gas component and hence, they might be in the so-called "slow-quenching phase", or, they might be in the early phase of strong RPS of their disk gas, i.e., in the early stage of the "rapid-quenching phase".\\
To conclude, the observational results of this study favour a scenario in which pre-processing is possibly responsible for the peculiar stellar morphologies, whereas cluster-specific hydrodynamical interactions such as RPS are probably leading to the inferred offsets from the mass--SFR, FMR, TFR and mass--$\rm{S_{0.5}}$ relations for a sub-sample of M0416 cluster galaxies.

\section{Summary and conclusion}
\label{concl}
In this paper, we present the morpho-kinematic analysis of 17 emission-line galaxies located close to the virial radius of the M0416 cluster and 5 field galaxies at  $\rm{z}\sim0.4$. 88\%  of the cluster galaxies from our sample have $\rm{log(M/M_{\odot})<9.5}$. The analysis of the physical properties and kinematics of the ionised gas is based on VLT/MUSE IFU spectroscopy from the MUSCATEL survey, whereas the structure and morphology of the galaxies are analysed using high-resolution HST and HAWK-I photometric observations. 
Our main findings can be summarized as follows:
\begin{enumerate}
\item  Ionising mechanisms: We used the BPT diagnostic diagram to differentiate between the main sources of ionisation within the cluster and field galaxies. Both populations follow the SF and composite sequence in the BPT diagram, with only one field galaxy securely classified as an Seyfert-II.
\item M-SFR relation: The majority of cluster and field galaxies in our sample can be classified as SF MS galaxies, with just 4 cluster galaxies being in the process of quenching. 
\item MZR and FMR: We observe significant scatter around the local SDSS relation in the MZ plane for the M0416 cluster and field galaxies, both towards lower and higher gas-phase metallicities. Following the prescriptions from the $\rm{Z(M, SFR)}$ of \cite{lilly13}, we find higher gas-phase metallicities for the lowest mass cluster galaxies, than predicted by models assuming primordial or metal-enriched gas inflow, while the higher mass systems are in agreement with the model predictions.  This discrepancy between observed and predicted metallicities indicates that the shutoff of low-metallicity gas inflow and/or the removal of the hot halo gas due to RPS/tidal stripping by the cluster potential might be responsible for the enhanced gas-phase abundances of the lowest mass cluster galaxies. 
\item Morpho-kinematics: $\sim60\%$ of the M0416 cluster members show a rather disturbed broad-band morphology,  hinting at gravitational interactions. Most probably, the galaxies showing the most irregular morphologies were pre-processed prior to the accretion into the cluster.\\
For $\sim65\%$   of the cluster galaxies, we infer $\rm{v_{max,\: gas}/\sigma_{gas}} < 1$, meaning that these galaxies have a substantial contribution of unordered motions to their velocity fields.  We emphasize that some of the systems showing  $\rm{v_{max,\: gas}/\sigma_{gas}} < 1$ are not intrinsically pressure-supported systems, but their $\rm{\sigma_{gas}}$ is most probably increased  by externally-driven environmental processes. Also, the spectral resolution of the MUSE instrument is limiting us in accurately measuring low values for $\rm{\sigma_{gas}}$, biasing our inferred $\rm{v_{max,\: gas}/\sigma_{gas}} < 1$ values for dwarfs with very low rotational velocities.\\ We quantitatively determine perturbations by analysing the differences between morphological and kinematical PAs. For $\sim$47\%  of  the cluster members, we measure $\Delta_{PA} > 25^{\circ}$, and the discrepancy between morphological and kinematical PAs can be attributed to both gravitational and/or hydrodynamical environmental processes.
\item TFR: galaxies showing $\rm{v_{max,\: gas}/\sigma_{gas}} >1$ follow the TFR, whereas the ones showing $\rm{v_{max,\: gas}/\sigma_{gas}} < 1$ depart from the relation towards low $\rm{v_{max}}$ values. Morphologically disturbed galaxies also mainly depart from the TFR, however, there does not seem to be a clear trend between morphological disturbance and offset from the scaling relation.  Many morphologically undisturbed, but kinematically disturbed cluster members also depart from the TFR, hinting at hydrodynamical interactions as the mechanisms driving the offsets.
\item stellar mass-$\rm{S_{0.5}}$ relation: In the mass-$\rm{S_{0.5}}$ plane, we observe a much tighter relation, with less scatter than in the TFR. However, we observe a population of galaxies strongly departing (>$4\sigma$) from this relation. The outliers show very high $\rm{\sigma_{gas}}$ values for their low stellar masses. We interpret these results to be consistent with a scenario in which the combined effect of RPS and gravitational interactions is boosting the  $\rm{\sigma_{gas}}$ of these cluster members, making them outliers in the  mass - $\rm{S_{0.5}}$ plane.
\end{enumerate}
This paper demonstrates the power of the VLT/MUSE IFU spectrograph, which allows us to study the spatially resolved gas kinematics even for low mass galaxies at intermediate redshifts.  We note, however, that a big caveat of this pilot study is the small sample size. In the future, we want to increase the sample size by analysing more spatially resolved late-type cluster galaxies at different redshifts in terms of their stellar morphologies, gas kinematics and physical properties of the ionised gas.

\begin{acknowledgements}
We would like to express our deep gratitude to the anonymous referee for providing constructive comments and help in improving the manuscript.
This study used ESO archival data from observations conducted by the MUSCATEL team,  programme ID: 1104.A-0026 and PI: Wisotzki, L.. This research made use of the following \texttt{PYTHON} packages: \texttt{Astropy} \citep{astropy}, \texttt{numpy} \citep{numpy}, \texttt{matplotlib} \citep{plot}, \texttt{MPDAF} \citep{mpdaf}.
\end{acknowledgements}


\begin{appendix}
\section{Description of the individual galaxies and their morpho-kinematic maps}
\label{a1}

In this section, we describe the morpho-kinematic maps for all galaxies analysed in this study. \\
The scale of 1'' corresponds to  $\sim 5$ kpc at the redshift of z=0.39 of the M0416 galaxy cluster.  
The MUSE data has a spatial sampling of 0.2"/px in the rest frame wavelength range  3417-6727 $\:\AA$ and a spectral resolution of $\sim2.5 \:\AA$. The HST F81W data probes the rest wavelength range of $\sim 4900-6875\:\AA$, with a plate scale of 0.03"/px. The HAWK-I Ks band imaging probes the rest-frame wavelength of 355-2130 nm with a plate scale of 0.106"/px at the redshift of the cluster.\\
The differences in extent between the observed and model flux, velocity and velocity dispersion maps are due to the different SNR thresholds used. For the observed maps, we discard spaxels with SNR lower than 10 to 15, whereas the \texttt{GalPaK3D} tool fits spaxels with SNR>3. Due to limitations imposed by the \texttt{MPDAF} Gaussian fitting routine, we can not acurately measure fluxes with this tool in spaxels with SNR<10 accurately.\\
The x- and y-axis scales, as well as the colour-bar limits,  are the same for all panels in one figure, but are different from one figure to the other.\\

\begin{figure}[t]
  \centering
  \captionsetup{width=0.5\textwidth}
  \includegraphics[width=0.5\textwidth,angle=0,clip=true]{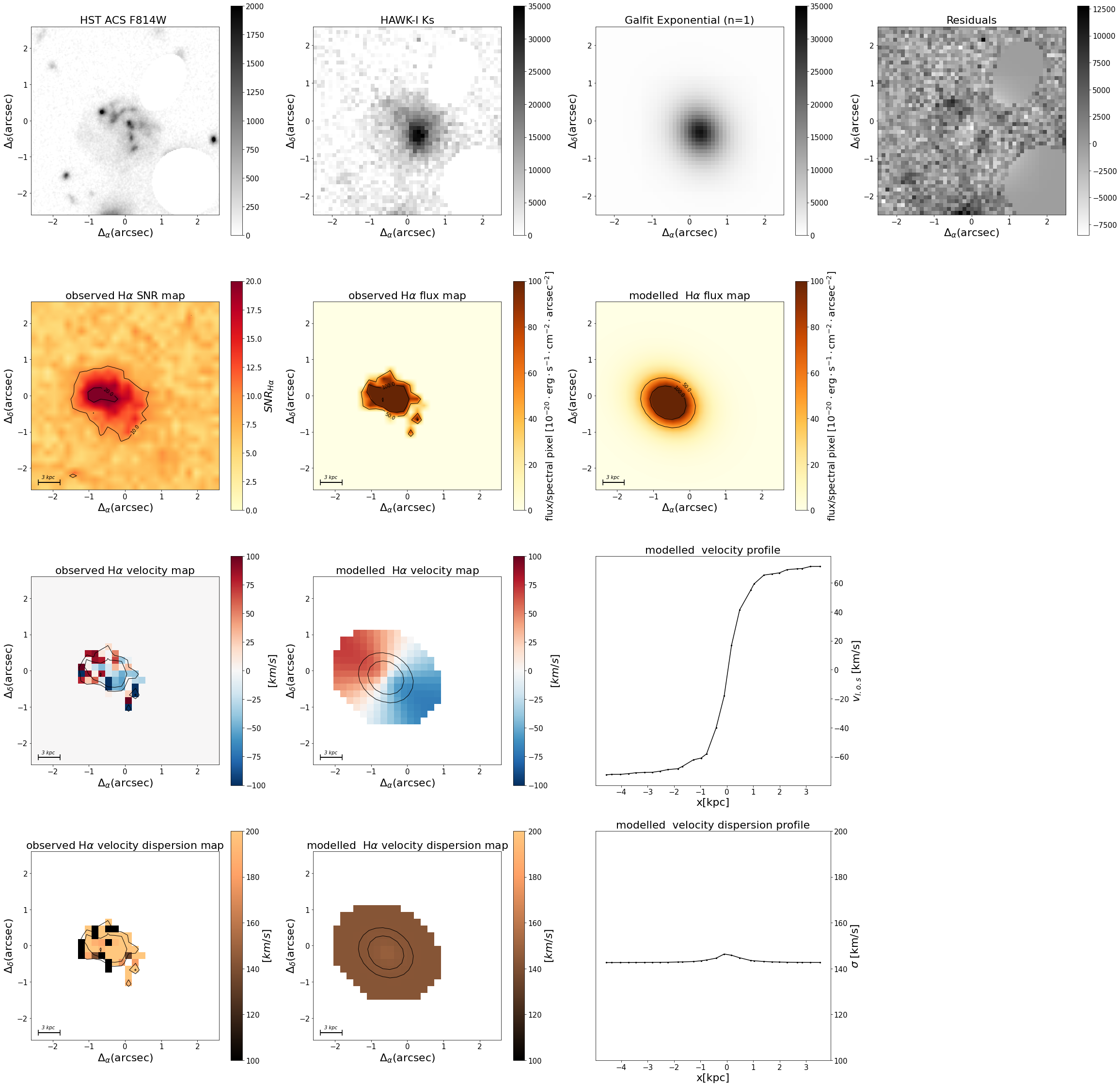}
    \caption{Morpho-kinematic analysis for cluster galaxy  \#1.
 \textit{top row left panel:} HST/ACS F814W image;  \textit{top row second panel:}  HAWK-I Ks image; \textit{top row third panel:} \texttt{Galfit} model (exponential profile); \textit{top row right panel:} \texttt{Galfit} residuals.  The rest is the same caption as in Fig. \ref{ID10}. 
This galaxy is classified as being morphologically disturbed and kinematically regular. It has the following physical parameters: $\rm{log(M/M_{\odot})}=8.39$; $\rm{v_{max}} = 127$ km/s; $\rm{v_{max,\: gas}/\sigma_{gas}}=0.9$; offset from SF MS: $\rm{\Delta_{MS}}=0.32$; offset from TFR:  $\rm{\Delta M_{TFR}}=1.08$; offset from stellar mass -- $\rm{S_{0.5}}$:  $\rm{\Delta M_{S_{0.5}}}=2.5$.}  
\label{ID1}
\end{figure}

\emph{\#1:} Fig. \ref{ID1} shows the morpho-kinematic analysis of the cluster galaxy with ID1 at a redshift z=0.38. 
Galaxy \#1 shows an irregular, patchy morphology at optical wavelengths, with many bright spots, which are probably SF regions, and a multi-component decomposition into bulge and disk can not properly fit this object's spatial light distribution.  The morphology of this dwarf galaxy is similar to that of the so-called "clump-cluster galaxies" (\citealt{clump}), however, such objects are generally more massive and found at higher redshifts.  
We performed the morphological analysis for this object, by fitting a single exponential profile on the HAWK-I IR image, owing to its smoother appearance at redder wavelengths.
For the stellar disk, we measure an inclination of  i=$36^{\circ}$ and a $\rm{r_{d}}$=1.8 kpc.\\
The $\rm{H\alpha}$ disk modelling reproduces well the observed velocity field, however, the inclination has been fixed to the value obtained for the stellar disk with \texttt{GalPaK3D} because of convergence problems. For the gas disk, we measure a $\rm{r_{d}}=3.3$ kpc, and thus, it is more extended than the stellar disk.  The kinematic and photometric PAs of this object differ by $\sim 33^{\circ}$.\\
The velocity field derived from the $\rm{H\alpha}$ line shows a regular rotation pattern within $\pm$ 70 km/s , and the maximum velocity reached at the plateau is $\rm{v_{max}} = 127$ km/s. 
The recovered velocity dispersion is very high for such a low mass object, in the order of $\rm{\sigma_{gas}} = 142$ km/s. The velocity dispersion map is flat, except for the centre.  
This object has an extremely clumpy and irregular morphology, which can also contribute to the high velocity dispersion that we measure, as observations have shown that high values for  $\rm{\sigma_{gas}}$ are expected for clumpy disks (\citealt{Bournaud}).
The gas dynamics of this galaxy are dominated by random motion, with a ratio  $\rm{v_{max,\: gas}/\sigma_{gas}}=0.9$, and hence, this cluster member strongly departs from the TFR.\\

\begin{figure}[t]
  \centering
  \captionsetup{width=0.5\textwidth}
    \includegraphics[width=0.5\textwidth,angle=0,clip=true]{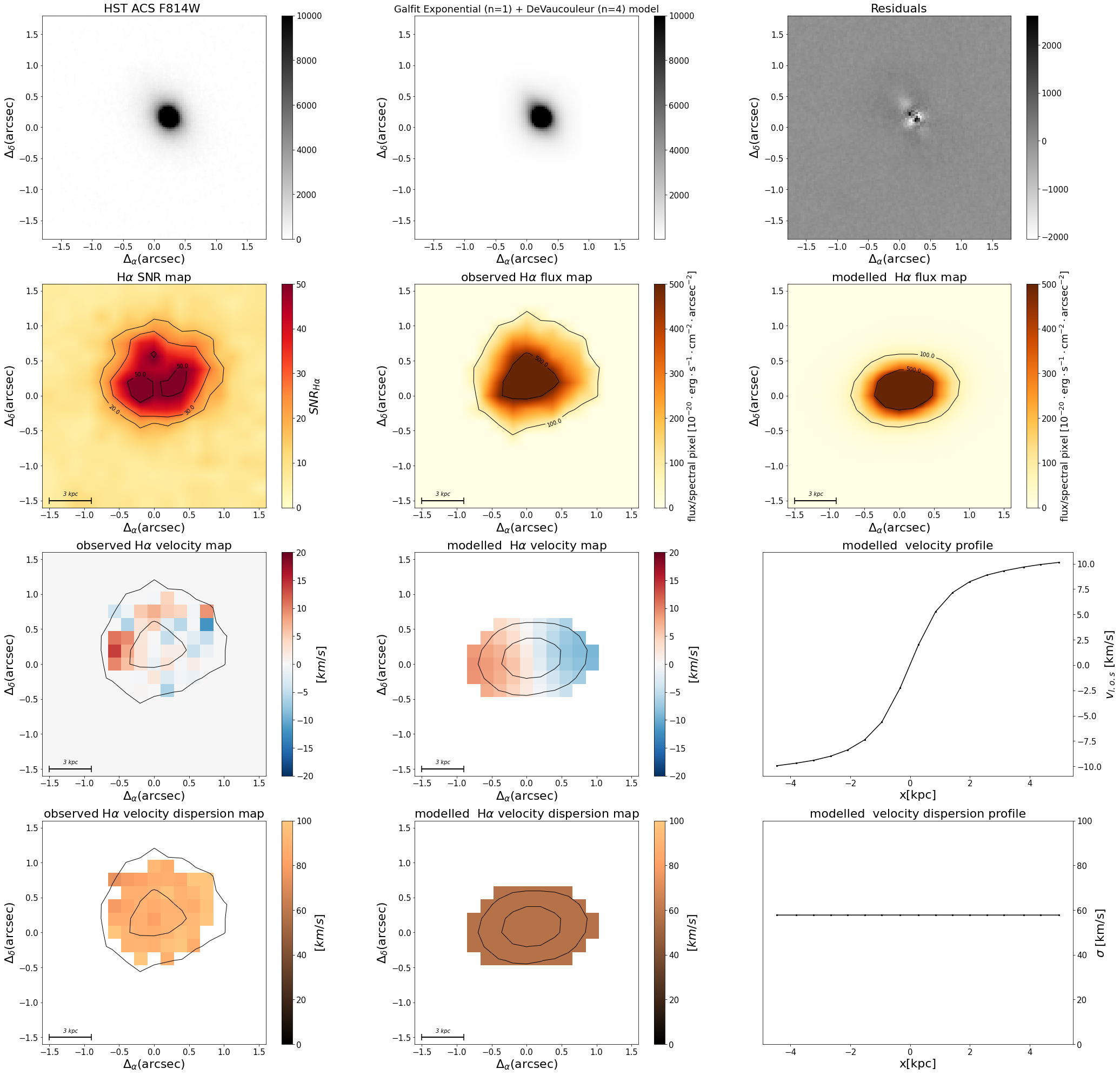}   
    \caption{Morpho-kinematic maps for galaxy \#2. Same caption as Fig. \ref{ID10}.  This galaxy is classified as being morphologically regular and kinematically disturbed.   It has the following physical parameters: $\rm{log(M/M_{\odot})}=9.46$; $\rm{v_{max}} = 15$ km/s; $\rm{v_{max,\: gas}/\sigma_{gas}}=0.27$; offset from SF MS: $\rm{\Delta_{MS}}=0.42$; offset from TFR:  $\rm{\Delta M_{TFR}}=-3.94$; offset from stellar mass -- $\rm{S_{0.5}}$:  $\rm{\Delta M_{S_{0.5}}}= -0.14$.}  
\label{ID2}
\end{figure}

\emph{\#2:} Fig.  \ref{ID2} shows the morpho-kinematic analysis of the galaxy with ID2, which is a cluster member at  z=0.39. This stellar disk is moderately inclined, with  an inclination of  i=$ 64^{\circ}$ and a $\rm{r_{d}}$= 0.74 kpc,  making it one of the smallest dwarfs of our sample. \\
For the gas $\rm{H\alpha}$ disk, we measure an inclination of  i=$47^{\circ}$,  in good agreement with the one recovered for the stellar disk using the HST data, and a $\rm{r_{d}}=1.79$ kpc. 
The kinematic and photometric PAs of this object differ by $\sim 63^{\circ}$.
The disk kinematic modelling reproduces more or less the observed $\rm{H\alpha}$ velocity field, which is somewhat convoluted. The gas velocity field shows a weak gradient from $\pm$ 10 km/s and is, thus, almost flat, with a $\rm{v_{max}}=15$ km/s reached at the plateau. The measured velocity dispersion is high, in the order of $\rm{\sigma_{gas}} = 58$ km/s and the dispersion map is also flat. Thus, the gas dynamics of this galaxy are dominated by random motion, with a ratio  $\rm{v_{max,\: gas}/\sigma_{gas}}=0.27$.\\

\begin{figure}[t]
  \centering
  \captionsetup{width=0.5\textwidth}
    \includegraphics[width=0.5\textwidth,angle=0,clip=true]{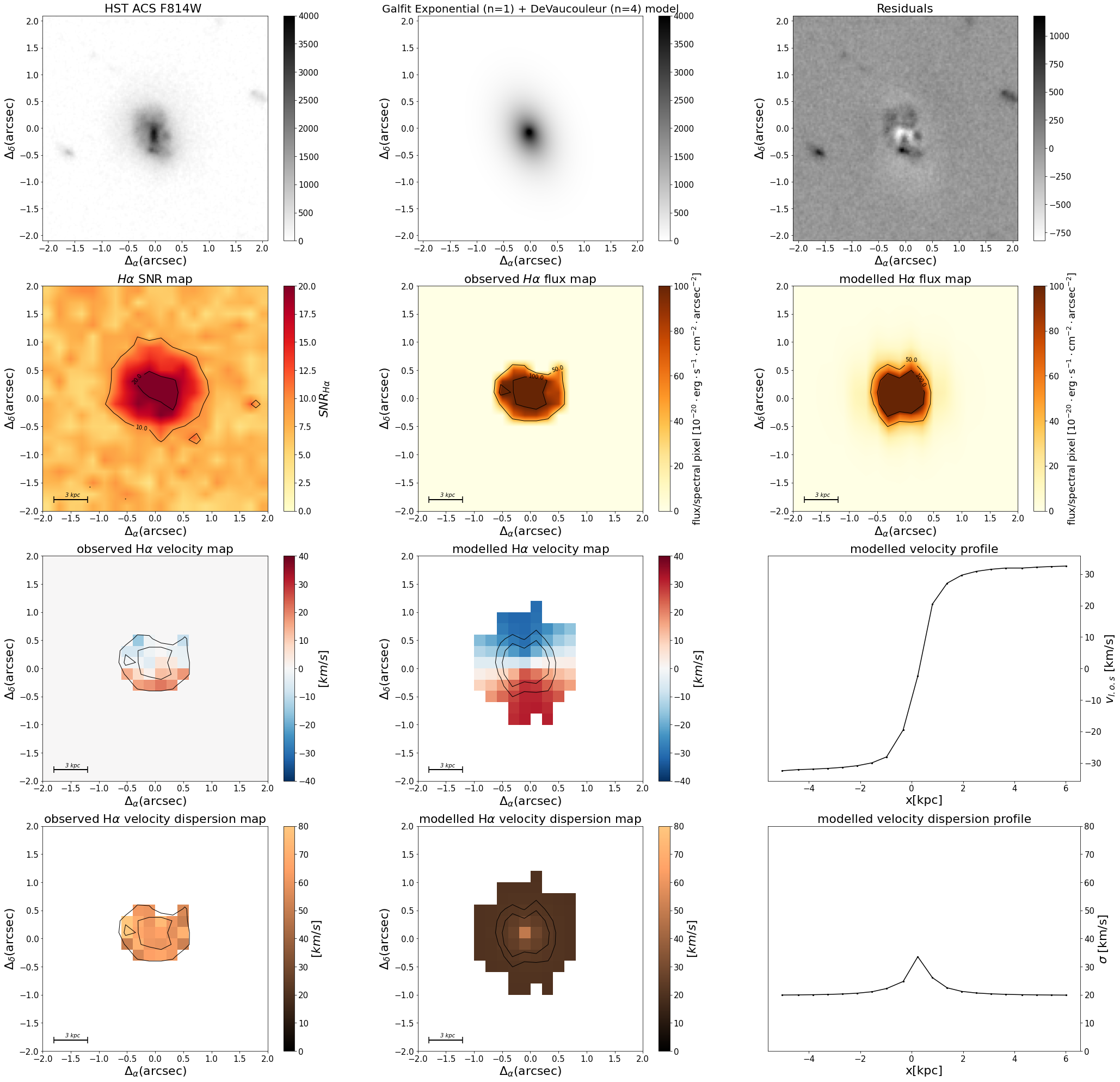}
    \caption{ Morpho-kinematic maps for galaxy \#3. Same caption as Fig.  \ref{ID10}. This galaxy is classified as being morphologically and kinematically regular.   It has the following physical parameters: $\rm{log(M/M_{\odot})}=9.0$; $\rm{v_{max}} = 99$ km/s; $\rm{v_{max,\: gas}/\sigma_{gas}}=5.03$; offset from SF MS: $\rm{\Delta_{MS}}=-0.18$; offset from TFR:  $\rm{\Delta M_{TFR}}= 0.02$; offset from stellar mass -- $\rm{S_{0.5}}$:  $\rm{\Delta M_{S_{0.5}}}=0.65$.  }  
\label{ID3}
\end{figure}

\emph{\#3:} Fig.  \ref{ID3} shows the morpho-kinematic analysis of the object with ID3,  a cluster galaxy with z=0.391. From the morphological analysis of the HST image, we measure an inclination of  i=$ 49.3^{\circ}$ and a $\rm{r_{d}} $= 1.6 kpc for the stellar disk. 
 In the residual map, some bright SF clumps become visible.\\
The $\rm{H\alpha}$ disk modelling reproduces well the observed velocity field, but the measured inclination of the gas disk does not agree with the one of the stellar disk, hinting at the fact that the gas does not follow the gravitational potential of the stars. Hydrodynamic interactions with the ICM can lead to differences in the normal vector of the stellar and gas disk.  
The inclination of the gas disk yielded by \texttt{GalPaK3D}  is i=$ 20^{\circ}$  and the disk scale length is $\rm{r_{d}} $= 1.8 kpc.   The kinematic and photometric PAs of this object agree, with $\Delta_{\rm{PA}}=12^{\circ}$.
The gas velocity field shows a weak gradient from $\pm$ 30 km/s, with a $\rm{v_{max}}=99$ km/s reached at the plateau. The measured velocity dispersion is low, in the order of $\rm{\sigma_{gas}} =19$ km/s and the dispersion map shows a peak at the centre. The gas dynamics of this cluster member are dominated by rotation, with a ratio  $\rm{v_{max,\: gas}/\sigma_{gas}}=5$.\\\\

\begin{figure}[t]
  \centering
  \captionsetup{width=0.5\textwidth}
    \includegraphics[width=0.5\textwidth,angle=0,clip=true]{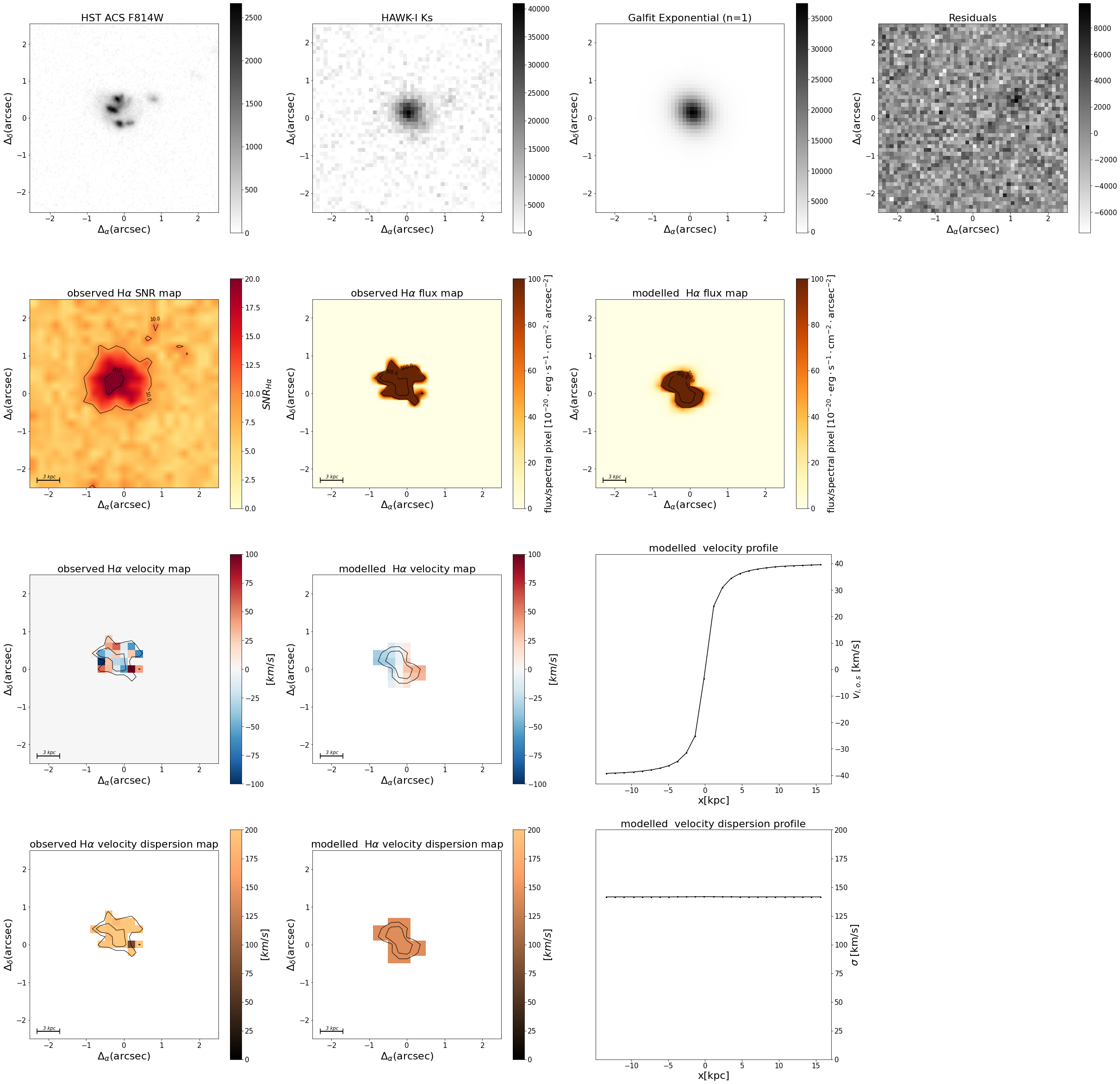}
    \caption{Morpho-kinematic maps for galaxy \#4.  \textit{top row left panel:} HST/ACS F814W image;  \textit{top row second panel:}  HAWK-I Ks image; \textit{top row third panel:} \texttt{Galfit} model (exponential profile); \textit{top row right panel:} \texttt{Galfit} residuals.  The rest is the same as in Fig. \ref{ID10}. This galaxy is classified as being morphologically and kinematically disturbed.  It has the following physical parameters: $\rm{log(M/M_{\odot})}=8.55 $; $\rm{v_{max}} =64$ km/s; $\rm{v_{max,\: gas}/\sigma_{gas}}=0.46 $; offset from SF MS: $\rm{\Delta_{MS}}=0.22$; offset from TFR:  $\rm{\Delta M_{TFR}}= -0.36$; offset from stellar mass -- $\rm{S_{0.5}}$:  $\rm{\Delta M_{S_{0.5}}}= 2.2$. }  
\label{ID4}
\end{figure}

\emph{\#4:} Fig.  \ref{ID4} shows the morpho-kinematic analysis of the galaxy with ID4, a cluster member at a redshift of  z=0.381. This dwarf galaxy displays an asymmetric, peculiar morphology in the HST image, with many clumps which form a semi-circle. The morphology of this dwarf galaxy is also similar to that of the "clump-cluster galaxies" (\citealt{clump}).
We measure a stellar disk inclination of  i$=39^{\circ}$ and $\rm{r_{d}}= 0.85$ kpc from the  IR HAWK-I data.\\ The observed $\rm{H\alpha}$ flux map is  highly skewed,  as is the observed velocity field. Given the strong morphological asymmetries in the HST image and emission-line flux map, the  3D kinematic modelling has difficulty converging. 
For this reason, we have run \texttt{GalPaK3D} by keeping the inclination and disk scale length fixed to the values obtained for the stellar component from the IR photometry. By doing so, we recover a $\rm{v_{max}}=64$ km/s, and a velocity dispersion of  $\rm{\sigma_{gas}}=141$km/s, a  very high value for such a low  mass object.  Its clumpy and irregular morphology possibly contribute to the high velocity dispersion that we measure.
The gas dynamics of this cluster member are dominated by random motion, with a ratio  $\rm{v_{max,\: gas}/\sigma_{gas}}=0.46$. The kinematic and photometric PAs differ by $40^{\circ}$.  The $\rm{H\alpha}$ velocity field shows a weak gradient from $\pm$ 10 km/s and the dispersion map is flat. \\\\

\begin{figure}[t]
  \centering
  \captionsetup{width=0.5\textwidth}
    \includegraphics[width=0.5\textwidth,angle=0,clip=true]{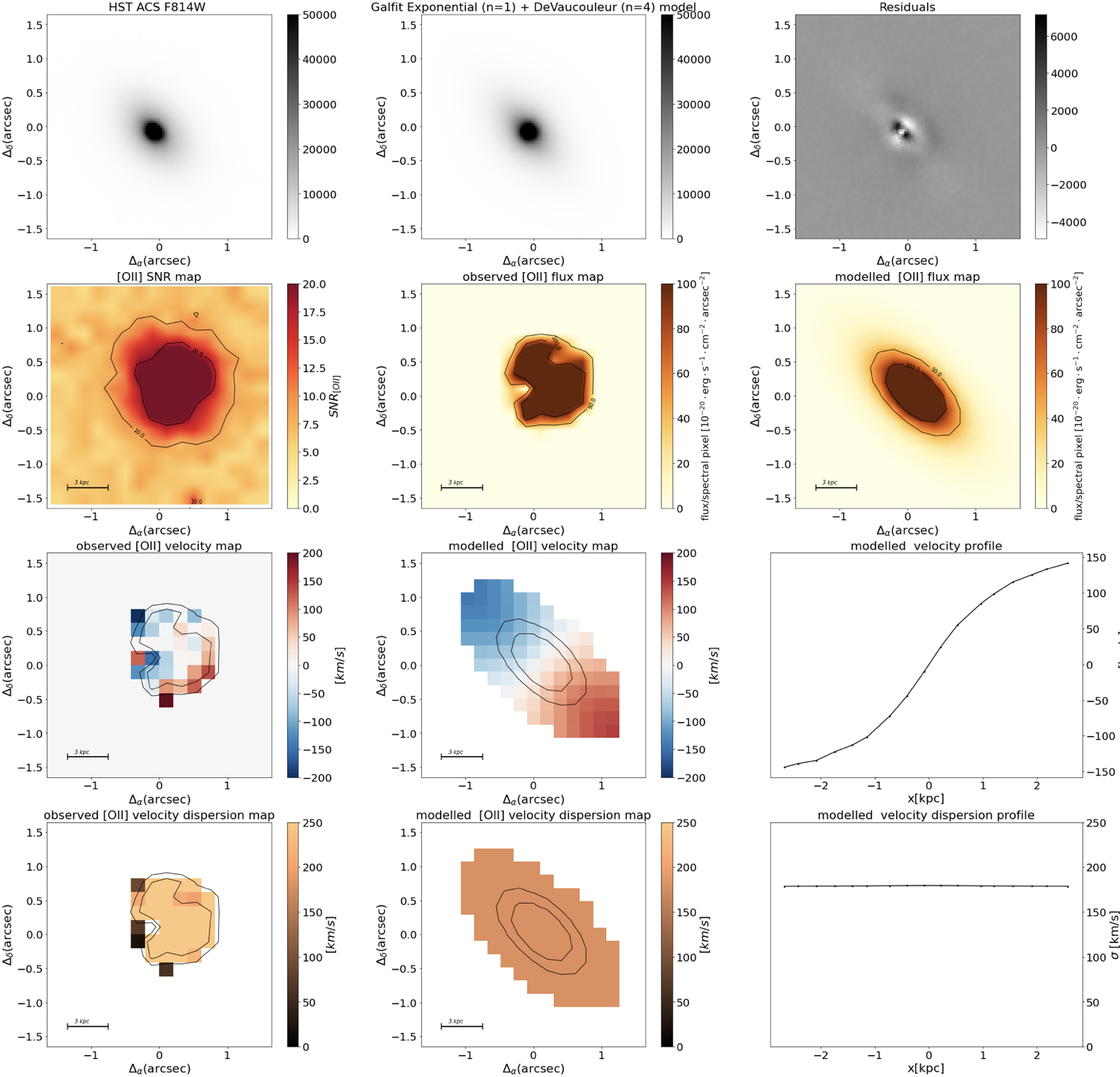}
   \centering   
    \caption{ Morpho-kinematic maps for galaxy \#5. Same caption as Fig.  \ref{ID10}. This galaxy is classified as being morphologically and kinematically regular.  It has the following physical parameters: $\rm{log(M/M_{\odot})}=10.59 $; $\rm{v_{max}} =212$ km/s; $\rm{v_{max,\: gas}/\sigma_{gas}}=1.19 $; offset from SF MS: $\rm{\Delta_{MS}}=-3.6$; offset from TFR:  $\rm{\Delta M_{TFR}}= -0.14$; offset from stellar mass -- $\rm{S_{0.5}}$:  $\rm{\Delta M_{S_{0.5}}}= 0.86$.}  
\label{ID5}
\end{figure}

\emph{\#5:} Fig.  \ref{ID5} shows the morpho-kinematic analysis of the galaxy with ID5, the brightest and most massive cluster member from our sample, with  z=0.384.  The stellar disk of this galaxy is moderately inclined, with an inclination of  i=$ 61^{\circ}$ and a $\rm{r_{d}} $= 1.5 kpc. 
The integrated spectrum of this object shows strong absorption features, and only fairly weak [OII] and $\rm{H\alpha}$ emission, typical for an earlier-type galaxy. \\
The disk modelling reproduces well the observed velocity field, however, the inclination has been fixed to the value obtained for the stellar disk with \texttt{GalPaK3D} because of convergence issues. For the gas [OII]$\lambda 3727 \:\AA$ disk, we find  $\rm{r_{d}}=3.1$ kpc, and hence, the stellar disk is more compact than the [OII] disk. 
The kinematic and photometric PAs of galaxy \#5 agree,  with $\Delta_{\rm{PA}}=6^{\circ}$. The gas velocity field shows a strong gradient from $\pm$ 150 km/s,  and we measure  $\rm{v_{max}}=212$ km/s. The velocity dispersion is the order of $\rm{\sigma_{gas}} = 178$ km/s,  and the dispersion map is rather flat. The dynamics of the [OII] disk of this galaxy are dominated by rotation, with a ratio  $\rm{v_{max,\: gas}/\sigma_{gas}}=1.2$. This object is probably transitioning from a rotation-dominated to a dispersion-dominated phase, due to the quenching of its SFR, as this system lies below the SF MS in the stellar mass-SFR plane.  \\\\

\begin{figure}[t]
  \centering
  \captionsetup{width=0.5\textwidth}
    \includegraphics[width=0.5\textwidth,angle=0,clip=true]{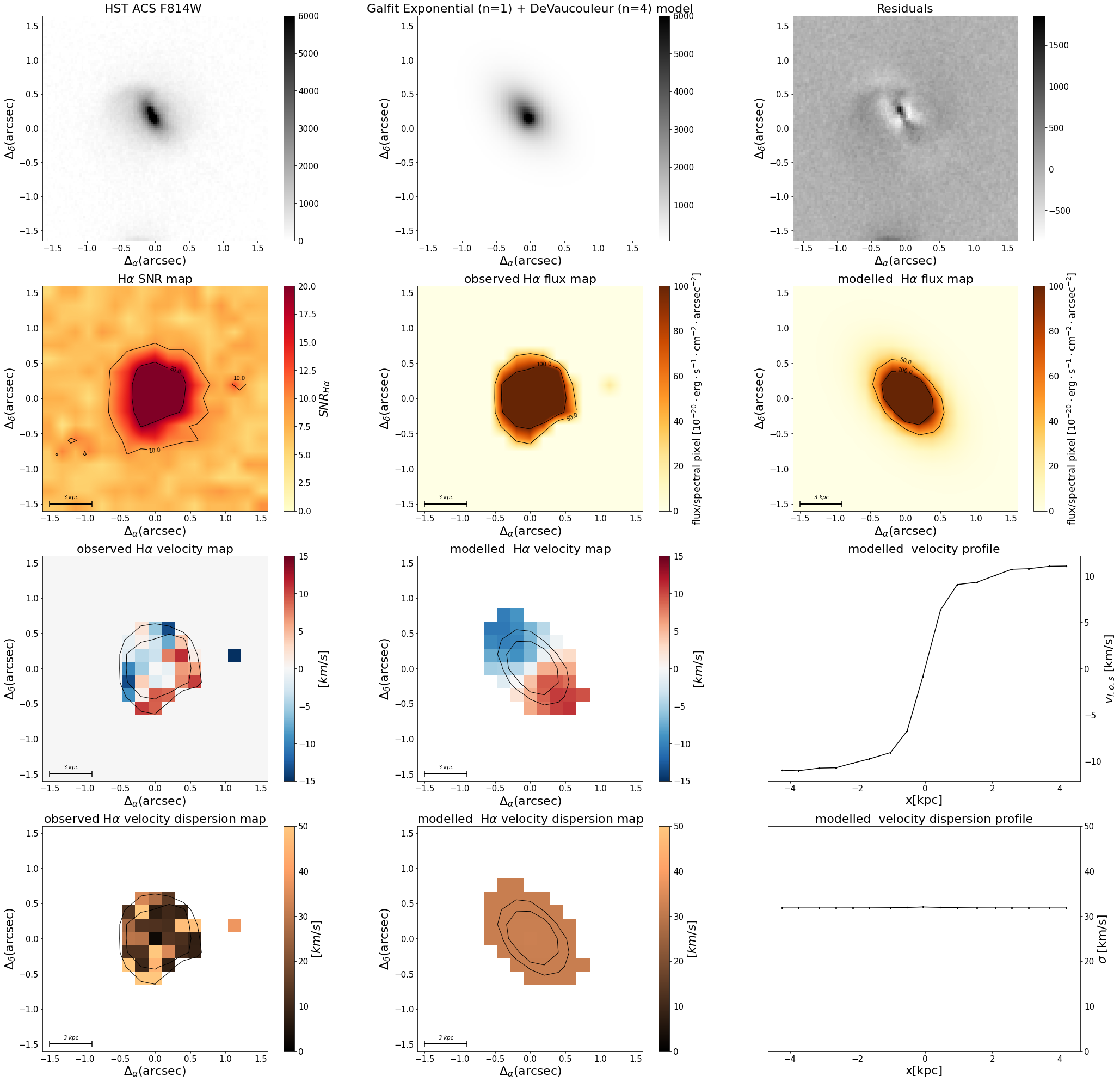}
   \centering   
    \caption{ Morpho-kinematic maps for galaxy \#6. Same caption as Fig.  \ref{ID10}. This galaxy is classified as being morphologically disturbed and kinematically regular.  It has the following physical parameters: $\rm{log(M/M_{\odot})}=9.26 $; $\rm{v_{max}} =15$ km/s; $\rm{v_{max,\: gas}/\sigma_{gas}}=0.46 $; offset from SF MS: $\rm{\Delta_{MS}}=0.03$; offset from TFR:  $\rm{\Delta M_{TFR}}= - 3.88$; offset from stellar mass -- $\rm{S_{0.5}}$:  $\rm{\Delta M_{S_{0.5}}}= -0.82$. }  
\label{ID6}
\end{figure}

\emph{\#6:} Fig.  \ref{ID6} shows the morpho-kinematic analysis of the galaxy with ID6, a cluster member at z=.4001. From the morphological analysis of the HST data, we measure  i=$ 53^{\circ}$ and a $\rm{r_{d}}=1.13$ kpc for the stellar disk. 
In the residual map, it becomes evident that this object shows an elongated core, which might be a bar, as well as an extended spiral arm to the N-W with no analogue to the S-E. This might hint at the fact that this galaxy is either a merger remnant or has experienced tidal interactions in the past.\\
The $\rm{H\alpha}$ disk modelling reproduces well the observed velocity field, however, due to convergence issues for the turnover radius, the  inclination was fixed to the value of the stellar disk. 
By doing so, we measure a  $\rm{r_{d}} $= 1.9 kpc for the gas disk, which is more extended than the stellar disk.   The kinematic and photometric PAs of this object agree perfectly, with $\Delta_{\rm{PA}}<1^{\circ}$.
The velocity field of the gas disk shows a weak gradient from $\pm$ 10 km/s, with a $\rm{v_{max}}=15$ km/s. The velocity dispersion of the disk is in the order of $\rm{\sigma_{gas}} =31$ km/s and the dispersion map is rather flat. The gas dynamics of this dwarf are dominated by random motion, with a ratio $\rm{v_{max,\: gas}/\sigma_{gas}}=0.46$.\\\\

\begin{figure}[t]
  \centering
  \captionsetup{width=0.5\textwidth}
    \includegraphics[width=0.5\textwidth,angle=0,clip=true]{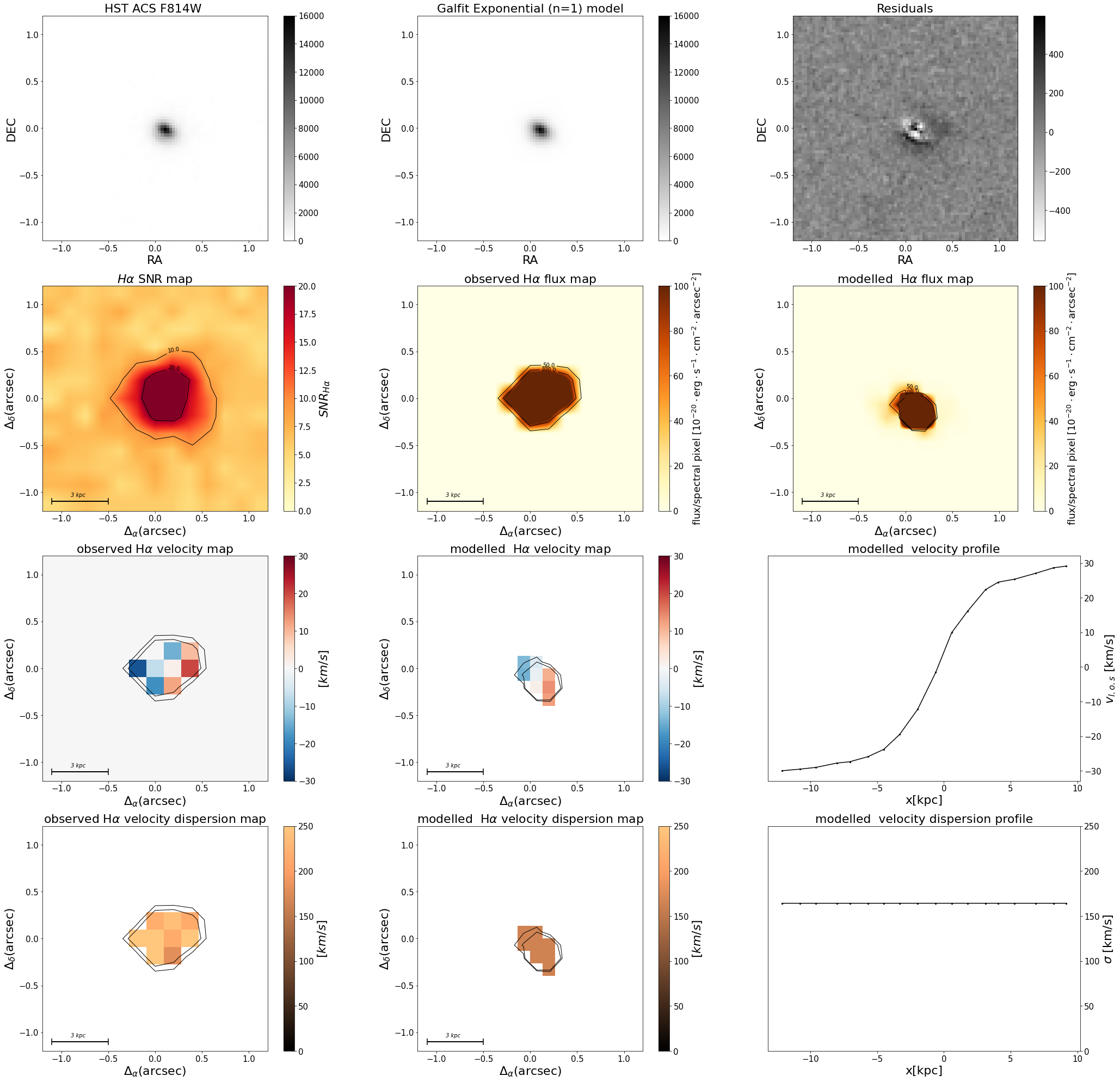}
   \centering   
    \caption{ Morpho-kinematic maps for galaxy \#7. Same caption as Fig.  \ref{ID10}.  This galaxy is barely resolved in the modelled maps, and it is considered to be morphologically regular and kinematically disturbed.  It has the following physical parameters: $\rm{log(M/M_{\odot})}=8.28 $; $\rm{v_{max}} =42$ km/s; $\rm{v_{max,\: gas}/\sigma_{gas}}=0.26 $; offset from SF MS: $\rm{\Delta_{MS}}=0.3$; offset from TFR:  $\rm{\Delta M_{TFR}}= -0.86$; offset from stellar mass -- $\rm{S_{0.5}}$:  $\rm{\Delta M_{S_{0.5}}}= 2.65$. }  
\label{ID7}
\end{figure}

\emph{\#7:} Fig.  \ref{ID7} shows the morpho-kinematic analysis of the galaxy with ID7, a cluster dwarf at z=0.3807.  The stellar disk of this dwarf is fairy inclined with  i=$ 62^{\circ}$ and very small, with  $\rm{r_{d}}\sim0.3$ kpc. The morphology of this object is well replicated by a single-exponential profile, even though from the residual map, $\rm{r_{d}}$ seems slightly underestimated.\\
The kinematic modelling of the $\rm{H\alpha}$ disk roughly reproduces the observed velocity field, which is somewhat convoluted. We recover an inclination of i$=50.9^{\circ}$ for the gas disk,  in good agreement with the one obtained for the stellar disk, and a $\rm{r_{d}}=0.8$ kpc,  larger than that of the stellar component. The kinematic and photometric PAs agree well, with a $\Delta_{\rm{PA}}\sim 8^{\circ}$. \\ The velocity field of the $\rm{H\alpha}$ disk shows a gradient from $\pm$ 30 km/s, with a $\rm{v_{max}}=42$ km/s. We note, that this cluster member is barely resolved in the MUSE data, and hence, the derived $\rm{v_{max}}$ value is highly uncertain. The velocity dispersion map displays very high values for such a low stellar mass, in the order of  $\rm{\sigma_{gas}} =164 $ km/s. This might hint at the fact, that this system is perturbed, or more probably, it has an unresolved velocity gradient. Also, the effect of beam smearing, which is hard to correct for such a small and barely resolved galaxy,  can artificially inflate the measured dispersion, as it combines regions with different line-of-sight velocities into a single spatial pixel. The dynamics of this dwarf, as derived from the MUSE data, are dominated by non-circular, random motions, with a ratio $\rm{v_{max,\: gas}/\sigma_{gas}}=0.26$.\\\\

\begin{figure}[t]
  \centering
  \captionsetup{width=0.5\textwidth}
    \includegraphics[width=0.5\textwidth,angle=0,clip=true]{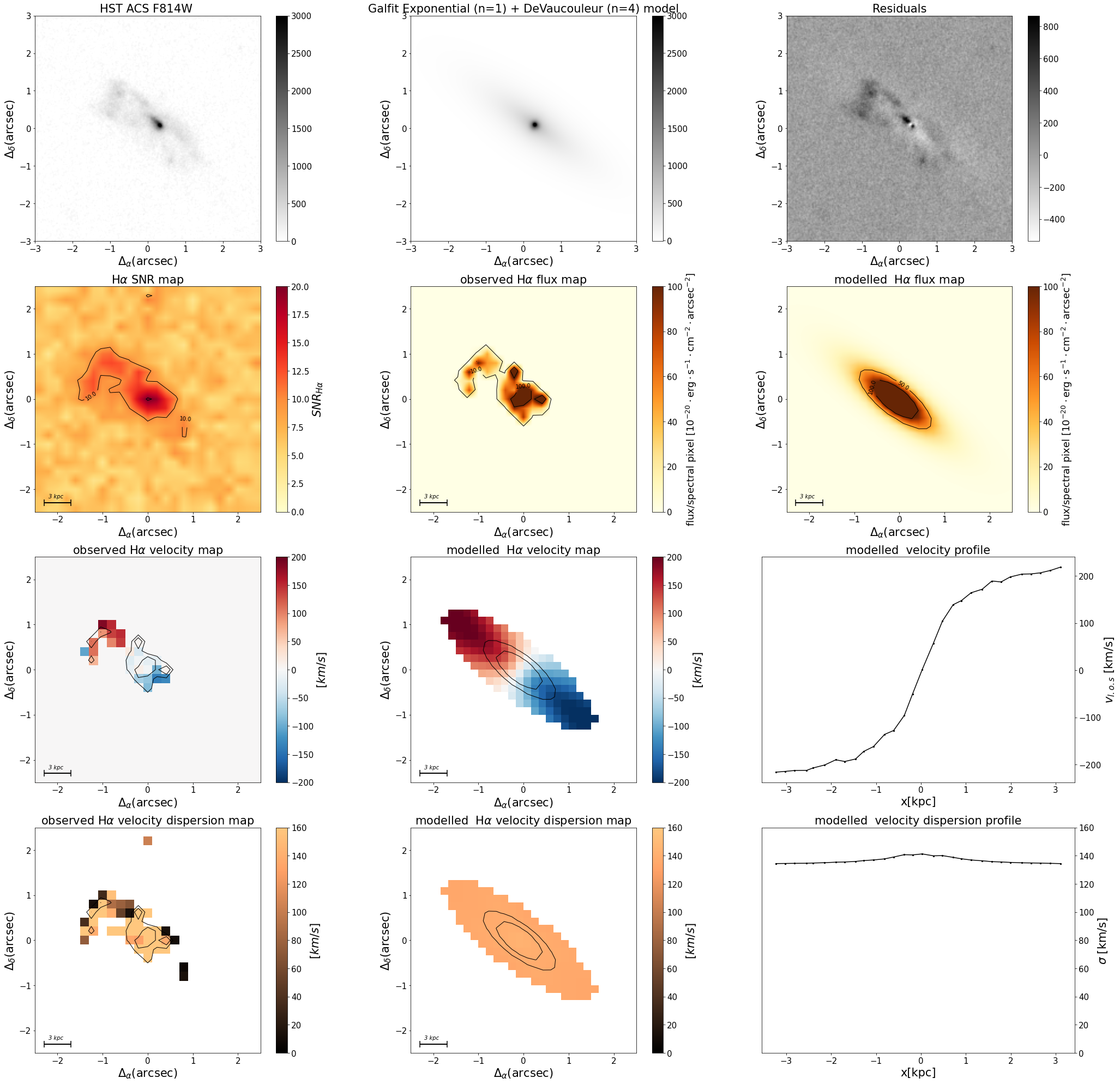}
   \centering   
    \caption{ Morpho-kinematic maps for galaxy \#8. Same caption as Fig.  \ref{ID10}. This galaxy is classified as being morphologically disturbed and kinematically regular.  It has the following physical parameters: $\rm{log(M/M_{\odot})}= 7.98 $; $\rm{v_{max}} = 250$ km/s; $\rm{v_{max,\: gas}/\sigma_{gas}}=1.82 $; offset from SF MS: $\rm{\Delta_{MS}}=-0.67$; offset from TFR:  $\rm{\Delta M_{TFR}}=  2.79$; offset from stellar mass -- $\rm{S_{0.5}}$:  $\rm{\Delta M_{S_{0.5}}}= 3.42$.    }  
\label{ID8}
\end{figure}

\emph{\#8:} Fig.  \ref{ID8} shows the morpho-kinematic analysis of the galaxy with ID8, an extended cluster member at z=0.3819.
From the HST image, we measure a  stellar disk  inclination  i=$ 73^{\circ}$ and a  $\rm{r_{d}}\sim4.74$ kpc. This object shows a somewhat skewed morphology, with a diffuse, extended disk-like structure. The bulge component seems more elongated to the NW. The observed  $\rm{H\alpha}$  SNR and flux maps are asymmetric, showing substantially more $\rm{H\alpha}$ emission to the NW of the centre. Such lopsided gas disks can arise due to RPS. When galaxies move through a cluster (nearly) edge-on, the gas is compressed on the leading side while it is stretched along the tail on the opposite side (\citealt{boselli}).  Most probably, tidal interactions with both other galaxies and the gravitational potential of the dark matter halo, as well as hydrodynamical interactions with the ICM are responsible for the peculiar stellar and gaseous morphology of this galaxy. We also note that this system lies below the MS of SF galaxies in the stellar mass-SFR plane, being currently in the process of quenching.\\ 
The modelling of the gas disk kinematics reproduces the observed velocity field. The $\rm{H\alpha}$ disk has an inclination of i$=62^{\circ}$,  in good agreement with the one of the stellar disk.  The $\rm{H\alpha}$ disk scale length is larger than the stellar one, with 1 kpc. The kinematic and photometric PAs agree, with a $\Delta_{\rm{PA}}\sim 16^{\circ}$. \\
The gas velocity field shows a gradient from $\pm$ 200 km/s, and the measured velocity dispersion is in the order of  $\rm{\sigma_{gas}} =138 $ km/s, with a peak at the centre.  The gas dynamics of this cluster member are  dominated by circular motions, with a ratio $\rm{v_{max,\: gas}/\sigma_{gas}}=1.8$. We speculate, however, given the morphology of this object  that the velocity peaks in the outer parts are not from rotation but are rather tidally and/or RPS induced, as is the gas velocity dispersion. \\\\

\begin{figure}[t]
  \centering
  \captionsetup{width=0.5\textwidth}
    \includegraphics[width=0.5\textwidth,angle=0,clip=true]{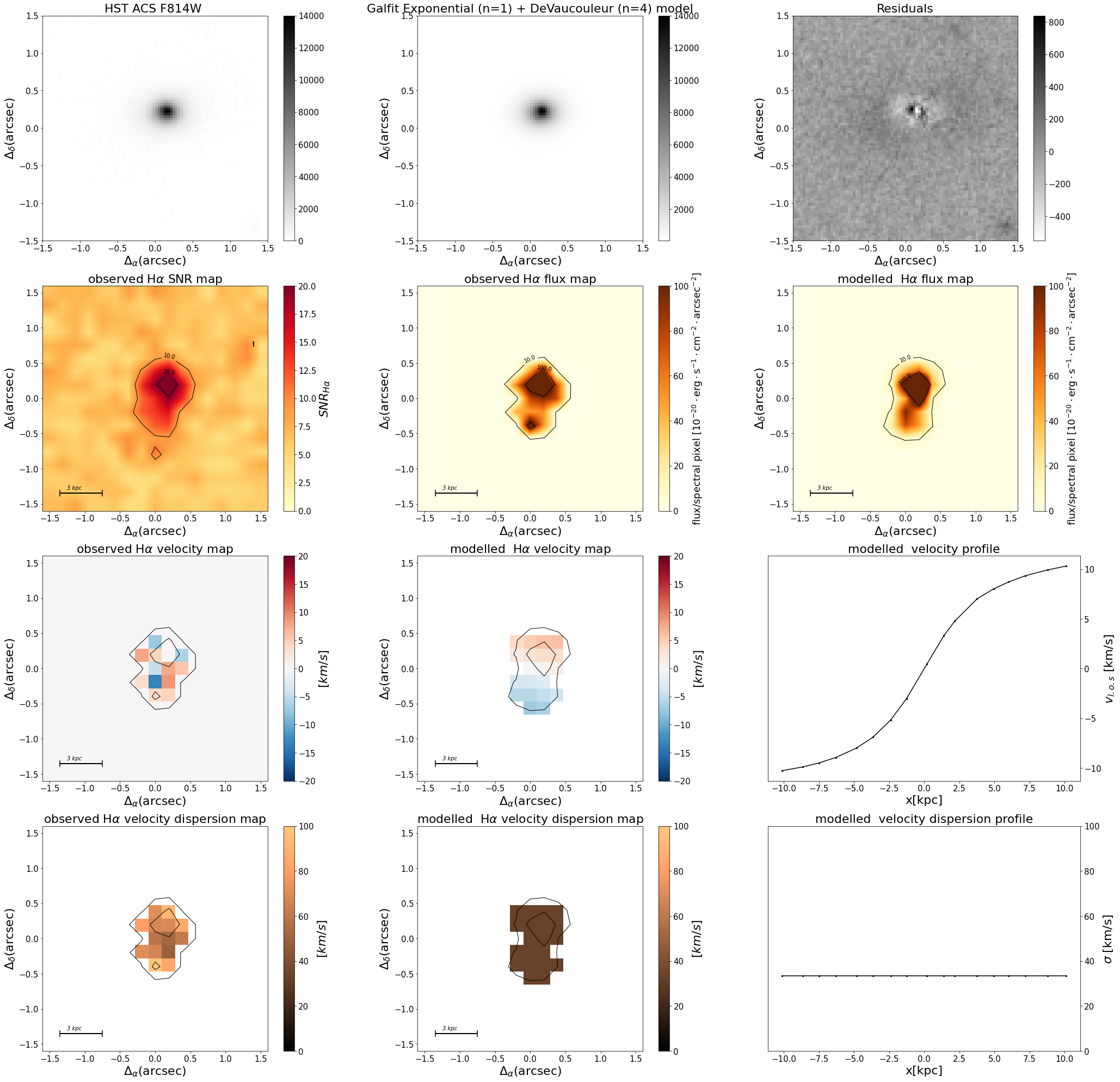}
   \centering   
    \caption{ Morpho-kinematic maps for galaxy \#9. Same caption as Fig.  \ref{ID10}. This galaxy is classified as being morphologically regular and kinematically disturbed.  It has the following physical parameters: $\rm{log(M/M_{\odot})}= 9.03 $; $\rm{v_{max}} = 17$ km/s; $\rm{v_{max,\: gas}/\sigma_{gas}}=0.53 $; offset from SF MS: $\rm{\Delta_{MS}}=-1.02$; offset from TFR:  $\rm{\Delta M_{TFR}}=  -3.27$; offset from stellar mass -- $\rm{S_{0.5}}$:  $\rm{\Delta M_{S_{0.5}}}= -0.49$.    }  
\label{ID9}
\end{figure}

\emph{\#9:} Fig.  \ref{ID9} shows the morpho-kinematic analysis of the cluster member with ID9, at z=0.384. 
The stellar disk has an inclination of i$=47^{\circ}$, and a  $\rm{r_{d}}\sim0.51$ kpc. In the observed $\rm{H\alpha}$ flux map, an additional peak to the S is visible, which is also present in the modelled flux map.  The kinematic modelling of the gas disk with \texttt{Gakpak3D} does not reproduce the observed velocity field, which is very unordered. However, the recovered structural parameters of the gas disk agree with the ones of the stellar disk, except for the PAs ($\Delta_{\rm{PA}}\sim72^{\circ}$). 
The gas velocity field shows a  weak gradient from $\pm$ 10 km/s, while the velocity dispersion is in the order of  $\rm{\sigma_{gas}} =33 $ km/s. Hence, the gas dynamics of this cluster member are dominated by random motions, with a ratio $\rm{v_{max,\: gas}/\sigma_{gas}}=0.5$. We also note that this galaxy falls exactly at the separation line between AGNs and LINERs in the BPT diagram, meaning that its ISM is ionised by both star formation and possibly an AGN and/or mechanisms that give rise to LINER-like emission. It is also worth mentioning that this object lies below the SF MS in the Mass-SFR plane, and hence, it is currently in the process of quenching.  However, the SFR value we infer for this object should be considered an upper limit, as the $\rm{H\alpha}$ flux possibly has a fractional contribution from other sources, and not only from SF.  \\\\

\begin{figure}[t]
  \centering
  \captionsetup{width=0.5\textwidth}
    \includegraphics[width=0.5\textwidth,angle=0,clip=true]{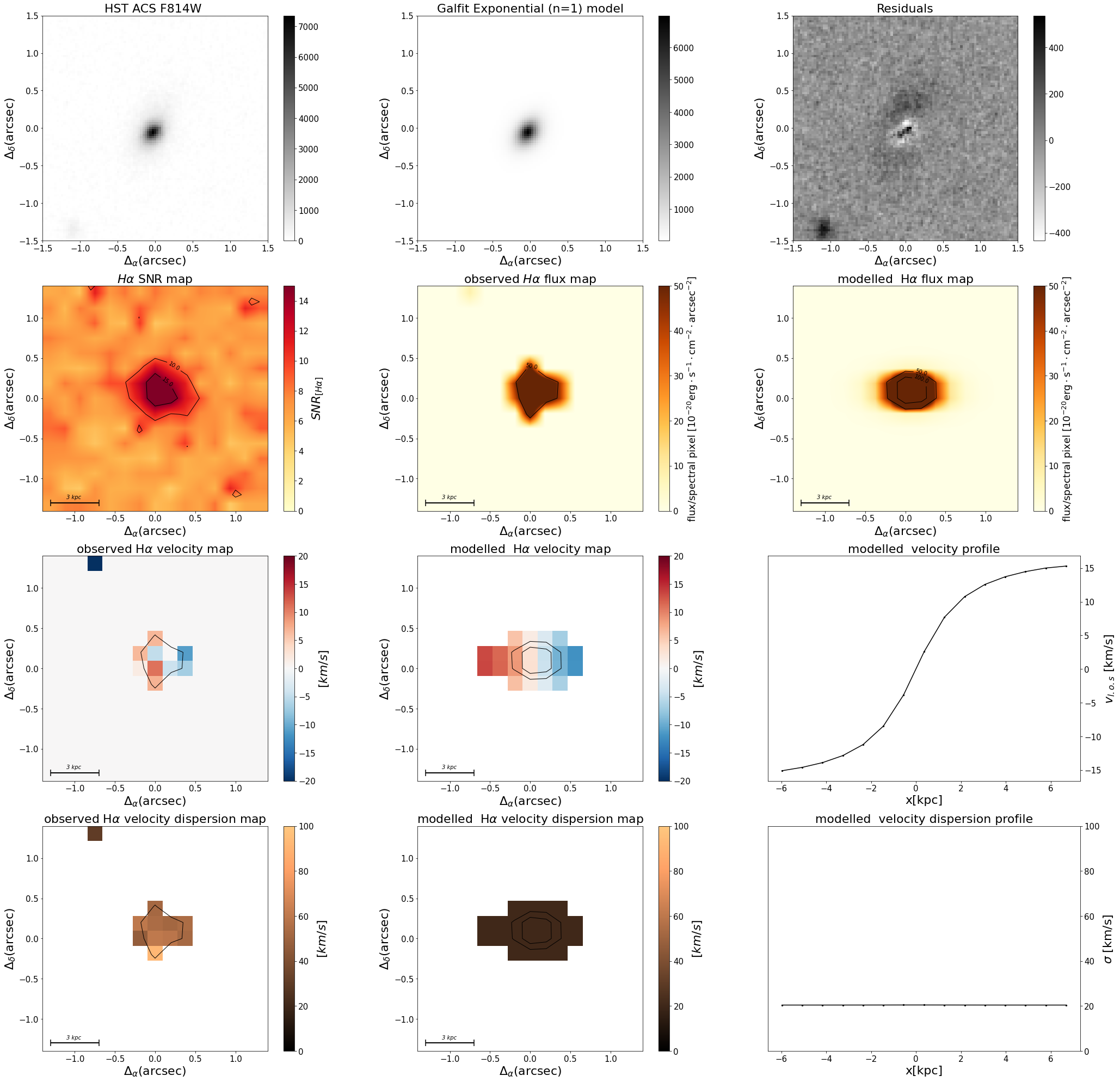}
   \centering   
    \caption{ Morpho-kinematic maps for galaxy \#11. Same caption as Fig.  \ref{ID10}. This galaxy is barely resolved in the observed maps, and it is classified as being morphologically regular and kinematically disturbed.  It has the following physical parameters: $\rm{log(M/M_{\odot})}= 8.5 $; $\rm{v_{max}} = 20.29$ km/s; $\rm{v_{max,\: gas}/\sigma_{gas}}=0.99  $; offset from SF MS: $\rm{\Delta_{MS}}=-0.3 $; offset from TFR:  $\rm{\Delta M_{TFR}}=  -2.48$; offset from stellar mass -- $\rm{S_{0.5}}$:  $\rm{\Delta M_{S_{0.5}}}= -0.51$.   }  
\label{ID11}
\end{figure}

\emph{\#11:} Fig.  \ref{ID11} shows the morpho-kinematic analysis of the cluster member with ID11, which is a very small and faint dwarf at z= 0.40037. The morphology of this galaxy is better reproduced by a single exponential component, as this dwarf galaxy does not show a significant bulge component. For the stellar disk, we infer an inclination of i$=53^{\circ}$, and a  $\rm{r_{d}}\sim0.36$ kpc. In the residuals map, a fain extended component from NE-SW becomes visible.\\
The modelling of the $\rm{H\alpha}$ disk kinematics somewhat reproduces the observed velocity field, which is quite convoluted. Due to the faint nature of this galaxy and its low SNR, flux measurements of the $\rm{H\alpha}$ line with \texttt{MPDAF} are problematic. 
The gas disk has an inclination of i$=60^{\circ}$,  in good agreement with the one derived for the stellar disk based on the HST image, and a disk scale length of  $\rm{r_{d}}\sim1.15$ kpc. The kinematic and photometric PAs are different, with a $\Delta_{\rm{PA}}\sim 55^{\circ}$, however, due to the round and compact morphology of this dwarf, it is more difficult to define a major axis and therefore to determine the morphological position angle. \\
The $\rm{H\alpha}$ velocity field shows a  weak gradient from $\pm$ 15 km/s, whereas the velocity dispersion is in the order of  $\rm{\sigma_{gas}} =20 $ km/s, leading to a  $\rm{v_{max,\: gas}/\sigma_{gas}}=1$. We note, however, that this cluster galaxy is barely resolved in the MUSE data, making the kinematic modelling fairly uncertain. \\\\

\begin{figure}[t]
  \centering
  \captionsetup{width=0.5\textwidth}
    \includegraphics[width=0.5\textwidth,angle=0,clip=true]{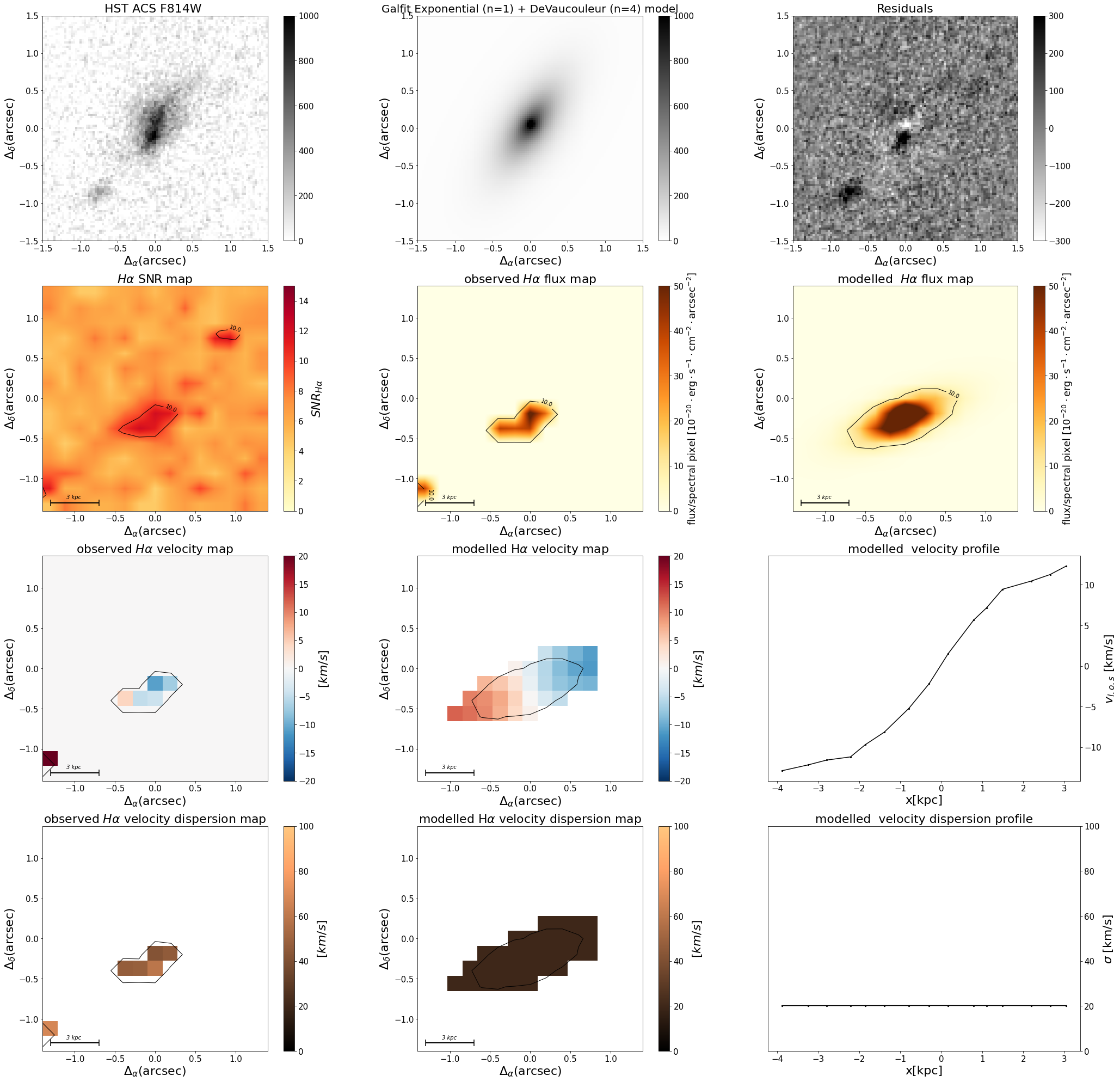}
   \centering   
    \caption{ Morpho-kinematic maps for galaxy \#12. Same caption as Fig.  \ref{ID10}.  This galaxy is barely resolved in the observed maps, and it is classified as being morphologically and kinematically disturbed.  It has the following physical parameters: $\rm{log(M/M_{\odot})}= 8.43 $; $\rm{v_{max}} = 16$ km/s; $\rm{v_{max,\: gas}/\sigma_{gas}}=0.81  $; offset from SF MS: $\rm{\Delta_{MS}}=-3.03 $; offset from TFR:  $\rm{\Delta M_{TFR}}=   -2.82$; offset from stellar mass -- $\rm{S_{0.5}}$:  $\rm{\Delta M_{S_{0.5}}}= - 0.55$.}  
\label{ID12}
\end{figure}

\emph{\#12:} Fig.  \ref{ID12} shows the morpho-kinematic analysis of the cluster member with ID12, a very faint dwarf at z=0.414. 
For the stellar disk component, we measure an inclination of i$=72^{\circ}$  and  a $\rm{r_{d}}\sim2.1$ kpc.  An extended, low surface brightness region can be seen in the residual map,  SW from the central region. Such a bent stellar disk is probably the result of tidal interactions. We note that this cluster member has the lowest SFR value from our sample, and is currently in the process of quenching.  Because this dwarf galaxy is faint, and the SNR of the MUSE data is low, flux measurements of the $\rm{H\alpha}$ line with \texttt{MPDAF} are difficult.  
The observed velocity field is skewed, showing no rotation. For \texttt{Gakpak3D}  to converge, we constrained the inclination and disk scale length to the values obtained for the stellar disk. The photometric and kinematic  PAs differ by $\Delta_{\rm{PA}}\sim37^{\circ}$. The modelled $\rm{H\alpha}$ velocity field shows a  weak gradient from $\pm$ 15 km/s. The recovered velocity dispersion is in the order $\rm{\sigma_{gas}} =20$ km/s. The gas dynamics of galaxy \#12 are dominated by random motions, with a ratio $\rm{v_{max,\: gas}/\sigma_{gas}}=0.8$. We note that the kinematic modelling is quite uncertain for this cluster galaxy, owing to its rather faint nature. \\\\

\begin{figure}[t]
  \centering
  \captionsetup{width=0.5\textwidth}
    \includegraphics[width=0.5\textwidth,angle=0,clip=true]{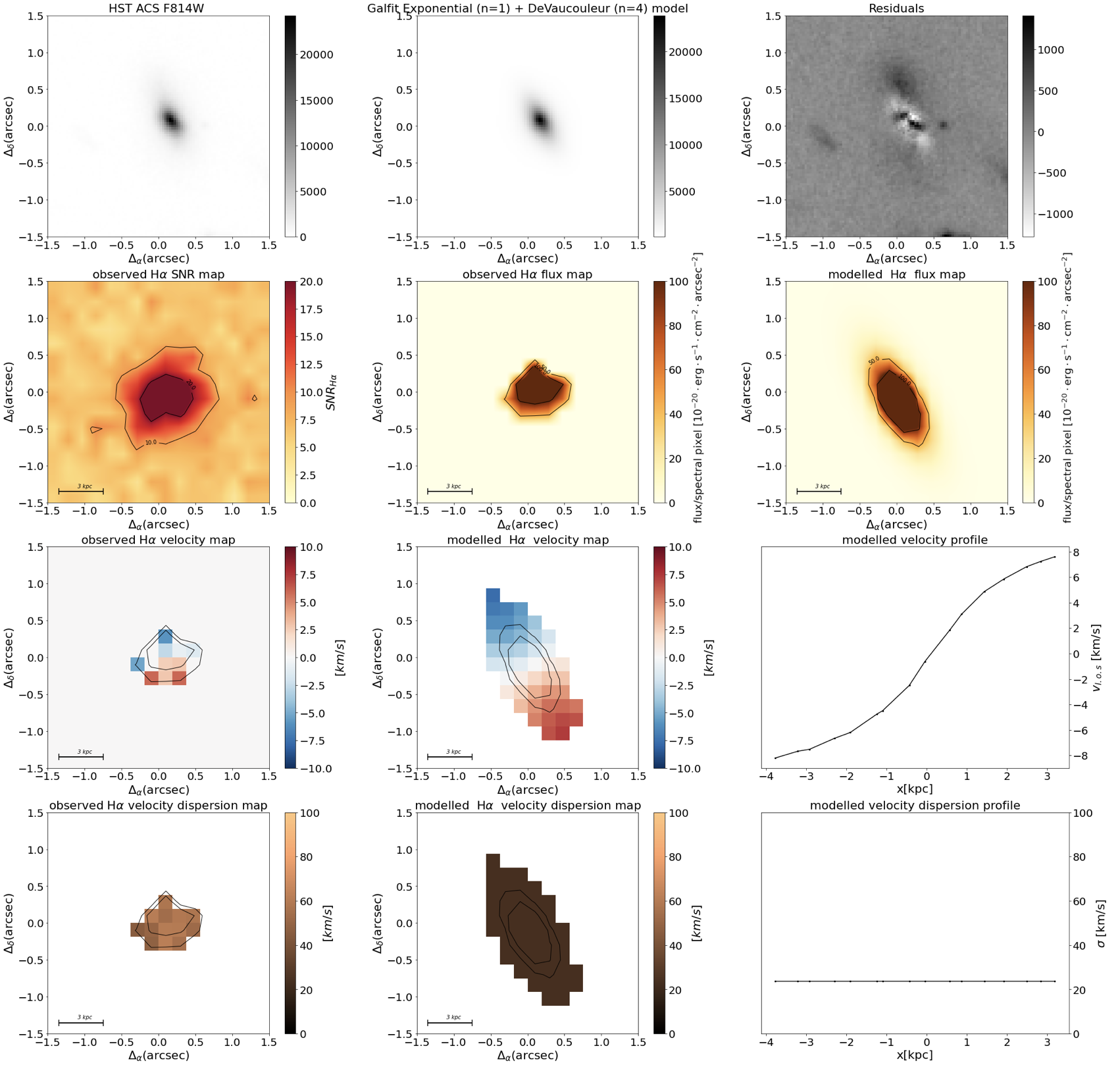}
   \centering   
    \caption{ Morpho-kinematic maps for galaxy \#13. Same caption as Fig.  \ref{ID10}. This galaxy is classified as being morphologically disturbed and kinematically regular.  It has the following physical parameters: $\rm{log(M/M_{\odot})}= 9.05 $; $\rm{v_{max}} = 13$ km/s; $\rm{v_{max,\: gas}/\sigma_{gas}}=0.56  $; offset from SF MS: $\rm{\Delta_{MS}}=0.02 $; offset from TFR:  $\rm{\Delta M_{TFR}}=   -3.85$; offset from stellar mass -- $\rm{S_{0.5}}$:  $\rm{\Delta M_{S_{0.5}}}= -1.03$.}  
\label{ID13}
\end{figure}

\emph{\#13:} Fig.  \ref{ID13} displays the morpho-kinematics of the galaxy with ID13, a  z=0.3952 cluster member.  This dwarf galaxy shows an asymmetric broadband morphology as revealed by the model residuals. This system is elongated to the N, and a bright clump becomes visible in the residual map to the S-E of the main body, which is probably an SF region. The stellar disk has an inclination of i$=67^{\circ}$  and a $\rm{r_{d}}\sim0.63$ kpc.\\The observed and modelled $\rm{H\alpha}$ velocity fields are consistent, however, for the 3D modelling, we constrained the kinematical PA to be close to the morphological PA, i.e. close to $30^{\circ}$. 
We measure an inclination of the $\rm{H\alpha}$ disk, which agrees with the stellar one, and a disk scale length of $\rm{r_{d}}=2.2$ kpc. The $\rm{H\alpha}$ velocity field shows a rather flat gradient from $\pm 10$ km/s and a flat velocity dispersion map. The dynamics of this dwarf are dominated by random motions, with $\rm{v_{max,\: gas}/\sigma_{gas}}=0.56$.\\\\

\begin{figure}[t]
  \centering
  \captionsetup{width=0.5\textwidth}
    \includegraphics[width=0.5\textwidth,angle=0,clip=true]{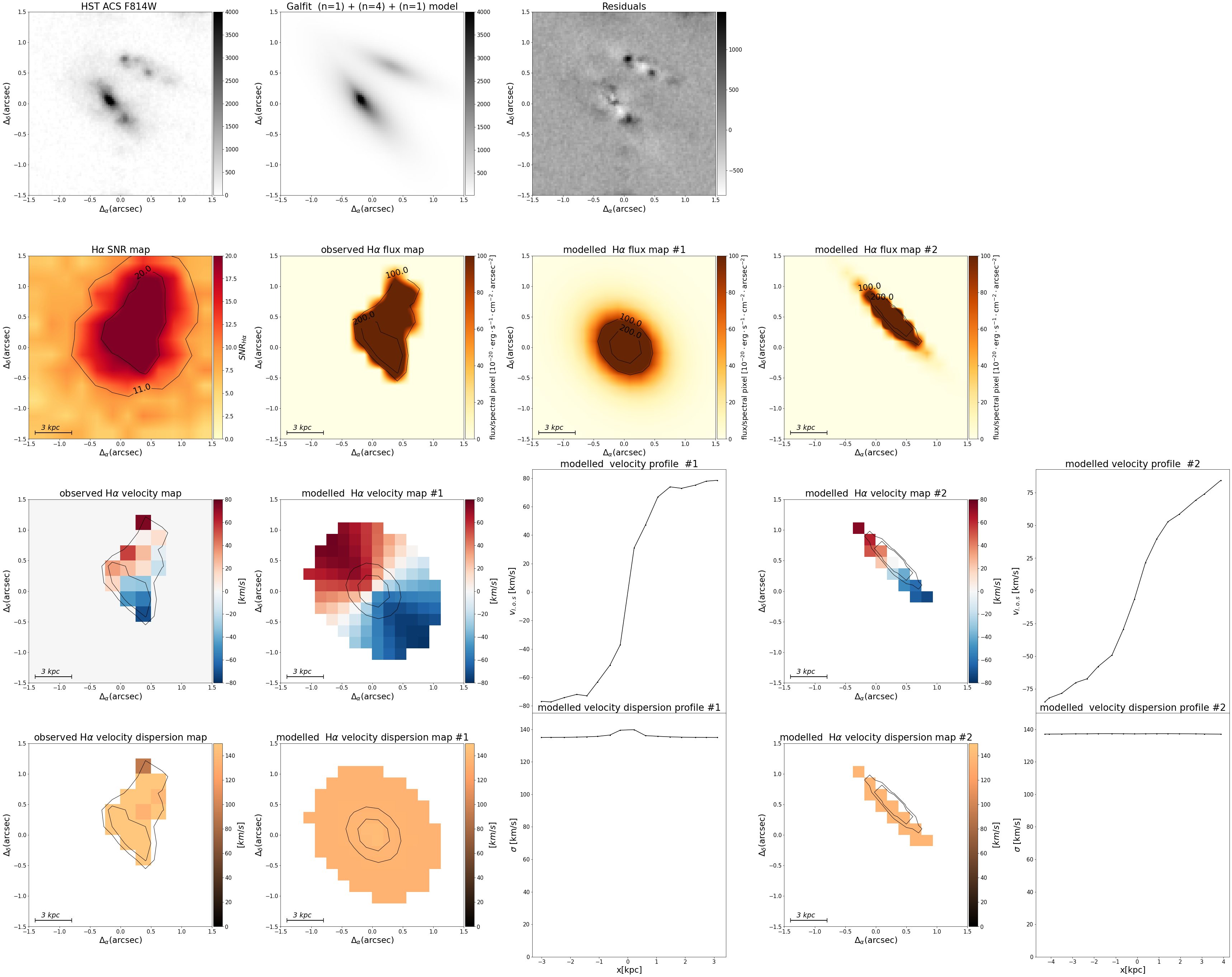}
   \centering   
    \caption{ Morpho-kinematic maps for components ID14-1 and ID14-2. Same caption as Fig.  \ref{ID10}. This interacting/disrupted object is kinematically and morphologically disturbed.  The remnant has a stellar mass of  $\rm{log(M/M_{\odot})}=  8.68$ and an offset from SF MS of $\rm{\Delta_{MS}}=0.39 $.  Component ID14-1 has the following physical parameters: $\rm{v_{max}} = 128$ km/s; $\rm{v_{max,\: gas}/\sigma_{gas}}=0.95  $, whereas  component ID 14-2 has  $\rm{v_{max}} = 105$ km/s and $\rm{v_{max,\: gas}/\sigma_{gas}}=0.77  $. The inffered  offsets fron the different scaling relations for the remnant are: offset from TFR:  $\rm{\Delta M_{TFR}}=   0.82$; offset from stellar mass -- $\rm{S_{0.5}}$:  $\rm{\Delta M_{S_{0.5}}}=  2.21$.}
\label{ID14}
\end{figure}

\emph{\#14:} Fig.  \ref{ID14} displays the morpho-kinematics of the interacting/disrupted system with ID14. Both components are at z=0.38. We have modelled the morphology of the larger object (ID14-1) with an exponential and a de Vaucouleurs  profile, and that of the smaller one (D14-2), with a single exponential profile. For component ID14-1 we measure a disk scale length of 1.7 kpc and an inclination of  $74^{\circ}$. For the object with ID14-2, we measure a disk scale length of 1.5 kpc and a higher inclination of $83^{\circ}$. However,  we could estimate the stellar mass only for the whole system, and not the two different components. In the residual map, many bright clumps become visible, which are probably SF regions. \\ In the observed $\rm{H\alpha}$ flux map, the two components are connected through a gas bridge. 
The $\rm{H\alpha}$ disk kinematic modelling reproduces the observed velocity fields.
 For component  ID14-1, \texttt{GalPaK3D} yields an inclination of i$=39^{\circ}$, and a $\rm{r_{d}}= 2.53$ kpc for the  $\rm{H\alpha}$  disk. The inclination is underestimated with respect to the one of the stellar disk. However, when keeping the inclination fixed in \texttt{GalPaK3D} to the one obtained for the stellar component, the yielded velocity field does not agree with the observed one. The photometric and kinematic PAs agree, with $\Delta_{\rm{PA}}<1^{\circ}$. For this system, the $\rm{H\alpha}$ velocity field shows a gradient from $\pm 80$ km/s, and we measure a $\rm{v_{max}}=128$ km/s. The velocity dispersion map shows a peak at the centre, with $\rm{\sigma_{gas}}=134$ km/s.  Hence, the gas dynamics of this galaxy are dominated by random motions.
 From the $\rm{H\alpha}$ kinematical modelling of the smaller object with ID14-2,  we measure an inclination of $87^{\circ}$, in accordance with the inclination of the stellar disk, and an $\rm{r_{d}}=1.89$ kpc. The photometric and kinematic PAs agree for this system, with $\Delta_{\rm{PA}}=21^{\circ}$. The maximum velocity  is $\rm{v_{max}}=105$ km/s and  $\rm{\sigma_{gas}}=136$ km/s. However, as these two components are interacting,  the modelled  velocity fields might not accurately reflect the gravitational potential.\\\\

\begin{figure}[t]
  \centering
  \captionsetup{width=0.5\textwidth}
    \includegraphics[width=0.5\textwidth,angle=0,clip=true]{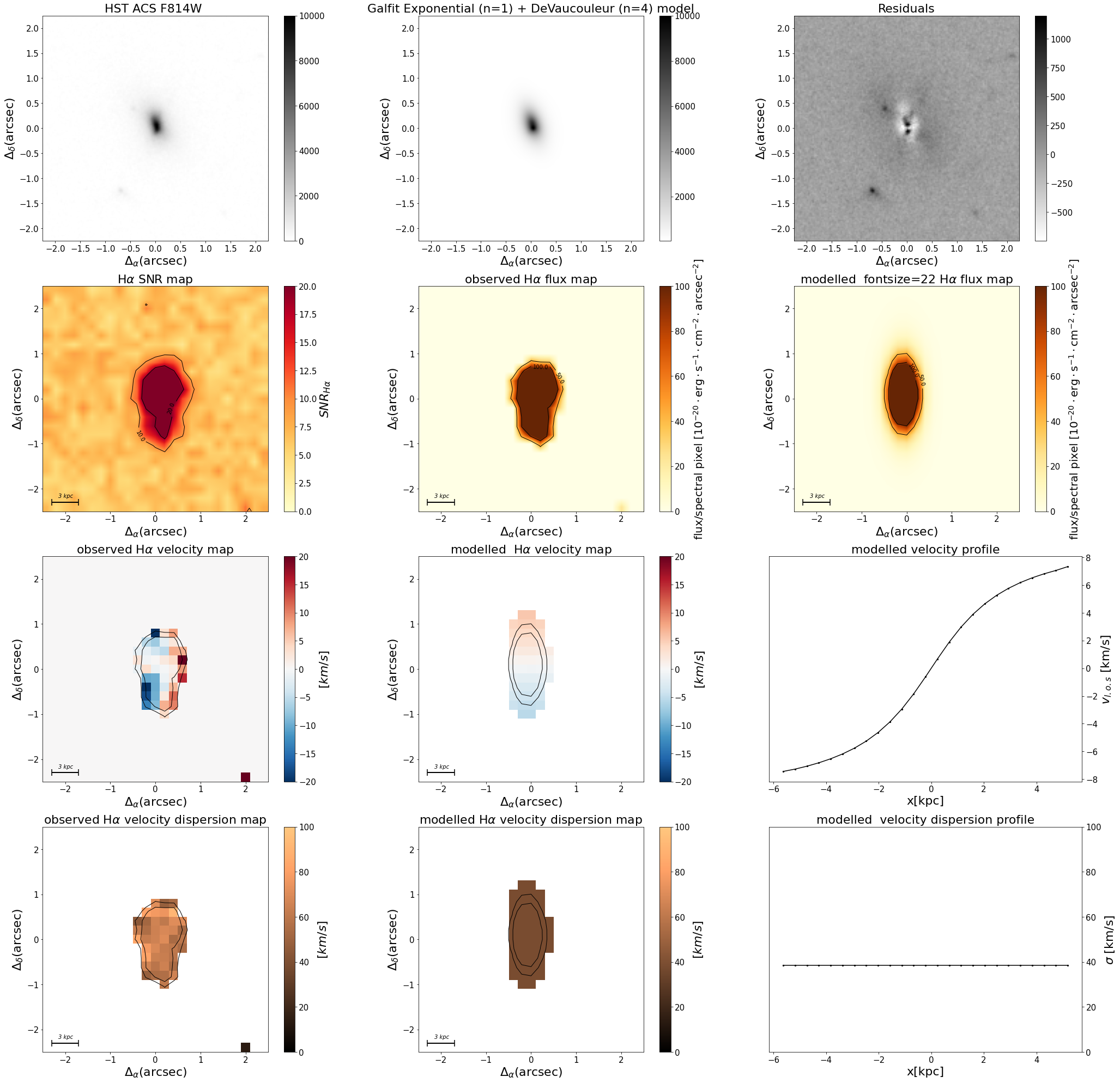}
   \centering   
    \caption{ Morpho-kinematic maps for galaxy \#15. Same caption as Fig.  \ref{ID10}. This galaxy is classified as being morphologically and kinematically disturbed.  It has the following physical parameters: $\rm{log(M/M_{\odot})}= 9.27 $; $\rm{v_{max}} = 10$ km/s; $\rm{v_{max,\: gas}/\sigma_{gas}}=0.27  $; offset from SF MS: $\rm{\Delta_{MS}}=-0.4  $; offset from TFR:  $\rm{\Delta M_{TFR}}=   -4.48$; offset from stellar mass -- $\rm{S_{0.5}}$:  $\rm{\Delta M_{S_{0.5}}}=  -0.57$. }  
\label{ID15}
\end{figure}

\emph{\#15:} Fig.  \ref{ID15} shows the morpho-kinematic analysis of the cluster member with ID15,  at z= 0.41.  Its broadband HST morphology appears slightly asymmetric, as is evident from the residual map. This galaxy seems to have an elongated, bent central region, which is not properly fitted with a bulge component.
For the stellar disk component, we measure an inclination of i$=62^{\circ}$  and a $\rm{r_{d}}\sim1.4$ kpc.     The kinematic modelling of the gas disk with \texttt{Gakpak3D} does not reproduce the observed velocity field, which shows minor axis rotation. 
However, the structural parameters of the $\rm{H\alpha}$ disk agree with the ones of the stellar disk. The $\rm{H\alpha}$ disk has an inclination of $66^{\circ}$ and a $\rm{r_{d}}\sim2.3$ kpc. The photometric and kinematic PAs agree for this galaxy, with $\Delta_{\rm{PA}}=17^{\circ}$.  The modelled $\rm{H\alpha}$ velocity profile shows a very weak gradient from $\pm 5$ km/s, whereas the measured velocity dispersion is in the order of $\rm{\sigma_{gas}}=39$ km/s, and hence, the dynamics of this object are dominated by random motion. \\\\

\begin{figure}[t]
  \centering
  \captionsetup{width=0.5\textwidth}
    \includegraphics[width=0.5\textwidth,angle=0,clip=true]{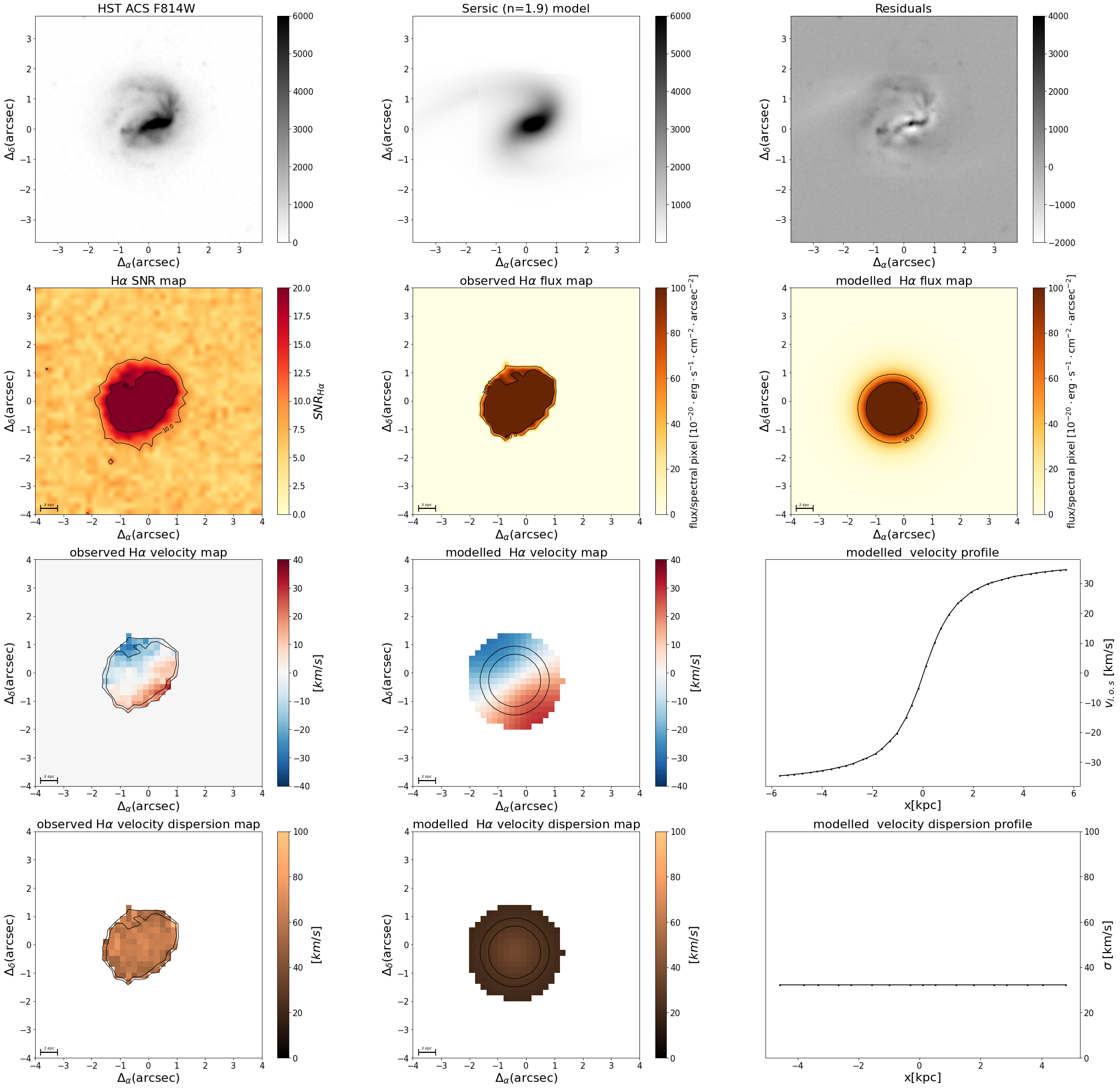}
   \centering   
    \caption{ Morpho-kinematic maps for galaxy \#16.  Same caption as Fig.  \ref{ID10}.  This galaxy is classified as being morphologically disturbed and kinematically regular.  It has the following physical parameters: $\rm{log(M/M_{\odot})}= 9.75 $; $\rm{v_{max}} = 346$ km/s; $\rm{v_{max,\: gas}/\sigma_{gas}}=26.84  $; offset from SF MS: $\rm{\Delta_{MS}}=0.36  $; offset from TFR:  $\rm{\Delta M_{TFR}}=   1.63$; offset from stellar mass -- $\rm{S_{0.5}}$:  $\rm{\Delta M_{S_{0.5}}}=  1.79$. }  
\label{ID16}
\end{figure}

\emph{\#16:} Fig. \ref{ID16} shows the morpho-kinematic analysis of the object with ID16, a  cluster member at a z=0.384. This barred spiral galaxy has a disturbed broadband morphology, with one extended spiral arm to the N, with no analogue to the S. Such a morphology could be the result of tidal interactions and harassment. We have modelled the structure and morphology of this object with a Sersic profile (n=1.9), and Fourier modes were used to modify the traditional ellipses to better fit the tidal features. For the stellar disk, we measure an inclination of i$=69^{\circ}$  and a $\rm{r_{d}}\sim5.7$ kpc.  
In the residual map,  the bar as well as many patchy regions become visible, which are most probably SF regions.\\
The observed $\rm{H\alpha}$ kinematics of this cluster member are completely dominated by the bar, with the velocity gradient along the bar being almost null. The modelled velocity field reproduces the observed one perfectly. For  $\rm{H\alpha}$, we measure a gradient from $\pm 30$ km/s, however, the retrieved maximum velocity is $\rm{v_{max}}=346$ km/s. The strong morphological asymmetries in the HST image could explain why \texttt{GalPaK3D} converges at a very high rotation velocity. This value most probably does not reflect the gravitational potential of the galaxy.
 The yielded velocity dispersion from \texttt{GalPaK3D} is low, in the order of $\rm{\sigma_{gas}}=13$ km/s. The structural parameters of the $\rm{H\alpha}$ disk do not agree with the ones measured for the stellar component. The gas disk has an inclination of i$=7^{\circ}$, a $\rm{r_{d}}=3.56$ kpc. The photometric and kinematic PAs differ by $\Delta_{\rm{PA}}=80^{\circ}$.  All these results hint at the fact that the gas component does not follow the gravitational potential of the stellar component in this galaxy.\\\\

\begin{figure}[t]
  \centering
  \captionsetup{width=0.5\textwidth}
    \includegraphics[width=0.5\textwidth,angle=0,clip=true]{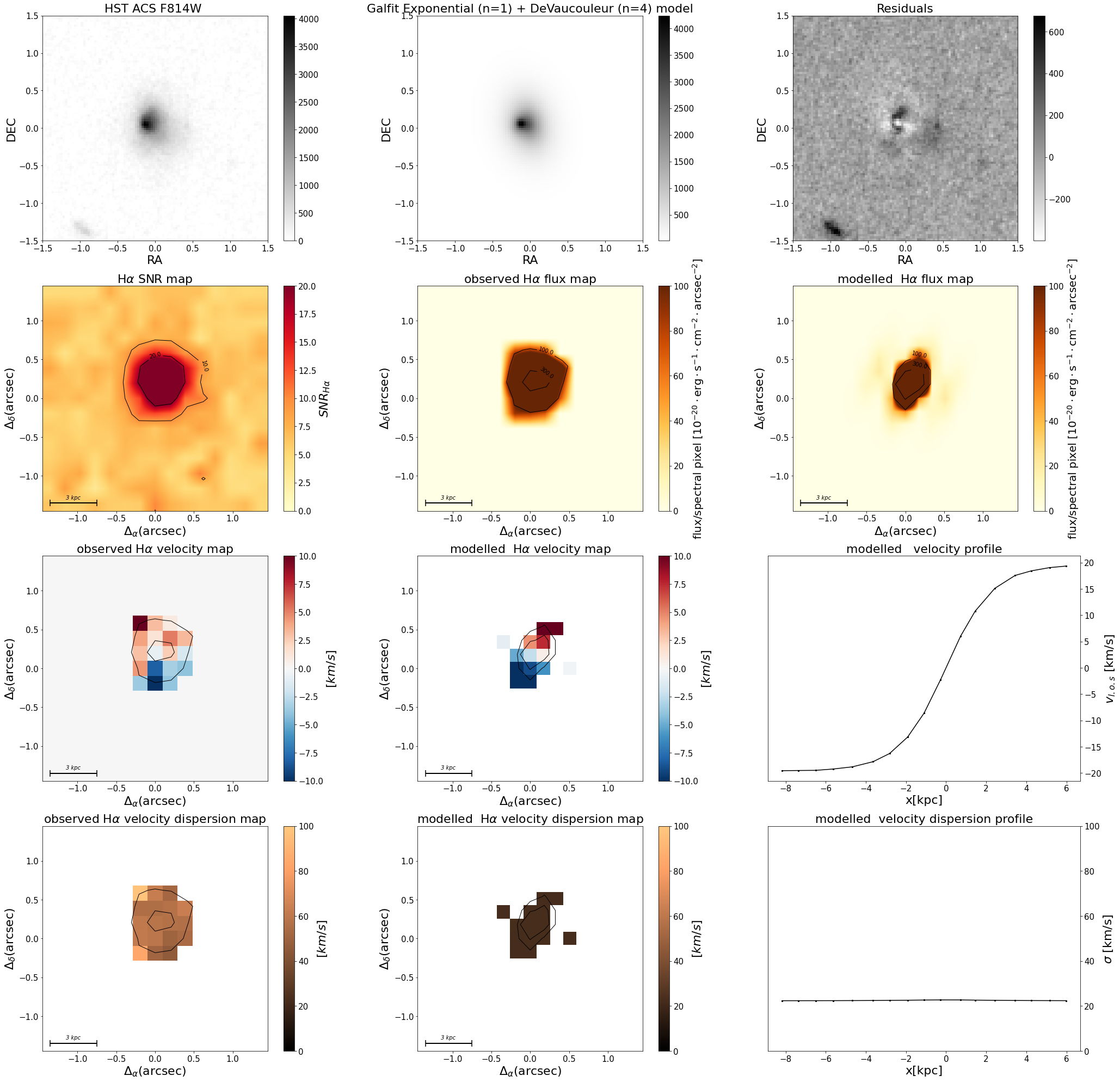}
   \centering   
    \caption{ Morpho-kinematic maps for galaxy \#17. Same caption as Fig.  \ref{ID10}. This galaxy is classified as being morphologically disturbed and kinematically regular.  It has the following physical parameters: $\rm{log(M/M_{\odot})}= 9.26 $; $\rm{v_{max}} = 54$ km/s; $\rm{v_{max,\: gas}/\sigma_{gas}}=2.42 $; offset from SF MS: $\rm{\Delta_{MS}}=-0.12  $; offset from TFR:  $\rm{\Delta M_{TFR}}=  -1.39$; offset from stellar mass -- $\rm{S_{0.5}}$:  $\rm{\Delta M_{S_{0.5}}}=  -0.38$.   }  
\label{ID17}
\end{figure}

\emph{\#17:} Fig. \ref{ID17} shows the morpho-kinematic analysis of the object with ID17, a cluster member at z=0.395. The morphology of this dwarf is well reproduced by a bulge and a disk, however, the two components do not share the same photometric centre. The diffuse disk component seems to be displaced to the E, with respect to the bulge component. For the stellar disk, we measure an inclination of  i$=51^{\circ}$  and a $\rm{r_{d}}\sim1$ kpc.  The $\rm{H\alpha}$ disk kinematic modelling reproduces the observed velocity field, however, the structural parameters of the gas disk do not agree with those of the stellar component. For the gas disk, \texttt{GalPaK3D} yields an inclination of i$=17^{\circ}$, and a  slightly larger  disk scale length, $\rm{r_{d}}\sim1.2$ kpc. The photometric and kinematic PA differ by $\Delta_{\rm{PA}}=37^{\circ}$. 
 These results suggest that the gas component of this galaxy does not follow the gravitational potential of the stellar component.The $\rm{H\alpha}$ velocity field shows a  regular pattern with $\pm20$ km/s, and we measure  a $\rm{v_{max}}=54$ km/s. The velocity dispersion map is quite flat and shows low values, in the order of $\rm{\sigma_{gas}}=22$ km/s. The gas dynamics of this system are dominated by rotation, with $\rm{v_{max,\: gas}/\sigma_{gas}}=2.4$.\\
\\\

\begin{figure}[t]
  \centering
  \captionsetup{width=0.5\textwidth}
    \includegraphics[width=0.5\textwidth,angle=0,clip=true]{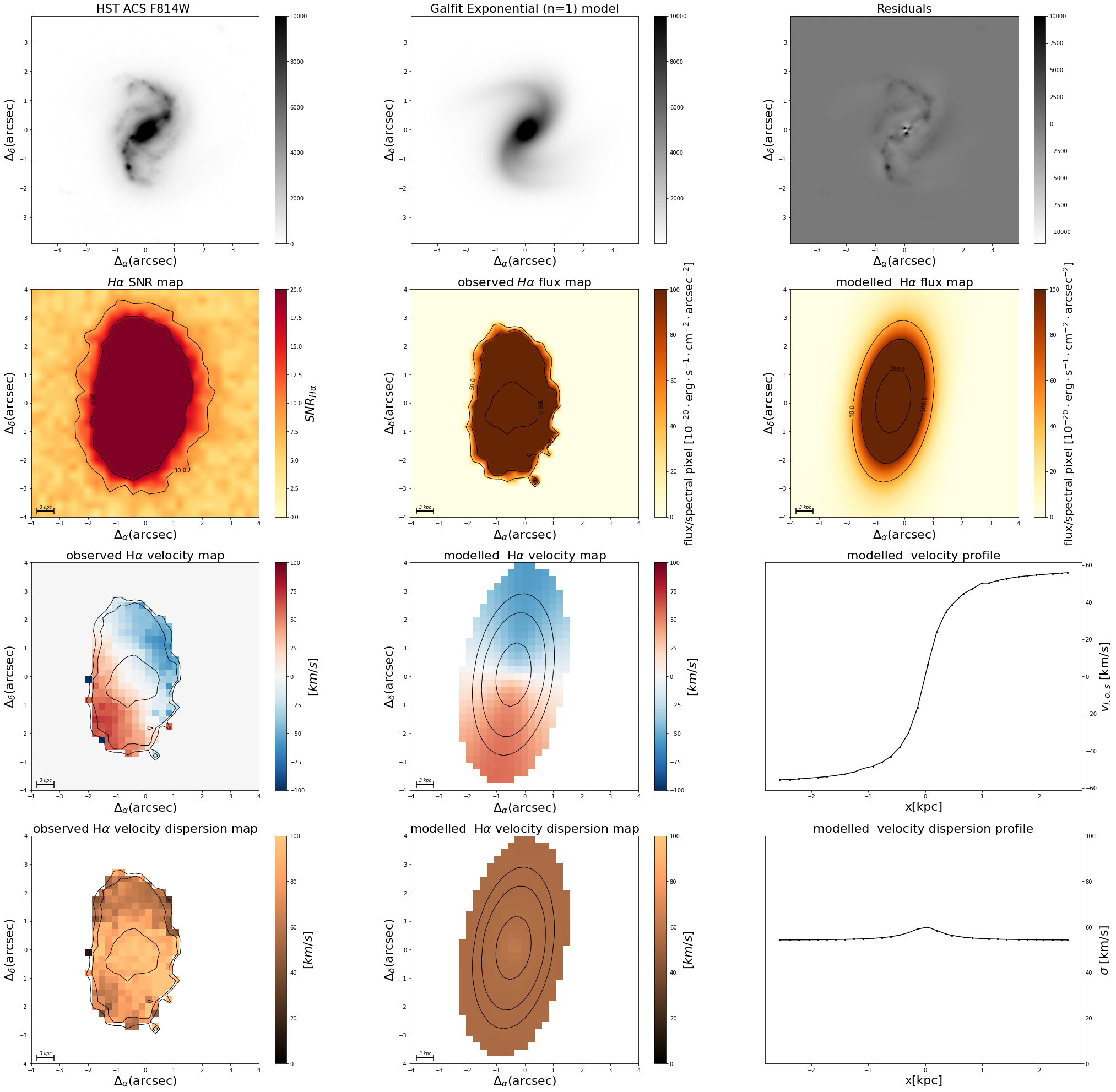}
   \centering   
    \caption{ Morpho-kinematic maps for field galaxy \#1. Same caption as Fig.  \ref{ID10}. This field galaxy has the following physical parameters: $\rm{log(M/M_{\odot})}= 10.65 $; $\rm{v_{max}} = 66$ km/s; $\rm{v_{max,\: gas}/\sigma_{gas}}=1.23 $; offset from SF MS: $\rm{\Delta_{MS}}=- 0.41   $; offset from TFR:  $\rm{\Delta M_{TFR}}=  -2.39$; offset from stellar mass -- $\rm{S_{0.5}}$:  $\rm{\Delta M_{S_{0.5}}}=   -1.02$.  }  
\label{field1}
\end{figure}

\emph{field \#1:} Fig. \ref{field1} shows the morpho-kinematic analysis of the field galaxy \#1, at a redshift of z=0.351.  This barred spiral galaxy has a complex "S-shaped" broadband morphology, which was fitted with a bulge and a disk, and Fourier modes were used to modify the traditional ellipses to better fit the spiral structure.  Many bright spots are visible in the residual map, which correspond to SF regions. For the stellar disk, we measure   i$=61^{\circ}$  and  $\rm{r_{d}}\sim8.6$ kpc. The rather asymmetric morphology of this system hints at tidal interactions.  In Fig. \ref{hist}, there seems to be an overdensity of galaxies at z$\sim0.35$, hence, we do not rule out the possibility that this system is a member of a galaxy group, where gravitational interactions are stronger than in the cluster environment.\\ The $\rm{H\alpha}$ disk kinematic modelling roughly reproduces the observed velocity field,  however, the null velocity gradient is not aligned with the observed one. The measured structural parameters of the $\rm{H\alpha}$ disk are in very good agreement with the ones derived for the stellar disk.  
The $\rm{H\alpha}$ velocity field shows a gradient from $\pm60$ km/s, with $\rm{v_{max}}=66$ km/s, and the velocity dispersion map is rather flat, except for the most central spaxels, where it peaks with values of $\rm{\sigma_{gas}} \sim 60$ km/s. The dynamics of this object are barely dominated by rotation, with $\rm{v_{max,\: gas}/\sigma_{gas}}=1.23$. \\\\

\begin{figure}[t]
  \centering
  \captionsetup{width=0.5\textwidth}
    \includegraphics[width=0.5\textwidth,angle=0,clip=true]{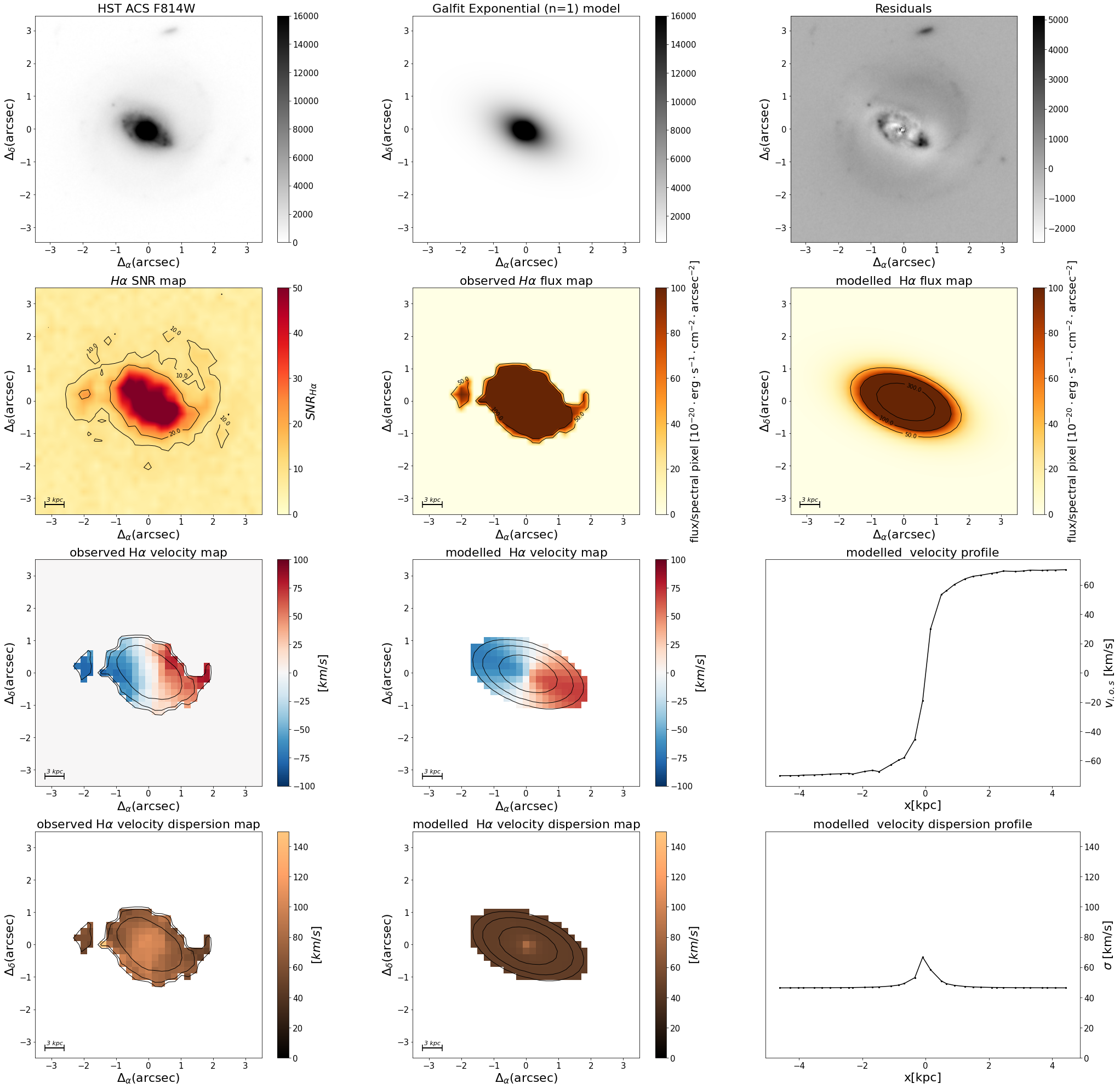}
   \centering   
    \caption{ Morpho-kinematic maps for field galaxy \#2. Same caption as Fig.  \ref{ID10}. This field galaxy has the following physical parameters: $\rm{log(M/M_{\odot})}= 10.79 $; $\rm{v_{max}} = 83$ km/s; $\rm{v_{max,\: gas}/\sigma_{gas}}=1.8 $; offset from SF MS: $\rm{\Delta_{MS}}=-0.23  $; offset from TFR:  $\rm{\Delta M_{TFR}}=   -2.11$; offset from stellar mass -- $\rm{S_{0.5}}$:  $\rm{\Delta M_{S_{0.5}}}=   -1.1$.   }  
\label{field2}
\end{figure}

\emph{field \#2:} Fig. \ref{field2} shows the morpho-kinematic analysis of the field galaxy \#2, at a redshift of z=0.3052. Its morphology is well reproduced by a bulge and a disk, and in the residual maps,  the spiral structure becomes visible, with some bright spots which correspond to SF regions. For the stellar disk, we determine an inclination of  $\rm{i}=57^{\circ}$  and a $\rm{r_{d}}\sim 2.24$ kpc. 
The modelling of the $\rm{H\alpha}$ disk reproduces the observed velocity field, and the structural parameters of the gas disk agree perfectly with those of the stellar disk. 
The $\rm{H\alpha}$ velocity field shows a strong gradient from $\pm70$km/s, with a $\rm{v_{max}}=83$ km/s reached at the plateau. The velocity dispersion map shows a peak at the centre, coincident with the location of the AGN, and we measure a mean value of $\rm{\sigma_{gas}}=46$ km/s. The dynamics of this system are dominated by rotation, with  $\rm{v_{max,\: gas}/\sigma_{gas}}=1.8$.\\\\

\begin{figure}[t]
  \centering
  \captionsetup{width=0.5\textwidth}
    \includegraphics[width=0.5\textwidth,angle=0,clip=true]{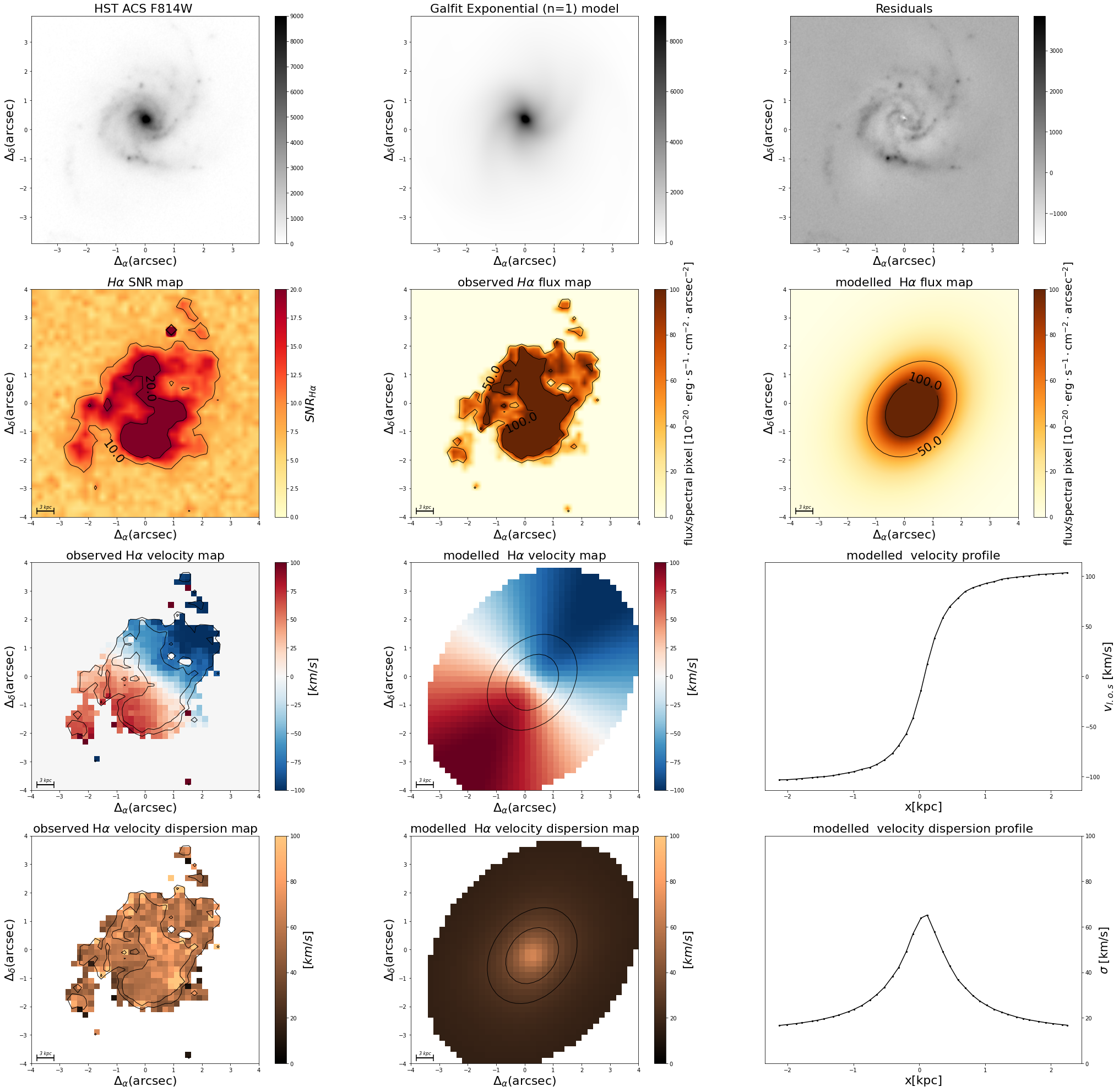}
   \centering   
    \caption{ Morpho-kinematic maps for field galaxy \#3. Same caption as Fig.  \ref{ID10}. This field galaxy has the following physical parameters: $\rm{log(M/M_{\odot})}= 10.75 $; $\rm{v_{max}} = 163$ km/s; $\rm{v_{max,\: gas}/\sigma_{gas}}= 12.5 $; offset from SF MS: $\rm{\Delta_{MS}}=-0.49  $; offset from TFR:  $\rm{\Delta M_{TFR}}=  -0.79 $; offset from stellar mass -- $\rm{S_{0.5}}$:  $\rm{\Delta M_{S_{0.5}}}=   -0.37$. }  
\label{field3}
\end{figure}

\emph{field \#3:} Fig. \ref{field3} shows the morpho-kinematic analysis of the field galaxy \#3, at a redshift of z=0.342.   This grand design spiral galaxy has a complex broadband morphology, which was fitted with a  de Vauculeorus and Sersic component. For the Sersic component, we have used Fourier modes to modify the traditional ellipses to better fit the spiral structure, which is somewhat asymmetric.
 For the stellar disk, we measure i$=40^{\circ}$  and a $\rm{r_{d}}\sim10$ kpc. 
The $\rm{H\alpha}$ disk kinematic modelling replicates the observed velocity field, and the yielded structural parameters of the gas disk  agree with the ones of the stellar disk, except for the PAs ($\Delta_{\rm{PA}}=32^{\circ}$).  
  The gas velocity field shows a gradient from $\pm110$ km/s, compatible with its morphology, and a $\rm{v_{max}}=163$ km/s.  The dispersion map shows a peak at the centre, with  $\sigma\sim 60$ km/s, however, the mean value of the velocity dispersion is $\rm{\sigma_{gas}}=13$ km/s.  The dynamics of this galaxy are, thus,  dominated by rotation, with $\rm{v_{max,\: gas}/\sigma_{gas}}=12.5$.\\\\

\begin{figure}[t]
  \centering
  \captionsetup{width=0.5\textwidth}
    \includegraphics[width=0.5\textwidth,angle=0,clip=true]{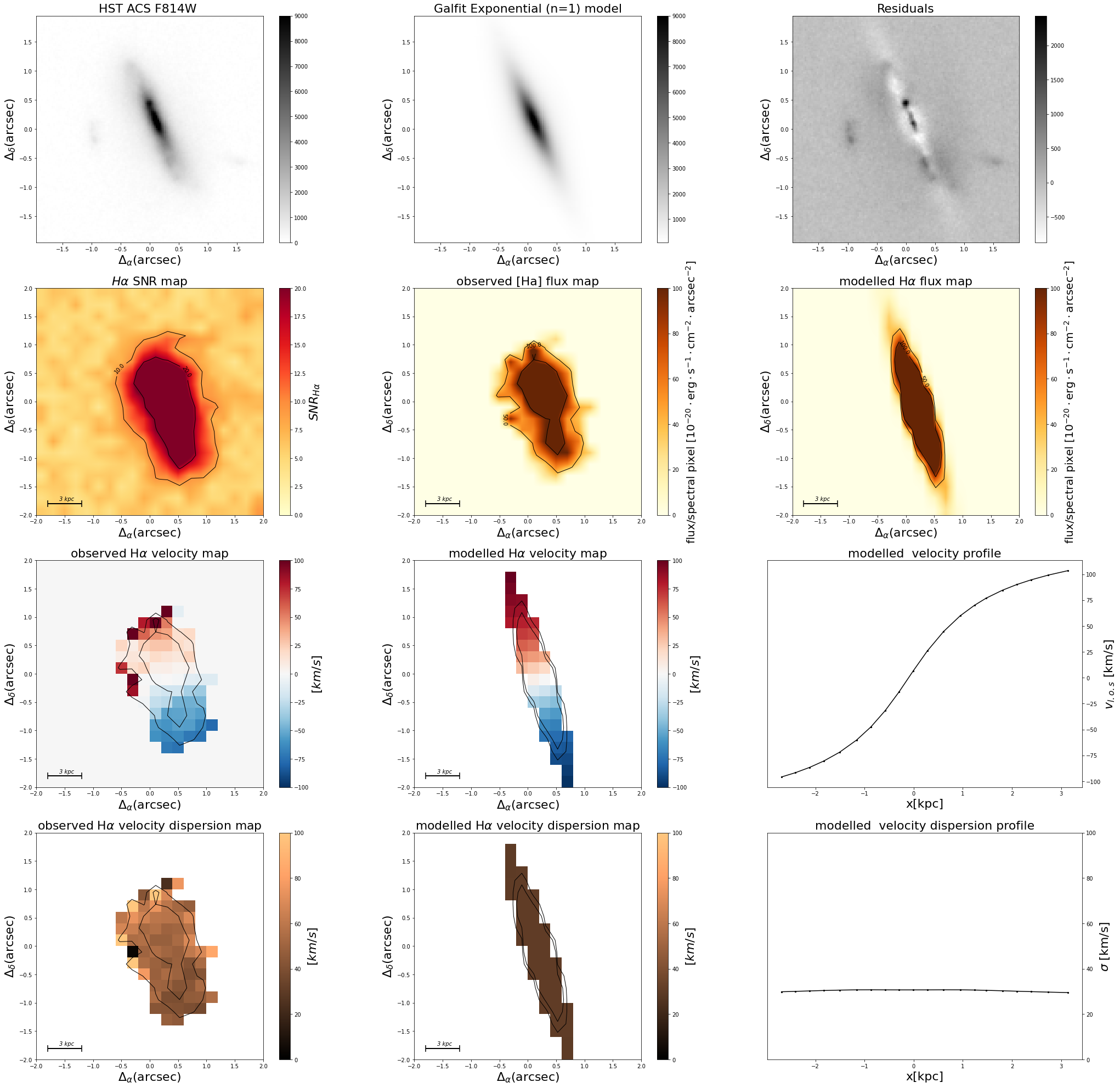}
   \centering   
    \caption{ Morpho-kinematic maps for field galaxy \#4. Same caption as Fig.  \ref{ID10}. This field galaxy has the following physical parameters: $\rm{log(M/M_{\odot})}= 9.65 $; $\rm{v_{max}} = 138$ km/s; $\rm{v_{max,\: gas}/\sigma_{gas}}= 4.84 $; offset from SF MS: $\rm{\Delta_{MS}}= -0.67  $; offset from TFR:  $\rm{\Delta M_{TFR}}=  -0.01 $; offset from stellar mass -- $\rm{S_{0.5}}$:  $\rm{\Delta M_{S_{0.5}}}=   0.52$.   }  
\label{field4}
\end{figure}

\emph{field \#4:} Fig. \ref{field4} shows the morpho-kinematic analysis of the field galaxy \#4, at a redshift of z=0.343. This galaxy is nearly edge on, with an inclination of $\rm{i}=83^{\circ}$ and a disk scale length of $\rm{r_{d}}=1.8$ kpc. Its broadband morphology was fitted with a single exponential disk. The $\rm{H\alpha}$ disk  kinematic modelling perfectly reproduces the observed velocity field, and the measured structural parameters of the gas disk agree with the ones measured for the stellar component, except for the disk scale length, which is larger. 
The gas velocity field shows a gradient from $\pm100$ km/s and a rather flat dispersion map, with $\rm{\sigma_{gas}}=28$ km/s. The gas dynamics of this galaxy are dominated by rotation, with $\rm{v_{max,\: gas}/\sigma_{gas}}=4.8$.\\\\

\begin{figure}[t]
  \centering
  \captionsetup{width=0.5\textwidth}
    \includegraphics[width=0.5\textwidth,angle=0,clip=true]{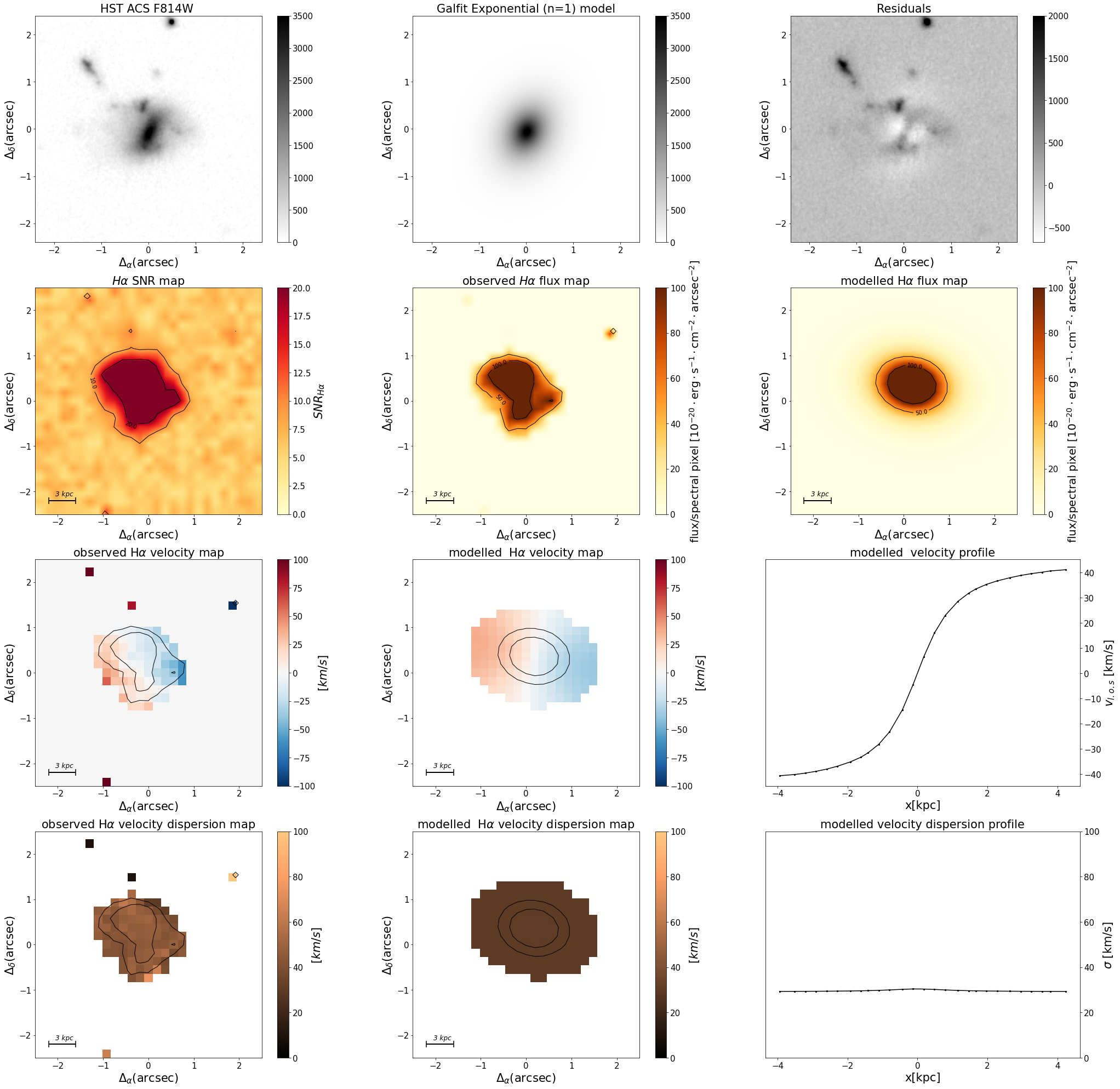}
   \centering   
    \caption{ Morpho-kinematic maps for field galaxy \#5. Same caption as Fig.  \ref{ID10}. This field galaxy has the following physical parameters: $\rm{log(M/M_{\odot})}= 9.28  $; $\rm{v_{max}} = 70$ km/s; $\rm{v_{max,\: gas}/\sigma_{gas}}=  2.4 $; offset from SF MS: $\rm{\Delta_{MS}}= -0.3  $; offset from TFR:  $\rm{\Delta M_{TFR}}=  -0.93 $; offset from stellar mass -- $\rm{S_{0.5}}$:  $\rm{\Delta M_{S_{0.5}}}=   0.01$.   }  
\label{field5}
\end{figure}

\emph{field \#5:} Fig. \ref{field5} displays the morpho-kinematic analysis of the field galaxy \#5, the lowest redshift galaxy of our sample, with z=0.298. This galaxy has a rather asymmetric broadband morphology, which was modelled with a single exponential profile. 
For the stellar disk, we measure an inclination of  i$=41^{\circ}$ and a disk scale length of $\rm{r_{d}}=2.3$. The kinematic modelling reproduces the observed $\rm{H\alpha}$ velocity field, however, due to convergence issues for the turnover radius, we have constrained the inclination to the value obtained for the stellar disk. The $\rm{H\alpha}$ disk has a scale length of $\rm{r_{d}}=3.3$ kpc, and the photometric and kinematic  PA differ by $73^{\circ}$. The $\rm{H\alpha}$ velocity field shows a gradient from $\pm 40$ km/s, with a $\rm{v_{max}}=70$ km/s reached at the plateau.  The velocity dispersion map is rather flat, with $\rm{\sigma_{gas}}=29$ km/s, and hence, the gas dynamics of this system are dominated by rotation, with  $\rm{v_{max,\: gas}/\sigma_{gas}}=2.4$\\\\

\begin{table*}[b]
\label{tab1}
\centering
\caption{Column (1):  galaxy ID. Column (2) and (3):   RA and DEC in [deg]. Column (4):  spectroscopic redshift. Columns  (5) to (7): inclination in $[^{\circ}]$,  $\rm{r_{d}}$ in [kpc] and PA in $[^{\circ}]$ of the stellar disk.  Columns  (8) to (10): inclination in $[^{\circ}]$,  $\rm{r_{d}}$ in [kpc] and PA in $[^{\circ}]$ of the gas disk. Columns (11) and (12): $\rm{v_{max,\:gas}}$ and $\rm{\sigma_{gas}}$ in [km/s]. Columns (13) and (14): Salpeter stellar masses and SFRs.  Column (15): ratio between $\rm{r_{d, \:gas\:disk}}$ and  $\rm{FWHM_{MUSE-PSF}}$. }
\begin{adjustbox}{width=1\textwidth}
\centering
\small
\begin{tabular}{lcccccccccccccl}
\hline
Galaxy ID  &  RA  &  DEC  & z&  i $[^{\circ}]$ stellar disk     &     $\rm{r_{d}}$ stellar disk [kpc]  &     PA $[^{\circ}]$ stellar disk  &   i $[^{\circ}]$ gas disk     &     $\rm{r_{d}}$ gas disk [kpc]  &     PA $[^{\circ}]$ gas disk & $ \rm{v_{max}}$ [km/s]  & $\rm{\sigma_{gas}}$ [km/s] & $\rm{log(M/M_{\odot})}$  & $\rm{log(SFR(M_{\odot}/yr)}$ & $\rm{r_{d, \:gas\:disk}/ FWHM_{MUSE-PSF}}$  \\                                                                                              
 \hline
\#1 & 64.13700 & -24.1367222 & 0.3803 & 36.2 &1.8& 23.5 & 36 &  3.45 & 236.8&  $127\pm13$ & $142.5 \pm 2.3$ & 8.39 & -0.434& 0.85  \\ 
\#2 & 64.1334167 & -24.1345278 & 0.3908&  64.05 &  0.74& 35.2& 47.1 & 1.79 & 278.6 & $15.5\pm2.0$ & $57.7 \pm 0.36$ & 9.45 &  0.472 & 0.44\\ 
\#3 & 64.1268333 & -24.1345278&  0.3909& 49.32&1.60 & 19.7& 19.7 & 1.80& 7.4& $99\pm9$ & $19.7 \pm1.66$ &8.99& -0.481& 0.45\\
\#4 & 64.137875& -24.1314444 &0.3811&39 & 0.85& 43.94&39 & 0.87&  83.4 & $64\pm16$ &  $141.5\pm1.7$ & 8.55&  -0.419& 0.21\\ 
\#5& 64.1279167& -24.1133056& 0.384& 61.8& 1.49& 41.5& 61.9&3.14&46.9&  $213\pm15$ & $178.4 \pm 1.4$ & 10.59  & -2.692& 1.06 \\
\#6&64.1325833 & -24.1231944& 0.4001&52.9&1.13&41.38&53&1.97&41.6&$14.5\pm3.2$&$31.7\pm1$&9.26&-0.073& 0.67\\
\#7&64.1447083& -24.0941944& 0.3807&62.37&0.306&58.33&50.9&0.84& 74.7& $ 43\pm34$ &  $164.15\pm 2.99$ &8.28&-0.548&  0.2\\
\#8&64.146625& -24.0943611& 0.3819&73.69&4.74&54.99&62.5&5.74&232.4&$251\pm9$& $138.09\pm3.24$&7.97&-1.742& 1.29\\
\#9& 64.1266667 & -24.1079167 &0.3847&  47.7&0.50 & 3.66&  45.8& 0.85& 166.9&   $18\pm7$&$33.43\pm1.78$&9.03&-1.299&0.34\\
\#10&64.1355417  &   -24.1065833 & 0.3957& 80.48&1.72&69.75&65.4&3.29&  158.9& $99\pm6$ &$32.59\pm1$&8.73&-0.405& 1.3\\
\#11&64.1352917 &-24.1171944 &0.4003& 53.8&0.363&54.38&60.4&1.15& 270.8&$20\pm8$&$20.45\pm 3.3$&8.49&-0.981& 0.45 \\
\#12&64.1382917& -24.1150833  & 0.4146&71.6 & 2.10&57.9&71.7&2.12&290.6&$16\pm7$&$20.18\pm3.5$&8.42&-3.76& 0.83 \\
\#13&64.1521250 & -24.1187778 & 0.3952&  67.0& 0.67&27.98&67.0&2.2&28.0&$13.2\pm1.6$&$23.66\pm 1.6$&9.05&-0.237&0.78\\
\#14-1&64.1520& -24.1193611 & 0.382 &72.1&1.74&  43.54&39.4&2.53&223.5& $129\pm11$&$134.83\pm2.3$&8.68&-0.1496&  0.90	\\
\#14-2&64.1520& -24.1193611 & 0.382 &78.8&1.55&65.65&87.6&1.89&223.9&$105\pm20$&$136.8\pm2.9$&8.68&-0.1496&0.67 	\\
\#15&64.1591667&  -24.1063056 &  0.4104&62.73& 1.43&17.3679& 66.7& 2.30& 180.4 &$10.6\pm1.5$& $38.49\pm0.6$& 9.26& -0.495& 0.85\\
\#16&64.1603750  &-24.1187222&  0.3847&69.26& 5.74& 4q.75&6.4& 3.56& 31.9& $346\pm6$& $12.9\pm1.5$& 9.74&0.631&	 1.20\\
\#17&64.1346667 & -24.0885833&  0.395&  50.7& 0.96&13.82&17.7& 1.17 &156.0& $54\pm14$& $22.4\pm1.4$& 9.26&-0.22 & 0.48 \\
\#1-f&64.1227917 &-24.1318889& 0.3517& 61.39 &8.66&62.85& 62.8&  6.52& 351.0& $66.7\pm1.6$& $54.11\pm0.4$& 10.65&1.36& 1.59\\
\#2-f&64.16366& -24.12165& 0.3052 &57.42&2.24&63.9&59.4&3.93&70.5&$83.1\pm0.7$&$46.27\pm0.3$&10.79&0.82& 1.31  \\
\#3-f&64.1434167& -24.1051389 &0.3428&39.77&10.07&109.75&42.6&9.30&321.9&$163.8\pm1.4$&$13.08\pm0.72$&$10.75$& 0.543&3.61  \\
\#4-f&64.1590833 &-24.0918056& 0.3439&83.42&1.84&23.21&89.6&3.43&196.5&$138.0\pm3.3$&$28.5\pm1$&9.650&-0.47& 0.76  \\
\#5-f&64.1283750 &-24.1023611& 0.2985&41.75&2.29&61.39&41.8&3.32&258.4&$70\pm5$&$29.25\pm0.55$&9.28&-0.387&1.03 \\
\hline

\end{tabular}
\end{adjustbox}
\label{tab}
\end{table*}

Table \ref{tab} summarizes the main physical properties of the sample of M0416 cluster and field galaxies. Column (1) displays the galaxy ID. Columns (2) and (3) show the  RA and DEC in [deg]. Column (4) displays the spectroscopic redshift, as derived using the \texttt{MARZ} tool from the MUSE data. Columns  (5) to (7) give the inclination in $[^{\circ}]$,  $\rm{r_{d}}$ in [kpc] and PA in $[^{\circ}]$ of the stellar disk, as derived with \texttt{Galfit} from the HST or HAWK-I data. Columns  (8) to (10) display the inclination in $[^{\circ}]$,  $\rm{r_{d}}$ in [kpc] and PA in $[^{\circ}]$ of the gas disk, as derived with \texttt{GalPaK3D} from the MUSE data. Columns (11) and (12) display the kinematical parameters,  $\rm{v_{max,\:gas}}$ and $\rm{\sigma_{gas}}$ in [km/s], as derived with \texttt{GalPaK3D} from the MUSE data. Columns (13) and (14)  display the Salpeter stellar masses and SFRs, as derived by \texttt{LePhare} from the two photometric catalogues. Column (15) shows the ratio between $\rm{r_{d, \:gas\:disk}}$, as derived with \texttt{GalPaK3D} from the MUSE IFU data,  and  $\rm{FWHM_{MUSE-PSF}}$, which differs from one pointing to the other, with a mean (seeing) value of  $\sim 0.6"$.

\end{appendix}


\begin{thebibliography}{}

\bibitem[Abril-Melgarejo et al.(2021)]{am}Abril-Melgarejo, V.,  Epinat, B., Mercier, W.,  Contini, T., et al. 2021, A\&A, 647A, 152A

\bibitem[Allen et al.(2008)]{allen08}Allen, M., G.,  Groves, B., A.,  Dopita, M., 2008, ApJS, 178, 20A

\bibitem[Anderson \& King (2000)]{epsf}Anderson, J.,  King, I. R.,  2000, PASP, 112, 1360A

\bibitem[Aquino-Ort{\'i}z et al.(2018)]{aq18}Aquino-Ort{\'i}z, E., Valenzuela, O., Sánchez, S. F., et al. 2018, MNRAS, 479, 2133

\bibitem[Aquino-Ort{\'i}z et al.(2020)]{aq}Aquino-Ort{\'i}z, E., S{\'a}nchez, S. F.,  Valenzuela, O., et al. 2020, ApJ, 900, 109A

\bibitem[Arnouts \& Ilbert (2011)]{lephare}Arnouts, S., \& Ilbert, O. 2011, Astrophysics Source Code Library, record ascl:1108.009


\bibitem[Bacon et al.(2010)]{bacon14}Bacon, R., Accardo, M., Adjali, L., et al. 2010, in SPIE Conf. Ser., 7735, 8

\bibitem[Bacon et al.(2016)]{mpdaf}Bacon, R., Piqueras, L., et al., 2016,ascl:1611.003

\bibitem[Bah{\'e} et al.(2013)]{bahe13} Bah{\'e}, Y. M., McCarthy, I. G., Balogh, M. L., \& Font, A. S., 2013, MNRAS, 430, 3017

\bibitem[Baldwin et al.(1981)]{bald81} Baldwin, J.~A., Phillips, M.~M., \& Terlevich, R.,  PASP, 93, 5


\bibitem[Balestra et al.(2016)]{balestra16} Balestra, I., Mercurio, A., Sartoris, B., et al. 2016,ApJ, S224, 33B

\bibitem[Bamford et al.(2005)]{Bamford}Bamford, S. P., Milvang-Jensen, B., et al.,  2005, MNRAS, 361, 109

\bibitem[Barat et al.(2019)]{barat}Barat, D., D’Eugenio, F., Colless, M., et al. 2019, MNRAS, 487, 2924

\bibitem[Behroozi et al.(2014)]{behroozi}Behroozi P. S.,  Wechsler R. H.,  Lu Y.,  Hahn O.,  Busha M. T.,  Klypin A.,  Primack J. R.. , ApJ, 2014, vol. 787 pg. 156

\bibitem[Belfiore et al.(2016)]{belfiore}Belfiore, F., Maiolino, R.,  Maraston, C.,  Emsellem, E., et al., 2016, MNRAS, 461, 3111B

\bibitem[Binney, Tremaine(1987)]{binney}Binney, J., \& Tremaine, S. 1987, Galactic Dynamics (Princeton: Princeton Univ. Press)

\bibitem[Bloom et al.(2017)]{bloom}Bloom, J. V., Croom, S. M., Bryant, J. J., et al. 2017, MNRAS, 472, 1809

\bibitem[Bouch{\'e} et al.(2015)]{galpak} Bouch{\'e}, N., Carfantan, H., Schroetter, I., et al., 2015, AJ, 150, 92

\bibitem[Bournaud et al.(2007) ]{clump}Bournaud, F ;  Elmegreen, B. et al., 2007, ApJ, 670, 237B

\bibitem[Bournaud et al.(2009)]{Bournaud} Bournaud, F., \& Elmegreen, B. G. 2009, ApJ, 694, L158

\bibitem[Boselli et al.(2006)]{boselli06}Boselli, A., Gavazzi, G,. 2006, PASP, 118, 517B

\bibitem[Boselli et al.(2022)]{boselli}Boselli, A.,  Fossati, M.,  Sun, M., 2022, A\&ARv, 30, 3B

\bibitem[Boselli et al.(2021b)]{boselli2}Boselli, A.,  Lupi, A., Epinat, B., Amram, P., Fossati, M., 2021, A\&A, 646A, 139B

\bibitem[B{\"o}hm et al.(2020)]{asmus}B{\"o}hm, A.,  Ziegler, B. L.,  P{\'e}rez-Mart{\'i}nez, J. M., et al., 2020, A\&A 633, A131 

\bibitem[Brammer et al.(2016)]{brammer}Brammer, G.l B.,  Marchesini, D.,  Labb{\'e}, I.,2016, ApJS, 226, 6B

\bibitem[Brinchmann et al.(2004)]{brinch}Brinchmann, J., Charlot, S., White, S. D. M., et al. 2004, MNRAS, 351, 1151

\bibitem[Brocklehurst et al.(1971)]{brock71} Brocklehurst, M.,1971, MNRAS.153, 471

\bibitem[Bruzual \& Charlot(2003)]{bc03}Bruzual, G.;  Charlot, S., 2003, MNRAS, 344, 1000B

\bibitem[Butcher \& Oemler(1978)]{BO}Butcher, H. ;  Oemler, A., Jr. 1978, ApJ, 219, 18B

\bibitem[Calzetti et al.(2001)]{calz} Calzetti, D.,  2001, PASP, 113, 1449C

\bibitem[Ciocan et al.(2020)]{eu}Ciocan, B. I.,  Maier, C.,  Ziegler, B. L.,  Verdugo, M., 2020, A\&A, 633A ,139C

\bibitem[Chabrier(2003)]{chabrier03} Chabrier G., 2003, PASP, 115, 763

\bibitem[Contini et al.(2016)]{contini16}Contini, T.,  Epinat, B.,  Bouche, N.,  Brinchmann, J, et al., 2016, A\&A, 591A, 49C

\bibitem[Cortese et al.(2014)]{cortese}Cortese L. et al. , 2014, ApJ, 795, L37 

\bibitem[Covington et al.(2010)]{Covington}Covington M. D. et al., 2010, ApJ, 710, 279

\bibitem[Cowie \& Songaila(1977)]{cowie}Cowie, L. L.,  Songaila, A., 1977, Natur, 266, 501C

\bibitem[den Brok et al.(2020)]{den Brok}den Brok, M., Carollo, C.-M.,   Erroz-Ferrer, S., Fagioli, M., et al., 2020, MNRAS. 491, 4089D

\bibitem[Desai et al.(2007)]{desai}Desai, V., Dalcanton, J. J., Arag{\'o}n-Salamanca, A., et al. 2007, ApJ, 660, 1151

\bibitem[Donahue et al.(2011)]{dh11}Donahue, M., de Messières, G., E.,  O'Connell, R., W.,  Voit, G. M., et al.,  2011, ApJ, 732, 40D

\bibitem[Dressler et al.(1980)]{dress} Dressler A. , ApJS, 1980, a42, 565

\bibitem[Dressler et al.(1997)]{dress2} Dressler, Alan; Oemler, Augustus, Jr et al. 1997, ApJ, 490, 577D.

\bibitem[Ebeling et al.(2001)]{macs}Ebeling H., Edge A.C., Henry J.P., et al., 2001, ApJ, 553, 668

\bibitem[Girardi et al.(1996)]{96}Girardi, L., Bressan, A., Chiosi, C., Bertelli, G., Nasi, E. 1996, A\&AS, 117,113

\bibitem[Gomes \& Papaderos(2017)]{fado}Gomes, J. M.;  Papaderos, P., 2017, A\&A,603A, 63G 

\bibitem[Gonzalez et al.(2020)]{gonzalez} Gonzalez, E. J.;  Chalela, M.;  Jauzac, M. ;  Eckert, D., et al., 2020, MNRAS, 494, 349G 

\bibitem[Gunn \& Gott (1972)]{rps}Gunn, J. E. \& Gott, III, J. R. 1972, ApJ, 176, 1

\bibitem[Han et al.(2018)]{han}Han, S.,  Smith, R., Choi, H.,  Cortese, L., et al., 2018, ApJ, 866, 78H

\bibitem[Haynes et al.(1999)]{Haynes}Haynes, M. P., Giovanelli, R., Chamaraux, P., et al. 1999, AJ, 117, 2039 

\bibitem[Heidmann et al.(1972)]{incl}Heidmann, J., Heidmann, N., \& de Vaucouleurs, G. 1972, MmRAS, 75, 85

\bibitem[Hern{\'a}ndez-Fern{\'a}ndez et al.(2012)]{her}Hern{\'a}ndez-Fern{\'a}ndez, Jonathan D. ;  V{\'i}lchez, J. M., et. al., 2012, ApJ, 751, 54H

\bibitem[Hinton et al.(2016)]{marz}Hinton, S. R., Davis,T. M., Lidman, C., Glazebrook, K.,  Lewis, G. F.,  2016, A\&C, 15, 61H

\bibitem[Hughes et al.(2013)]{hughes}Hughes, T. M., Cortese, L., Boselli, A., Gavazzi, G., Davies, J. I., 2013, A\&A, 550A, 115H

\bibitem[Hunter (2007)]{plot}Hunter, J. D. 2007, Computing In Science \& Engineering, 9, 90

\bibitem[Johnson et al.(2018)]{jh}Johnson, H. L.,  Harrison, C. M., Swinbank, A. M., et al., 2018, MNRAS, 474, 5076J

\bibitem[Kassin et al.(2007)]{kassin}Kassin, S. A., Weiner, B. J., Faber, S. M., et al. 2007, ApJL, 660, L35

\bibitem[Kassin et al.(2012)]{kassin12}Kassin S. A. et al., 2012, ApJ, 758, 106

\bibitem[Kauffmann et al.(2003)]{kaufm03} Kauffmann, G., Heckman, T. M., Tremonti. C. et al., 2003, MNRAS, 346, 1055

\bibitem[Kennicutt(1998b)]{ken98} Kennicutt, R., C., Jr. 1998, ARA\&A, 36, 189

\bibitem[Kewley et al.(2001)]{kewley01} Kewley, L.J. et al.,  2001, ApJS, 132, 37

\bibitem[Kewley et al.(2008)]{kew2008} Kewley, L. J., \& Ellison, S. L. 2008, ApJ, 681, 1183

\bibitem[Kewley et al.(2013)]{kewley13a} Kewley, L. J., Dopita, M. A., Leitherer, C., et al. 2013, ApJ, 774, 100

\bibitem[Kronberger et al.(2008)]{Kronberger}Kronberger, T.,  Kapferer, W., Unterguggenberger, S., Schindler, S.,  Ziegler, B. L, 2008, A\&A, 483, 783K 

\bibitem[Kuchner et al.(2017)]{uli}Kuchner, U.,   Ziegler, B.,  Verdugo, M., et al., 2017, A\&A, 604A, 54K

\bibitem[Kutdemir et al.(2008)]{kut2008}Kutdemir, E., Ziegler, B. L., Peletier, R. F., et al., 2008, A\&A, 488, 117K



\bibitem[Larson et al.(1980)]{larson}Larson, R. B., Tinsley, B. M., \& Caldwell, C. N. 1980, ApJ, 237, 692

\bibitem[Lilly et al.(2013)]{lilly13} Lilly et al., 2013, ApJ, 772, 119

\bibitem[Lin et al.(2017)]{lin}Lin Y.,T., Hsieh B.,C., Lin S.,C., et al. 2017 ApJ 851 139

\bibitem[Loubser et al.(2013) ]{l13} Loubser, S. I.; Soechting, I. K., 2013, MNRAS, 431.2933L


\bibitem[Lotz et al.(2017)]{lotz17}Lotz, J. M.;  Koekemoer, A.;  Coe, D.,  et al.,  2017, ApJ, 837, 97L 

\bibitem[Lu et al.(2012)]{lu}Lu T.,  Gilbank D. G.,  McGee S. L.,  Balogh M. L.,  Gallagher S.. , MNRAS, 2012, vol. 420 pg. 126 

\bibitem[Maier et al.(2016)]{maier16}Maier, C. ;  Kuchner, U. ;  Ziegler, B. L., 2016, A\&A,590A,108M

\bibitem[Maier et al.(2019)]{maier19}Maier, C.,  Ziegler, B. L., Haines, C. P., Smith, G. P., 2019, A\&A, 621A, 131M

\bibitem[Maier et al.(2022)]{maier22}Maier, C.,  Haines, C. P.,   Ziegler, B. L., 2022, A\&A, 658A.190M

 \bibitem[Mannucci et al.(2010)]{mannu10} Mannucci et al. 2010, MNRAS, 408, 2115


\bibitem[McDonald et al.(2012)]{mc12}McDonald, M., Veilleux, S., \& Rupke, D. S. N. 2012, ApJ, 746,153 


\bibitem[McGaugh et al.(2010)]{mcgaugh}McGaugh, S., S.,  Wolf, J., 2010, ApJ, 722, 248M

\bibitem[McGee et al.(2009)]{mcgee}McGee S. L., Balogh M. L., Bower R. G., Font A. S., McCarthy I. G., 2009, MNRAS, 400, 937


\bibitem[Miller et al.(2011)]{miller11}Miller, S. H., Bundy, K., Sullivan, M., Ellis, R. S., \& Treu, T. 2011, ApJ, 741,115

\bibitem[Miller et al.(2014)]{miller14}Miller, S. H., Ellis, R. S., Newman, A. B., \& Benson, A. 2014, ApJ, 782, 115 


\bibitem[Moore et al.(1998)]{harassment}Moore, B., Lake, G., \& Katz, N. 1998, ApJ, 495, 139

\bibitem[Moran et al.(2007)]{Moran}Moran, S. M., Miller, N., Treu, T., Ellis, E. S., \& Smith, G. P. 2007, ApJ, 659, 1138


\bibitem[Olsson et al.(2010)]{olsson10}Olsson, E., Aalto, S., Thomasson, M., \& Beswick, R. 2010, A\&A, 513, A11

\bibitem[Papaderos et al.(2013)]{polis13}Papaderos, P.,  Gomes, J. M.,  Vílchez, J. M.,  Kehrig, C., et al.  2013, A\&A, 555, L1 

\bibitem[Peng et al.(2010a)]{galfit}Peng, C. Y., Ho, L. C., Impey, C. D., \& Rix, H.-W. 2010, AJ, 139, 2097

\bibitem[Peng al.(2010b)]{peng10} Peng, Y.-j., Lilly, S.~J., Kova{\v c}, K., et al., 2010, ApJ, 721, 193 

\bibitem[Peng et al.(2015)]{peng15}Peng, Y.,  Maiolino, R.,  Cochrane, R., 2015, Natur, 521, 192P


\bibitem[P{\'e}rez-Mart{\'i}nez et al.(2017)]{jose1}Pérez-Martínez, J. M., Ziegler, B. L., Verdugo, M., Böhm, A., \& Tanaka, M. 2017, A\&A, 605, A127

\bibitem[P{\'e}rez-Mart{\'i}nez et al.(2020)]{jose2}Pérez-Martínez, J. M., Ziegler, B. L.,  Böhm, A.,Verdugo, M.,2020, A\&A, 637A, 30P

\bibitem[P{\'e}rez-Mart{\'i}nez et al.(2021)]{jose}P{\'e}rez-Mart{\'i}nez, J. M., Ziegler, B.,  Dannerbauer, H. et al., 2021, A\&A, 646A, 53P

\bibitem[Pintos-Castro et al.(2019)]{pc}Pintos-Castro, I., Yee, H. K. C., Muzzin, A., Old, L., \&Wilson, G. 2019, ApJ, 876, 40,

\bibitem[Postman et al.(2012)]{postman12} Postman, M., Coe, M., Benitez, N., et al. 2012, ApJS, 199, 25

\bibitem[Pozzetti et al.(2007)]{pozzetti07} Pozzetti, L., Bolzonella, M., Lamareille, F., et al., 2007, A \& A, 474, 443


\bibitem [Roberts et al.(2019)]{roberts19} Roberts, I. D, Parker, L. C, Brown, T., et al., 2019, ApJ, 873, 42R

\bibitem[ Robitaille \& Tollerud(2013)]{astropy} Robitaille, T. P., Tollerud, E. J., et al. 2013, A\&A, 558, A33

\bibitem[Rodrigues et al.(2017)]{pastuff}Rodrigues, M., Hammer, F., Flores, H., Puech, M., \& Athanassoula, E. 2017,MNRAS, 465, 1157

\bibitem[Rosati et al.(2014)]{rosati14} Rosati, P., Balestra, I., Grillo, C., et al.,  2014, The Messenger, 158, 48

\bibitem[Ryder \& Dopita(1994)]{rd}Ryder, Stuart D.,  Dopita, Michael A., 1994, ApJ, 430, 142R


\bibitem[Salpeter(1955)]{salpeter}Salpeter E. E., 1955, ApJ, 121, 161

\bibitem[Schawinski et al.(2007)]{sw17} Schawinski, K., Thomas, D., 2007, MNRAS, 382, 1415

\bibitem[Schommer et al.(1993)]{schommer}Schommer, R. A., Bothun, G. D., Williams, T. B., \& Mould, J. R. 1993, AJ, 105, 97

\bibitem[Sersic (1968)]{sers}Sersic, J. L. 1968, Atlas de galaxias australes (Cordoba, Argentina: Obs. Astron. Publ.)

\bibitem[Sharp \& Bland-Hawthorn(2010)]{sharp10}Sharp \& Bland-Hawthorn 2010, ApJ 711, 818

\bibitem[Shipley et al.(2018)]{shipley18}Shipley, H. V.,   Lange-Vagle, D., Marchesini, D., et al 2018, ApJ, S235, 14S 


\bibitem[Simons et al.(2015)]{simons}Simons, R. C., Kassin, S. A., Weiner, B. J., et al. 2015, MNRAS, 452, 986


\bibitem[Soto et al.(2016)]{soto}Soto K. T., Lilly S. J., et al., 2016, MNRAS, 458, 3210

\bibitem[Sparks et al.(2012)]{sparks12}Sparks, W. B., Pringle, J. E.,   Carswell, R. F.,  Donahue, M., et al.,  2012, ApJ, 750L, 5S 


\bibitem[Steinhauser et al.(2012)]{steinh}Steinhauser, D., Haider, M., Kapferer, W., \& Schindler, S. 2012, A\&A, 544, A54

\bibitem[Steyrleithner et al.(2020)]{st}Steyrleithner, P., Hensler, G., \& Boselli, A. 2020, MNRAS, 494, 1114


\bibitem[Tan et al.(2022)]{tan}Tan, V., Y., Y.,   Muzzin, A.,  Marsan, Z., C., et al. 2022, arXiv220507913T

\bibitem[Toloba et al.(2009)]{tobala}Toloba, E., Boselli, A., Gorgas, J., Peletier, R. F., et al, 2009, ApJ, 707L, 17T

\bibitem[Tremonti et al.(2004)]{trem04} Tremonti, C. A., Heckman, T. M., Kauffmann, G., et al., 2004, ApJ, 613, 898

\bibitem[Tully \& Fisher(1977)]{t-f}Tully, R. B., \& Fisher, J. R. 1977, A\&A, 54, 661

\bibitem[Tully et al.(1998)]{tully}Tully, R. B., Pierce, M. J., Huang, J.-S., et al. 1998, AJ, 115, 2264

\bibitem[van der Walt et al.(2011)]{numpy}van der Walt, S., Colbert, S. C., \& Varoquaux, G. 2011, Computing in Science Engineering, 13, 22


\bibitem[Verdugo et al.(2008)]{miguel}Verdugo, M., Ziegler, B. L., \& Gerken, B. 2008, A\&A, 486, 9

\bibitem[Verheijen(2001)]{Verheijen} Verheijen, M. A. W., 2001, ApJ, 563, 694V

\bibitem[Vijayaraghavan et al.(2013)]{Vijayaraghavan}Vijayaraghavan, R., Ricker, P. M., 2013, MNRAS, 435, 2713V

\bibitem[Vulcani et al.(2020)]{vulcani}Vulcani, B., Poggianti, B., M., Tonnesen, S.,  McGee, S., L., et al., 2020, ApJ, 899, 98V

\bibitem[Weilbacher et al.(2020)]{pipeline}Weilbacher P. M., Palsa, R., Streicher O., et al., 2020, arXiv:200608638W 

\bibitem[Weiner et al.(2006)]{weiner}Weiner, B. J., Willmer, C. N. A., Faber, S. M., et al. 2006, ApJ, 653, 1027

\bibitem[Wetzel et al.(2013)]{wetzel}Wetzel A. R. Tinker J. L. Conroy C. van den Bosch F. C . MNRAS., 2013, 432, 336

\bibitem[Whitaker et al.(2014))]{whitaker}Whitaker, K. E., Franx, M., Leja, J., et al. 2014, ApJ, 795, 104

\bibitem[Ziegler et al. (2003)]{bodo}Ziegler, B. L., B{\"o}hm, A., J{\"a}ger, K., et al., 2003, ApJ, 598,L87

\end{thebibliography}
\end{document}